\documentclass[superscriptaddress,nobibnotes,amsmath,amssymb,notitlepage,twocolumn,pra,reprint]{revtex4-1}

\usepackage[colorlinks=true,linkcolor=blue,citecolor=blue,urlcolor=blue]{hyperref}

\usepackage{setspace} 
\usepackage{graphicx} \graphicspath{{./img/}} 
\usepackage{amsmath}
\usepackage{color}
\usepackage{amsmath}
\usepackage{amssymb}
\usepackage{verbatim}
\usepackage{latexsym}
\usepackage{enumerate} 
\usepackage{bm} 
\usepackage{ulem} 

\begin{document}

\title{
Monolayer C$_{60}$ networks: A first-principles perspective
}



\newcommand{\TCM}{Theory of Condensed Matter Group, Cavendish Laboratory, University of Cambridge, J.\,J.\,Thomson Avenue, Cambridge CB3 0HE, UK}
\newcommand{\HarvardFAS}{Harvard University, Cambridge, Massachusetts 02138, USA}


 \author{Bo Peng}
 \email{bp432@cam.ac.uk}
 \affiliation{Theory of Condensed Matter Group, Cavendish Laboratory, University of Cambridge, Cambridge CB3 0HE, United Kingdom}

 \author{Michele Pizzochero}
 \email{mp2834@bath.ac.uk}
 \affiliation{Department of Physics, University of Bath, Bath BA2 7AY, United Kingdom}
 \affiliation{School of Engineering and Applied Sciences, Harvard University, Cambridge, Massachusetts 02138, United States}

\date{\today}

\begin{abstract}
Monolayer fullerene (C$_{60}$) networks combine molecular-level rigidity with crystalline connectivity, offering a promising platform for numerous applications. In this Feature article, we review the physical and chemical properties of fullerene monolayers, focusing on first-principles studies. We first explore the structural stability of monolayer phases and investigate their thermal expansion behaviours. We then outline criteria for photocatalytic water splitting and introduce theoretical predictions which are supported by recent experimental verification. Finally, we show how interlayer stacking, molecular size, and dimensional tuning (from 2D monolayers into 3D crystals, 1D chains, or nanoribbons) offer versatile approaches to modulate their chemical functionality. Together, these insights establish fullerene networks as a novel class of carbon-based materials with tailored properties for catalysis, photovoltaics, and flexible electronics. 
\end{abstract}

\maketitle



\section{Introduction}

\begin{figure}
    \centering
    \includegraphics[width=\linewidth]{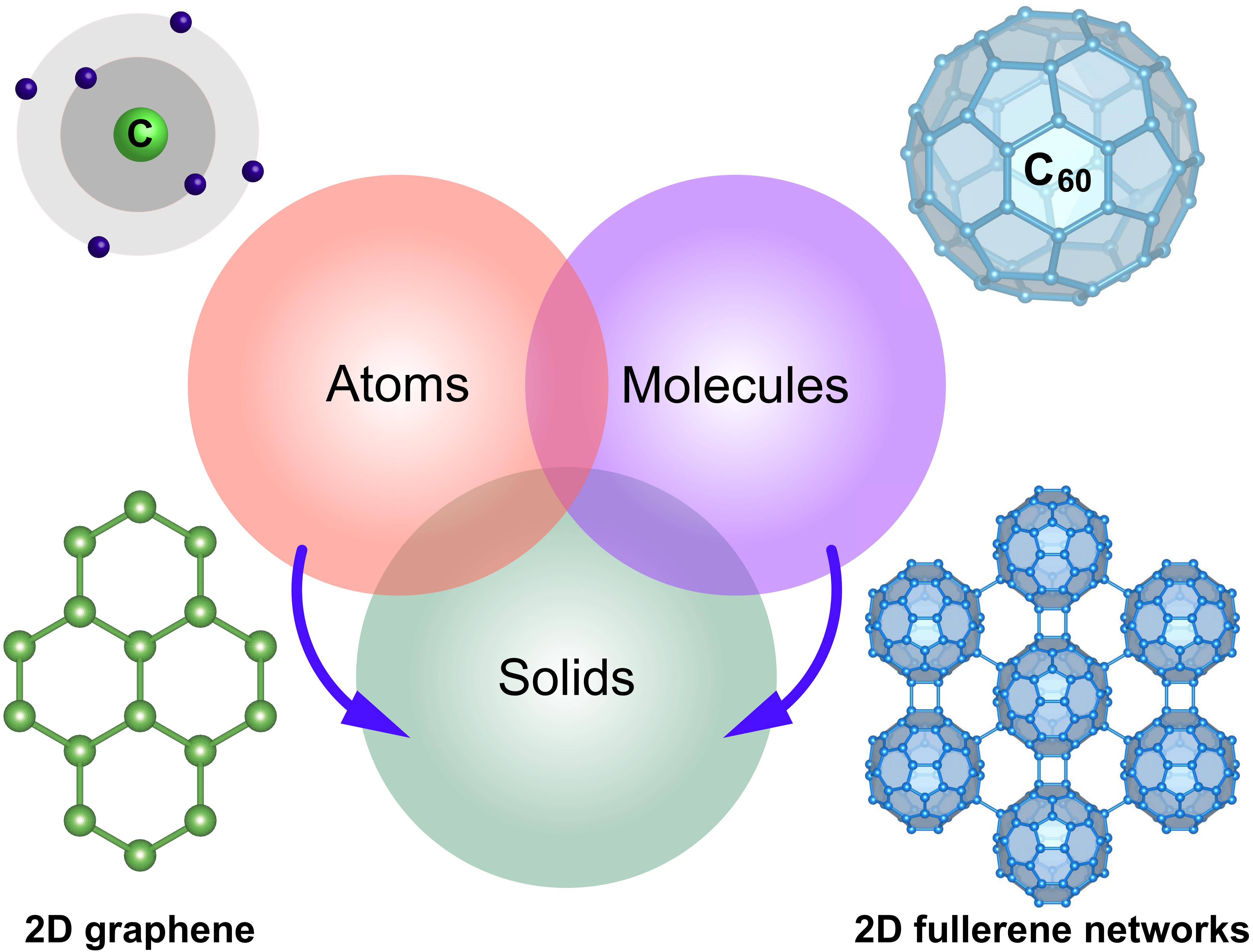}
    \caption{Carbon atoms and fullerene molecules as basic building units for graphene and monolayer fullerene networks respectively. 
    }
    \label{fig:schematics}
\end{figure}

Since the discovery of graphene in 2004\,\cite{Novoselov2004}, 2D materials have attracted tremendous interest due to their extraordinary electronic\,\cite{Novoselov2005,Zhang2005,Geim2009,Novoselov2012}, magnetic\,\cite{Gong2017,Huang2017,Deng2018,Pizzochero2020,Pizzochero2020a}, optical\,\cite{Zhang2015e,Peng2016d,Peng2018c}, topological\,\cite{Kane2005,Wada2011,Liu2011,Cai2015,Zhu2015}, and thermal\,\cite{Balandin2008,Seol2010,Peng2016b,Peng2016c,Peng2016e,Chen2019} properties. The basic building blocks of these 2D materials are atoms. Consider, for example, graphene: carbon atoms form a planar honeycomb lattice with $sp^2$ hybridisation from three planar $\sigma$ bonds and the out-of-plane $\pi$ bonding\,\cite{Sahin2009}. By varying constituent atoms (e.g., from C to Si or Sn), the lattice composition and crystalline symmetry can be controlled at the atomic scale\,\cite{Cahangirov2009,OHare2012,Roome2014,Peng2016a,Peng2016f,Peng2017,Peng2018b}, leading to tailored functionalities such as thermoelectricity\,\cite{Peng2018a,Peng2019a} and non-Abelian braiding\,\cite{Peng2022,Peng2022a}. Additionally, structural phase transitions can be induced by external external stimuli such as temperature and light\,\cite{Cho2015a,Peng2020,Gou2023,Peng2024}, enabling precisely tuneable physical and chemical properties. The crystal structures can be further manipulated by defect engineering\,\cite{Qiu2013,Peng2016g,Pizzochero2018,Pizzochero2020b}, offering further degrees of freedom for applications such as single-photon sources\,\cite{Li2022b,Trainer2022,Li2025}. Despite their advantages such as atom-by-atom precision, the designed 2D crystals from atomic building blocks are constrained by their intrinsic characteristics such as limited types of chemical bonds and restricted stability as freestanding monolayers\,\cite{Peng2017a}. 
Moreover, precise atomic construction remains challenging\cite{Mannix2017}. As an alternative, replacing atoms with molecules as building units of crystals (Fig.\,\ref{fig:schematics}) results in the combination of molecular rigidity and chemical tuneability, offering new opportunities in rational materials design for tailored electronic\,\cite{Niu2023,Qie2024}, optical\,\cite{Dubey2021}, chemical\,\cite{Niu2023a}, and magnetic\,\cite{Su2025,Fu2025} properties. 

The discovery of buckminsterfullerene C$_{60}$ in 1985 has provided an atom-like, stable building units\,\cite{Kroto1985,Kroto1988,Kroto1990,Lamb1993} to form various lattice structures, as demonstrated by experimental synthesis of solid C$_{60}$ in 1990-92 where fullerene molecules are bounded by van der Waals interactions with orientational ordering transition upon heating\,\cite{Kratschmer1990,Heiney1991,Fischer1991,David1991,Heiney1992}. A series of follow-up studies in 1993-96 have found that neighbouring C$_{60}$ cages can be connected through covalent [2\,+\,2] cycloaddition bonds as a result of photo- or pressure-induced polymerisation\,\cite{Rao1993,Iwasa1994,Nunez-Regueiro1995,Eklund1995,Xu1995,Springborg1995,Marques1996}, forming a rich phase diagram of fullerene polymers\,\cite{Giacalone2006,Murga2015}. These fullerene crystals exhibit distinct mechanical\,\cite{Ruoff1991}, optical\,\cite{Venkateswaran1996}, vibrational\,\cite{Rao1997}, electronic\,\cite{Forro2001,Makarova2001,Sun2005}, and polaronic\,\cite{Belosludov2003,Belosludov2006} properties, leading to numerous applications such as photovoltaics\,\cite{Kennedy2008,Dennler2009} and superconductivity\,\cite{Stephens1991,Chauvet1994,Stephens1994,Gunnarsson1997,Huq2001}. Recent breakthroughs in synthesis of monolayer fullerene networks in 2022 have demonstrated the feasibility of creating 2D materials based on C$_{60}$ molecules\,\cite{Hou2022}. Different from graphene with limited chemical bonding types of carbon atoms, fullerene molecules provide abundant intermolecular bonding positions with extra rotational degrees of freedom to form richer lattice geometries. As shown in Fig.\,\ref{fig:schematics}, neighbouring carbon cages can be linked through both intermolecular [2\,+\,2] cycloaddition bonds and C$-$C single bonds, resulting in a nearly-triangular lattice with impossible coordination environments for individual carbon atoms. Similarly, it is possible to realise 1D, nearly-square, and nearly-hexagonal lattices in monolayer C$_{60}$ networks. The diverse lattice geometries, delocalised $\pi$ electrons, and large surface area in C$_{60}$ monolayers endow them with promising electronic, optical, and chemical properties for energy applications\,\cite{Peng2022c,Jones2023,Wu2025,Shearsby2025}. Their sizable band gaps and high charge transport are critical for use in photocatalytic water splitting, organic photovoltaics, and flexible energy storage devices. Moreover, the presence of abundant surface active sites offers opportunities for catalytic reactions such as the hydrogen evolution reaction (HER) and nitrogen reduction reaction.



Despite the significant experimental progress in the synthesis and characterisation of 2D fullerene networks\,\cite{Hou2022,Meirzadeh2023}, a comprehensive understanding of their physical and chemical properties is still lacking. To this end, modern computational techniques can close this gap by providing atomistic resolution on chemical functionalities of monolayer polymeric C$_{60}$. Here, we highlight three key aspects from our recent research based on first-principles calculations: (1) Are monolayer C$_{60}$ networks stable? (2) Are they promising photocatalysts? (3) Are their chemical functionalities tuneable? To answer these questions, our Feature article is organized as follows: In Section\,\ref{sect:stable}, we summarise the structural properties of C$_{60}$ monolayers, their stability and strength, as well as their thermal expansion behaviours. In Section\,\ref{sect:chem}, we examine the criteria for 2D C$_{60}$ networks to be efficient photocatalysts and show how theoretical predictions lead to experimental verifications. In Section\,\ref{sect:tune}, we discuss possible strategies in tuning chemical functionalities of fullerene networks. Finally, we conclude by identifying potential avenues for future research in Section\,\ref{sect:conclusion}.

\section{Are monolayer C$_{60}$ networks stable?}\label{sect:stable}

In this Section, we address the structural stability of monolayer fullerene networks from first principles. We start with an introduction to the structural properties of the experimentally-reported crystal structures. Next, we discuss their stability and strength, including dynamic, thermodynamic, and mechanical properties to rationalise experimental observations. Lastly, we explore the dimensional stability of monolayer fullerene networks upon heating in the context of thermal expansion.

\subsection{Crystal Structures}

\begin{figure}
    \centering
    \includegraphics[width=\linewidth]{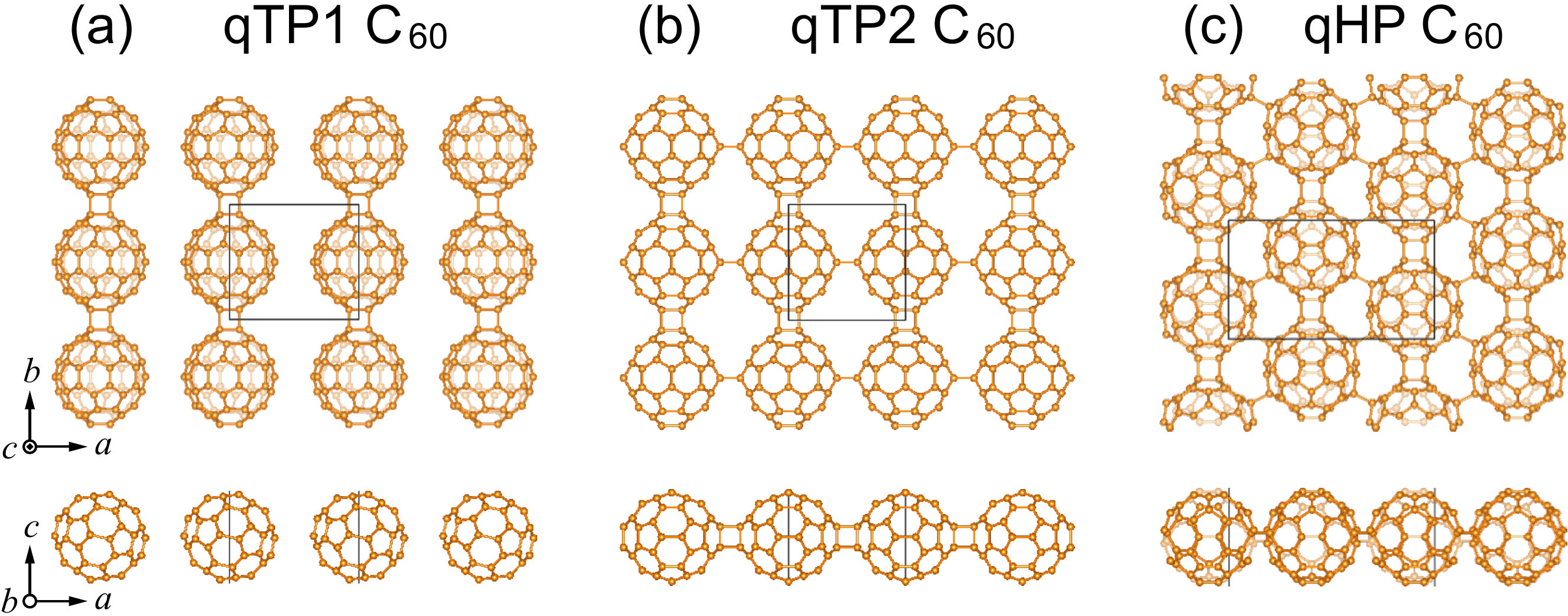}
    \caption{Crystal structures of (a) qTP1, (b) qTP2, and (c) qHP monolayer polymeric C$_{60}$ from top and front views\,\cite{Peng2022c}. 
    }
    \label{fig:crystal}
\end{figure}

The crystal structures of monolayer fullerene networks reveal a remarkable diversity in intermolecular bonding environments, stemming from the intrinsic rotational degrees of freedom of C$_{60}$ molecules and their multiple intermolecular bonding sites. Starting from the experimentally-reported structures\,\cite{Hou2022}, three distinct crystalline phases are obtained from geometry optimisation based on density functional theory (DFT)\,\cite{Hohenberg1964,Kohn1965}, as overviewed in Fig.\,\ref{fig:crystal}: A quasi-tetragonal phase with one-dimensional chains along $b$ where C$_{60}$ units are connected through [2\,+\,2] cycloaddition bonds (denoted as qTP1), a quasi-tetragonal phase in two-dimensional nearly-square lattices with vertical and horizontal [2\,+\,2] cycloaddition bonds along $a$ and $b$ respectively (denoted as qTP2), and a quasi-hexagonal phase in a nearly triangular lattice via [2\,+\,2] cycloaddition bonds along $b$ and C$-$C single bonds along the diagonal of the $a$ and $b$ lattice vectors (denoted as qHP). 

All three networks retain the molecular curvature of individual fullerenes, but differ significantly in packing density, bond connectivity, and unit cell geometry, leading to distinct phonon spectra and elastic behaviours. Covalent polymerisation of qTP1 C$_{60}$ molecules along $b$ yields quasi-one-dimensional crystalline networks without interchain bond formation. These arrangements result in anisotropic lattice parameters and weak interchain bonding. Monolayer qTP2 C$_{60}$ networks possess a more symmetric structure of space group $Pmmm$ (No.\,47) with enhanced in-plane isotropy and more robust connectivity. As the only networks with intermolecular C$-$C single bonds, the qHP monolayers exhibit nearly isotropic packing density but strong anisotropic intermolecular bonds, which impose anisotropic structural, elastic, and vibrational constraints on mechanical stiffness and thermal expansion, as discussed below.

\subsection{Stability and strength}

Comparing to the well-understood formation mechanisms and stability of fullerene molecules\,\cite{Kroto1987,Goroff1996,Bernal2019}, when assembling C$_{60}$ units into 2D configurations, it remains unclear whether ordered monolayer polymeric C$_{60}$ structures are stable under ambient conditions. Experimentally, only qHP C$_{60}$ has been obtained as both monolayers\,\cite{Hou2022} and few-layers\,\cite{Meirzadeh2023,Wang2023,Chen2024,Zhang2025}, while all the qTP C$_{60}$ flakes are few-layer\,\cite{Hou2022}. These results raise doubts regarding the stability of monolayer fullerene networks. Previous first-principles studies have examined a range of structural phases of monolayer C$_{60}$ and confirmed the mechanical stability of several configurations\,\cite{Yu2022,Tromer2022,Ying2023}. Additionally, molecular dynamics simulations have addressed their thermal stability, indicating that both qTP and qHP structures remain stable up to 800\,K\,\cite{Ribeiro2022}, partially corroborating experimental findings that qHP monolayers do not decompose at 600\,K\,\cite{Hou2022}. Nevertheless, the absence of experimental exfoliation of qTP monolayers raises unresolved questions, suggesting that thermal and mechanical stability alone may be insufficient to explain their phase behaviour. 

To access the structural stability, a systematic investigation on dynamic, thermodynamic, and mechanical stability is required\,\cite{Zhang2012,Peng2017a}. In the following, we show how first-principles simulations can be used to address the following questions\,\cite{Peng2023}: 

($i$) Are qTP and qHP C$_{60}$ monolayers dynamically stable? 

($ii$) What is their relative thermodynamic stability? 

($iii$) Do their mechanical properties support phase stability?

\subsubsection{Dynamic stability}

\begin{figure}
    \centering
    \includegraphics[width=\linewidth]{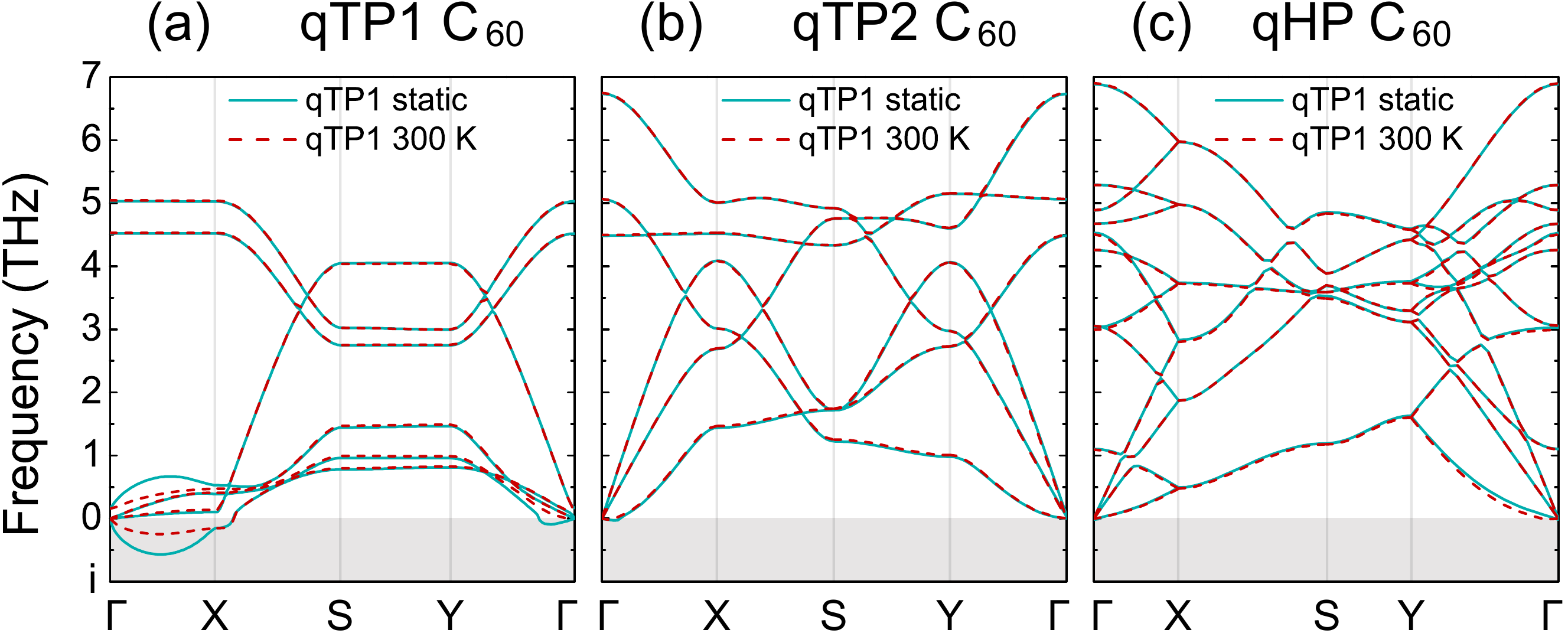}
    \caption{Low-frequency phonons of (a) qTP1, (b) qTP2, and (c) qHP polymeric C$_{60}$ using the static and room-temperature lattice constants\,\cite{Peng2023}. 
    }
    \label{fig:phonon}
\end{figure}

The dynamic stability is determined by the absence of imaginary phonon modes, which reflects whether the structure resides at a local minimum on the potential energy surface\,\cite{Malyi2019,Luo2022,Pallikara2022}. To assess the dynamic stability of monolayer fullerene networks, the phonon dispersion relations are shown in Fig.\,\ref{fig:phonon}. The dark blue solid curves correspond to phonons under the harmonic approximation based on density functional perturbation theory (DFPT)\,\cite{DFPT,Gonze1995,Gonze1995a}, while the wine dashed lines represent phonons computed from the quasi-harmonic approximation\,\cite{Dove1993,Togo2008,Togo2015,Huang2015a,Huang2016,Peng2019,Shaikh2025}.

For the qTP1 monolayers, small imaginary frequencies (less than 0.6i\,THz) are observed along the $\Gamma$--X direction when using static lattice constants, suggesting non-restorative atomic displacements\,\cite{Pallikara2022} to split the monolayers into individual 1D chains in the presence of interchain out-of-plane vibrations. The presence of a fourth, nearly zero-frequency, positive torsional mode along $\Gamma$--X further reflects the quasi-one-dimensional nature of the qTP1 structure\,\cite{Peng2018,Lange2022}. Incorporating finite-temperature effects, we find that at 300\,K, qTP1 retains this soft mode behaviour, albeit with a reduced magnitude (< 0.2i\,THz), thus confirming its weak dynamic stability. In contrast, the qTP2 and qHP lattices exhibit no imaginary phonon modes under either static or thermally-expanded lattice parameters, indicating that both structures reside at a local minimum on the potential energy surface and are dynamically stable over a wide temperature range.

\subsubsection{Thermodynamic stability}

\begin{figure}
    \centering
    \includegraphics[width=0.8\linewidth]{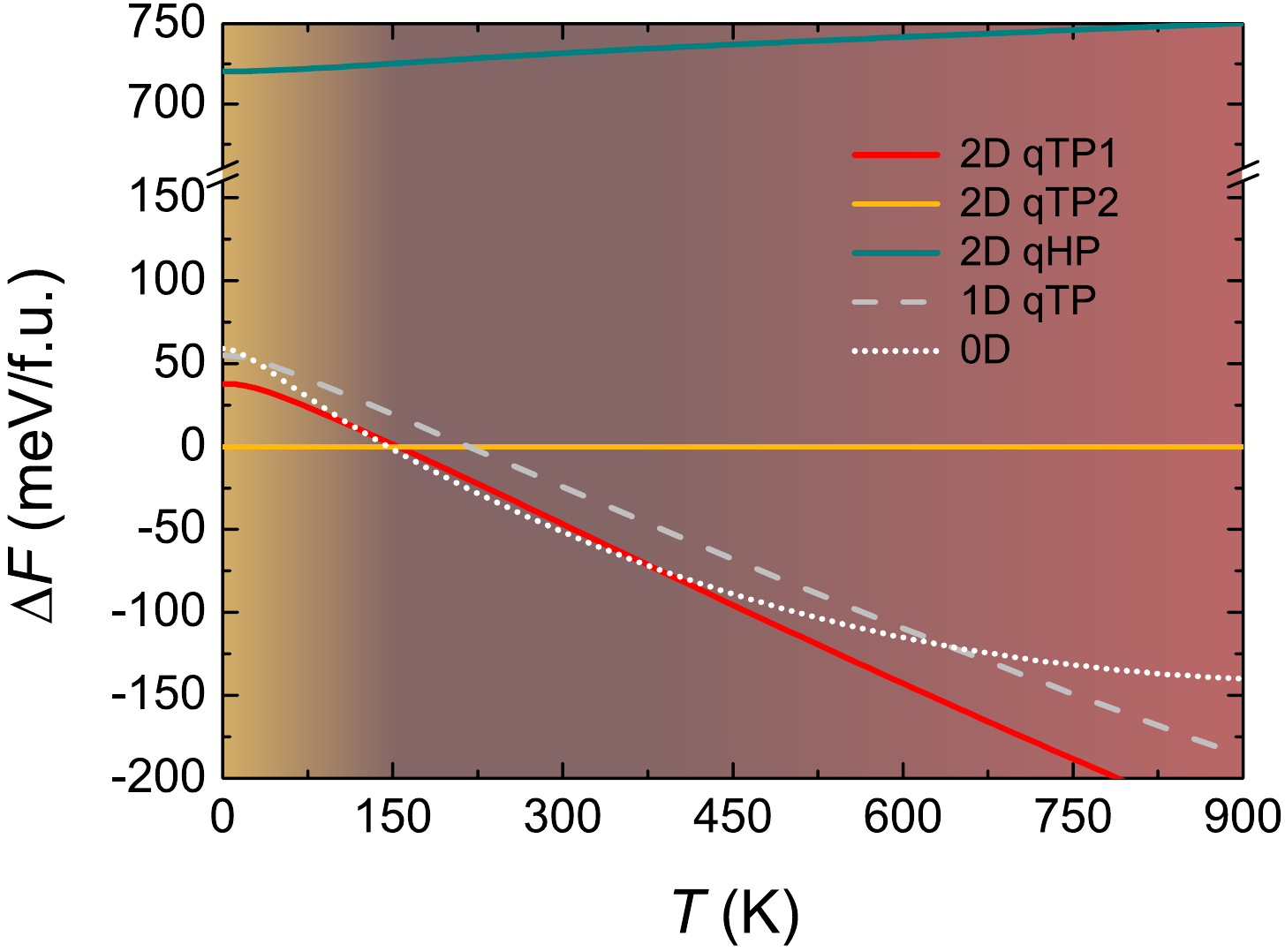}
    \caption{Relative thermodynamic stabilities of monolayer fullerene networks, a one-dimensional fullerene chain, and a zero-dimensional fullerene molecule, with the Gibbs free energy F of monolayer qTP2 C$_{60}$ set to zero to compare the relative stabilities\,\cite{Peng2023}. 
    }
    \label{fig:free}
\end{figure}

The relative thermodynamic stability is quantified through free-energy considerations at finite temperatures\,\cite{Pavone1998,Pavone2001,Masago2006,Setten2007,Stoffel2010,Zhang2011a,Deringer2014,Deringer2014a,White2015,Nyman2016,Skelton2017,Pallikara2021,Bartel2022}. The Gibbs free energy $F$ can be calculated from\,\cite{Dove1993,Togo2008,Togo2015}
	\begin{equation}
	\label{helmholtz}
			F= \mathop{\min}_{a,b} \bigg[
			E_{\mathrm{ tot}}
			+\frac{1}{2}\sum_{\lambda}\hbar \omega_{\lambda}
			+k_BT\sum_{\lambda}
			\ln{\big( 1- \mathrm{e} ^{-\hbar \omega_{\lambda}/k_BT} \big)
			}
			\bigg],
	\end{equation}
where $\mathop{\min}_{a,b} [\ ]$ changes the lattice constants $a$ and $b$ to find the unique minimum value of the Helmholtz free energy, $E_{\mathrm{ tot}}$ is the total energy of the crystal, $\hbar$ is the reduced Planck constant, $\omega_{\lambda}$ is the phonon frequency at mode $\lambda$, $k_B$ is the Boltzmann constant, and $T$ is the temperature. 
As shown in Fig.\,\ref{fig:free}, the temperature dependence of the Gibbs free energy reveals a crossover in phase stability. Although qTP2 is most stable at 0\,K, qTP1 becomes thermodynamically preferred above 150\,K due to its lower vibrational frequencies and hence greater entropy. At room temperature (300\,K), the free energy of qTP1 is 47\,meV/f.u. lower than that of qTP2, and this energetic preference increases at higher temperatures, reinforcing the relative thermodynamic stability of qTP1 at high temperatures.

As the qTP1 monolayer tends to split into individual chains, the relative thermodynamic stability of 1D qTP chain and 0D C$_{60}$ molecule is also shown in Fig.\,\ref{fig:free}. Notably, while the 2D qTP1 structure remains more stable than the 1D chain throughout the entire temperature range, the 1D chain becomes more stable than the qTP2 monolayer above 220\,K. Interestingly, the 0D C$_{60}$ molecule shows the steepest decrease in free energy below room temperature, becoming more stable than qTP1 and qTP2 monolayers at 120\,K and 150\,K, respectively. Above 380\,K, however, qTP1 regains its dominance as the most stable phase. Notably, the energy difference between the 1D and 2D qTP1 lattices is comparable to the thermal fluctuation energy ($k_B T$) at 300\,K of $\sim$\,26\,meV, implying the possibility for qHP1 monolayers to split into 1D chains upon thermal fluctuations. These results highlight a temperature dependence for the thermodynamically most stable phase: below 150\,K, qTP2 is preferred; between 150 and 380\,K, the 0D C$_{60}$ molecule is energetically favoured; and above 380\,K, the 2D qTP1 monolayer becomes thermodynamically dominant.

\subsubsection{Mechanical stability}

The mechanical stability of monolayer fullerene networks is assessed through the elastic constants,\cite{Peng2023} determined from the finite-difference method\,\cite{LePage2002,Wu2005} and summarised in Table\,\ref{tbl:elastic}. These values are which are consistent with previous computational studies\,\cite{Yu2022,Tromer2022}. According to the Born-Huang lattice dynamical theory\,\cite{Born1954,Wu2007}, the mechanical stability criteria are fulfilled for qTP2 and qHP with space group $Pmmm$ (No.\,47) and $Pmna$ (No.\,53) respectively, satisfying the conditions specific to orthorhombic lattices. In contrast, monoclinic qTP1 C$_{60}$ in space group $P2/m$ (No.\,10) exhibits a negative shear modulus ($C_{66} < 0$), indicating intrinsic mechanical instability against shear deformation. This is attributed to the weak interchain interactions in its quasi-one-dimensional structure, which are prone to bending and sliding under shear stress.

\begin{table}
    \caption{
    Elastic properties for monolayer qTP1, qTP2, and qHP C$_{60}$ with the elastic constants $C_{ij}$. The elastic constants $C_{11}$, $C_{22}$ and $C_{66}$ calculated from phonon speed of sound are listed in parentheses for comparison\,\cite{Peng2023}.
    } 
    \centering
    \begin{tabular}{ccccc}
    \hline
    Phase & $C_{11}$ & $C_{22}$ & $C_{12}$ & $C_{66}$ \\ 
    \hline
    qTP1 & 5.4 & 123.7 & --1.2 & --0.2 \\
 & (2.5) & (121.3) & - &  \\
    qTP2 & 149.9 & 148.7 & 22.9 & 53.4 \\
 & (150.5) & (141.2) & - & (54.5) \\
    qHP  & 150.8 & 186.8 & 22.5 & 60.6 \\
 & (142.4) & (172.7) & - & (61.7) \\
    \hline
    \end{tabular}  
    \label{tbl:elastic} 
\end{table}

Notably, the elastic constant $C_{11}$ in qTP1 is more than an order of magnitude smaller than $C_{22}$, reflecting the pronounced anisotropy and structural softness along the $a$-axis. This is consistent with its dynamic instability and the weak bonding along that direction. In comparison, qTP2 monolayers exhibit nearly isotropic in-plane stiffness, with $C_{11} \approx C_{22}$, due to the presence of comparable [2\,+\,2] cycloaddition bonds along both $a$ and $b$ lattice vectors. Among the three phases, qHP C$_{60}$ networks display the highest values of $C_{11}$, $C_{22}$, and $C_{66}$, consistent with its denser intermolecular bonding networks.

\subsubsection{Overall effects}

The conflicting dynamic, thermodynamic, and mechanical stability of monolayer fullerene networks is consistent with experimental observations. To date, only the qHP C$_{60}$ has been successfully exfoliated as a monolayer, whereas the qTP C$_{60}$ has been obtained in multilayer form\,\cite{Hou2022}. Interestingly, all the follow-up experimental studies have only reported few-layer qHP polymeric C$_{60}$\,\cite{Meirzadeh2023,Wang2023,Zhang2025}. Our free-energy analysis indicates that qTP1 becomes the most stable phase above 150\,K. However, thermodynamic stability alone does not ensure dynamic or mechanical stability. In qTP1, the presence of soft phonon modes associated with out-of-plane vibrations suggests weak dynamic stability along the interchain direction. Furthermore, its low elastic moduli, limited shear resistance, and reduced mechanical strength indicate that qTP1 C$_{60}$ is unlikely to exhibit intrinsic resilience under mechanical deformation. Notably, between 120 and 380\,K, the monolayer qTP1 network is thermodynamically less stable than the isolated 0D C$_{60}$ molecule, suggesting decomposition into individual chains or molecules under thermal fluctuations, interchain acoustic modes, or external strain. In contrast, the qHP monolayer displays both dynamic and mechanical stability, consistent with its successful exfoliation from bulk single crystals. These findings highlight the importance of jointly considering dynamic, thermodynamic, and mechanical stability when interpreting phase behaviour to guide the experimental realisation of monolayer fullerene networks\,\cite{Peng2023}.

\subsection{Thermal expansion}

Having understood the stability of various monolayer fullerene networks, the next step is to consider how these 2D frameworks respond to thermal effects. In low-dimensional materials, thermal expansion is not only a fundamental physical property but also a sensitive probe of bonding anisotropy and lattice flexibility. This motivates a detailed investigation of how temperature influences lattice constants and vibrational properties, with a particular focus on the role of intermolecular bonding geometry in driving positive or negative thermal expansion\,\cite{Shaikh2025}.

\begin{figure}
    \centering
    \includegraphics[width=0.9\linewidth]{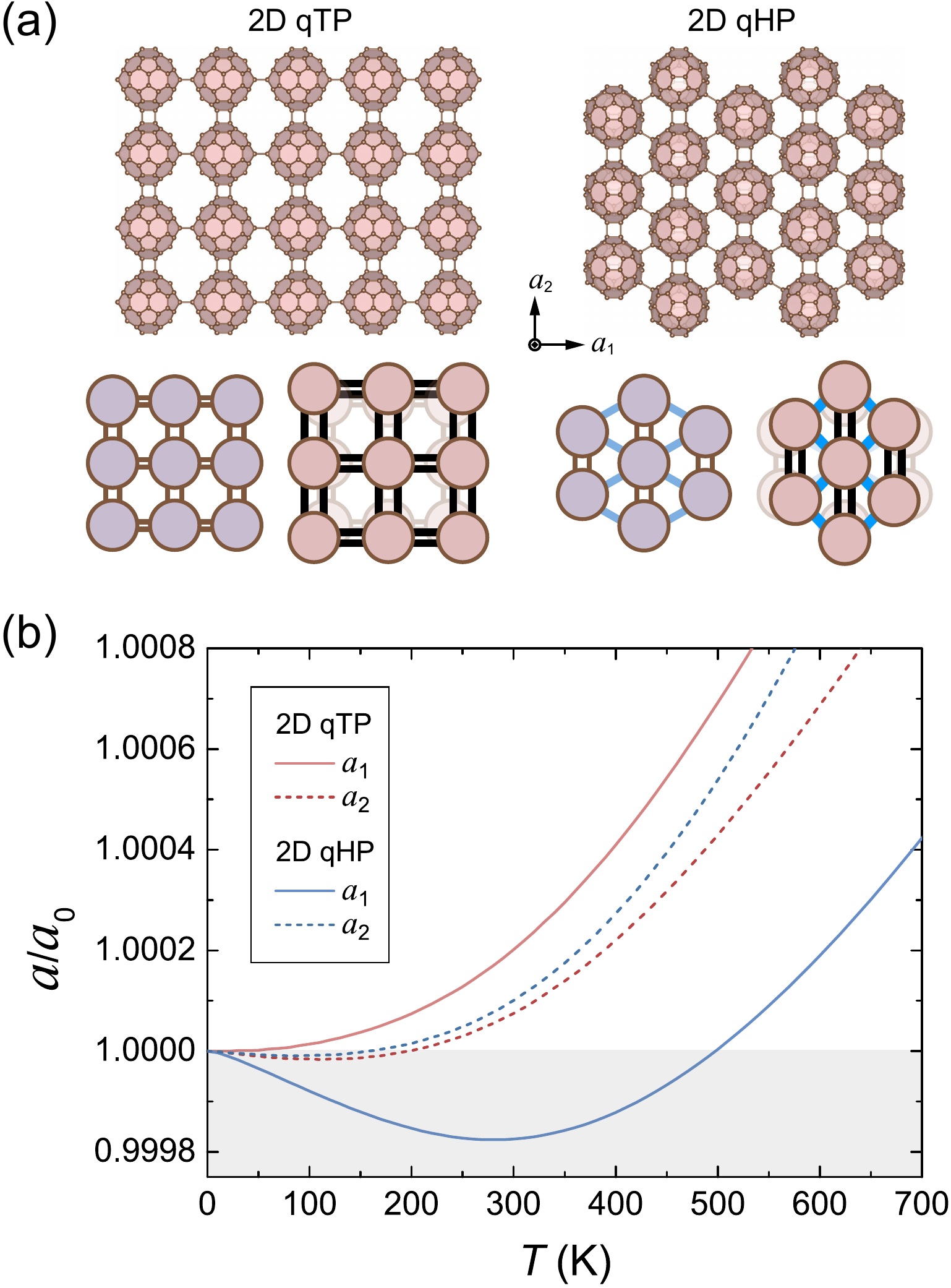}
    \caption{(a) Crystal structures and (b) thermal expansion of monolayer qTP and qHP C$_{60}$ networks. The schematics in (a) show the structural changes with increased temperature\,\cite{Shaikh2025}. 
    }
    \label{fig:expansion}
\end{figure}

As shown in Fig.\,\ref{fig:expansion}, the type of intermolecular bonds critically governs the anisotropic thermal response. In the qTP C$_{60}$ networks, neighbouring cages are connected via [2\,+\,2] cycloaddition bonds along both in-plane directions, and early isotropic positive thermal expansion is observed. In contrast, the qHP monolayers incorporate C$-$C single bonds along one axis and [2\,+\,2] cycloaddition bonds along the other, exhibiting pronounced anisotropy with positive expansion along the cycloaddition bond direction and negative thermal expansion along the single bond direction persisting up to 500\,K.

The unique combination of rigid [2\,+\,2] and flexible single bonds in qHP fullerene networks indicates strong anisotropic thermal expansion behaviour, offering an intriguing platform for engineering thermal expansion properties through molecular-scale structural design. This contrasting behaviour arises from distinct mechanical responses of different types intermolecular bonds. The [2\,+\,2] cycloaddition bonds are structurally rigid and resist transverse deformation. On the other hand, the C$-$C single bonds display considerable geometric flexibility, which allows hinge-like compression in response to perpendicular strain and therefore results in thermal contraction. Additionally, vibrational characteristics suggest large negative Gr\"uneisen parameters for low-frequency out-of-plane acoustic and optical vibrations associated with the single bonds. Conversely, vibrations associated with the cycloaddition bonds show positive or near-zero Gr\"uneisen values, contributing to thermal expansion.

The interplay between bond rigidity and flexibility provides a general strategy for tailoring thermal expansion in molecular materials beyond conventional approaches such as crystalline networks\,\cite{Goodwin2008,Dove2016}, rigid unit modes\,\cite{Pryde1996,Sleight1998,Tucker2005,Tan2024}, and transverse displacements of bridging atoms\,\cite{Hancock2004,Goodwin2005} or membranes\,\cite{Huang2016c,Koocher2021}. Through varied intermolecular bonds, one can rationally modulate both the magnitude and directionality of the thermal response. These design principles extend to other systems beyond monolayer polymeric C$_{60}$, where variations in molecular curvature and intermolecular bonding types further tune thermomechanical behaviour\,\cite{Shaikh2025}. Overall, these insights offer predictive guidelines for developing 2D fullerene-based materials with controllable thermal expansion for applications in flexible electronics, precision engineering, and energy-related technologies.

\section{Are 2D C$_{60}$ networks promising photocatalysts?}\label{sect:chem}

Photocatalytic water splitting harnesses solar energy to decompose water into hydrogen and oxygen, presenting an environmentally-friendly method for green hydrogen production. Since the discovery of photocatalytic TiO$_2$ in 1972\,\cite{Fujishima1972}, extensive research has been dedicated to developing efficient photocatalysts\,\cite{Norskov2004,Rossmeisl2007,Zhang2007,Zhang2011,Suzuki2012,Jiang2013,Xu2013b,Zhu2013,LeBahers2014,Zheng2015,Qiao2018,Yang2019,Ju2020,Nakada2021,Wang2022,Fu2022}. To enhance photocatalytic efficiency under visible light, materials must exhibit: ($i$) effective light absorption to generate sufficient electron-hole pairs; ($ii$) efficient separation and transport of these charge carriers to reactive sites; and ($iii$) appropriate band-edge positions to drive the redox reactions involved in water splitting. In this Section, we present recent computational studies addressing these criteria\,\cite{Peng2022c} and discuss the subsequent experiments that verify these predictions\,\cite{Wang2023}.

\subsection{Electronic structure from first principles}

Band alignment and optical absorption are central to photocatalytic processes. However, one main challenge in the theoretical description of electronic structure of monolayer C$_{60}$ networks is that, conventional DFT and HSE approaches underestimate the band gaps by at least 10\%\,\cite{Tromer2022,Yu2022} and fail to predict exciton binding energy. Therefore, an accurate theoretical framework is essential for fully exploiting the potential of these fullerene-based networks in photocatalytic water splitting.

\subsubsection{Appropriate theory for band gaps}

To reliably evaluate the electronic and optical properties of monolayer fullerene networks, we list the computed electronic and optical band gaps in Table\,\ref{tbl:gap}. Standard semilocal and even screened hybrid functionals, such as PBE/PBEsol\,\cite{Perdew1985,Perdew1996,Perdew2008} and HSE06/HSEsol\,\cite{HSE1,HSE2,HSE3,Schimka2011}, significantly underestimate the electronic band gaps by at least 10\%. Instead, the use of unscreened hybrid functional ($\mu = 0$, denoted as PBEsol0\,\cite{Perdew1996a,Becke1996,Adamo1999,Ernzerhof1999,Bernardi2020}) accurately reproduces the experimentally-measured electronic band gap.

\begin{table}
\centering
\caption{Calculated band gaps (eV) of qTP1, qTP2, and qHP C$_{60}$ using PBEsol, HSEsol and unscreened hybrid functional PBEsol0 with their corresponding screening parameter $\mu$ (\AA$^{-1}$). The optical band gaps (eV) for qHP C$_{60}$ from TDDFT or TDHF are shown in parentheses\,\cite{Peng2022c}. The $GW$\,+\,BSE\,\cite{Champagne2024} and measured\,\cite{Hou2022,Meirzadeh2023,Wang2023} band gaps are listed for comparison.
}
\begin{tabular}{c c c c c c}
\hline
 & PBEsol & HSEsol & PBEsol0 & $GW$ & Exp. \\ 
 $\mu$ & $\infty$ & 0.2 & 0 & - & - \\
  \hline
qTP1 & 1.09 & 1.65 & 2.31 & - & - \\
qTP2 & 0.94 & 1.48 & 2.18 & - & - \\
 qHP & 0.86 & 1.44 & 2.12 & 2.37 & 1.6$-$2.1 \\
     &(0.86)&(1.44)&(1.69)&(1.60)& (1.55) \\
\hline
\end{tabular}
\label{tbl:gap}
\end{table}

For optical band gaps, time-dependent DFT (TDDFT) or time-dependent Hartree-Fock (TDHF) calculations on top of the PBEsol or HSEsol band structures fail to capture excitonic effects. Instead, the PBEsol0\,+\,TDHF approach accurately reproduces both the optical band gap and exciton binding energy. 

Notably, a more recent study has applied the most accurate many-body perturbation theory $GW$ calculations for electronic band gaps, as well as the Bethe-Salpeter equation (BSE) for optical band gaps and exciton binding energy, which yield agreeable results with PBEsol0\,+\,TDHF\,\cite{Champagne2024}. 
Therefore, the PBEsol0\,+\,TDHF approach accurately describes low dielectric screening in fullerene networks and is essential to obtain physically meaningful band alignments for evaluating photocatalytic performance and to predict the excitonic absorption spectrum.

\subsubsection{Band edges}

\begin{figure}
    \centering
    \includegraphics[width=\linewidth]{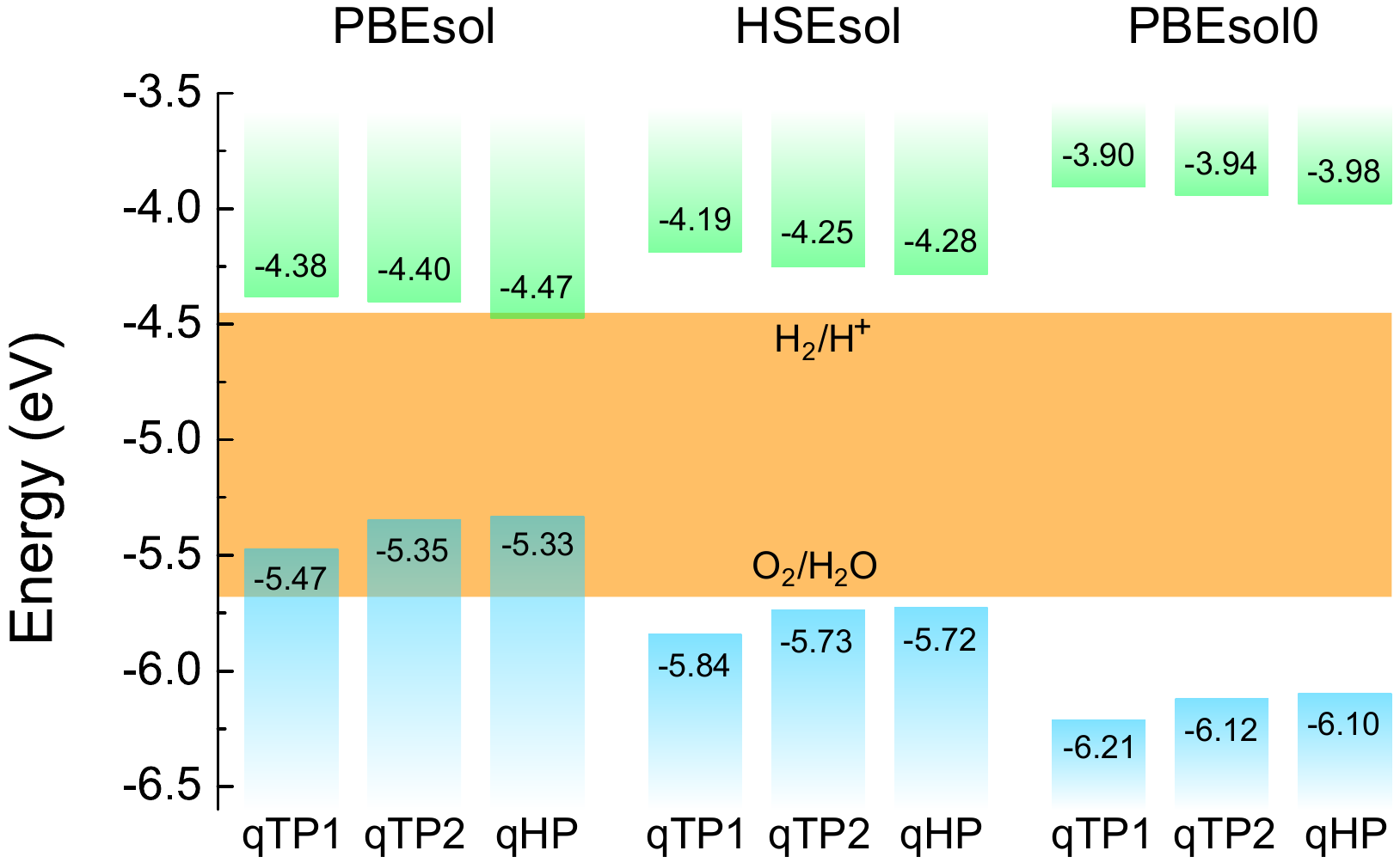}
    \caption{Band alignment of qTP1, qTP2, and qHP monolayers calculated with PBEsol, HSEsol, and PBEsol0\,\cite{Peng2022c}. 
    }
    \label{fig:redox}
\end{figure}

After determining the band gaps, we next assess the band edges in the context of photocatalysis for overall water splitting. For an overall water splitting reaction, the energy levels of the conduction band minimum (CBM) and valence band maximum (VBM) must straddle the redox potentials of water. In other words, the CBM (with respect to the vacuum level) should be higher than the hydrogen evolution potential of $-4.44 + {\rm pH} \times 0.059$\,eV, and the VBM should be lower than the oxygen evolution potential of $-5.67 + {\rm pH} \times 0.059$\,eV\,\cite{Chakrapani2007,Zhang2019d,Chen2022}. As shown in Fig.\,\ref{fig:redox}, unscreened hybrid functional calculations find that, in all three phases, the CBM lies above the hydrogen reduction potential, while the VBM falls below the oxygen evolution potential, providing the required thermodynamic driving force for water splitting\,\cite{Peng2022c}.

\subsection{Carrier dynamics}

We next discuss the carrier dynamics in photocatalysis. Specially, the photocatalysts need to: ($i$) generate sufficient electron-hole pairs upon photoexcitation; ($ii$) separate the photoexcited electrons and holes effectively; and ($iii$) transport carriers efficiently on the monolayer surface.

\subsubsection{Sufficient carrier generation}

Different structural phases of monolayer fullerene networks exhibit distinct optical absorbance\,\cite{Peng2022c}. For both qTP1 and qTP2 structures, the low optical transition probability between the top valence and bottom conduction bands leads to relatively weak absorbance below 2\,eV, especially when excitonic effects are taken into account. In contrast, qHP C$_{60}$ exhibits strong optical transitions from bright excitons with high binding energies, resulting in much larger absorbance in nearly the entire visible-light range. These features suggest that qTP C$_{60}$ is a likely electron acceptor, whereas qHP C$_{60}$ can generate a substantial number of carriers as a suitable electron donor.

\subsubsection{Effective carrier separation}

\begin{figure}[h]
    \centering
    \includegraphics[width=\linewidth]{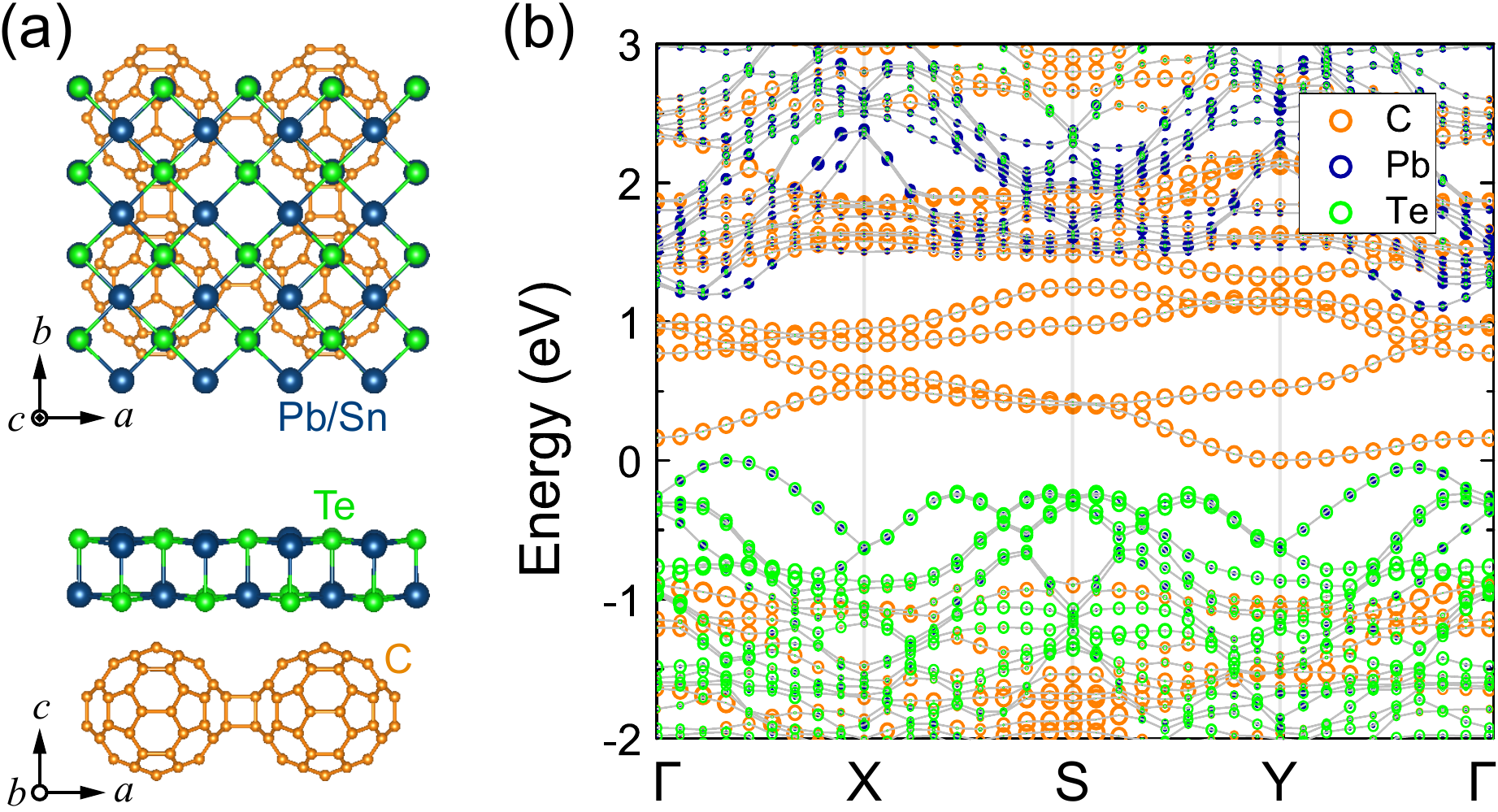}
    \caption{(a) Crystal structures and (b) PBEsol electronic structures of qTP2-PbTe heterostructures\,\cite{Peng2022c}. 
    }
    \label{fig:type-II}
\end{figure}

Efficient photocatalysis requires not only the generation of carriers but also their spatial separation to prevent recombination. In qTP monolayers, weak optical transitions suppress radiative recombination, making them well suited as electron acceptors. In particular, it is possible to form type-II band alignments by creating heterostructures of monolayer fullerene networks and other 2D materials. This enables efficient spatial separation of photoexcited electrons and holes across different layers\,\cite{Scanlon2013,Ozcelik2016,Hu2017,Yang2020a}. As an example, we show type-II alignments between qTP2 fullerene networks and monolayer PbTe/SnTe in Fig.\,\ref{fig:type-II}. These heterostructures have good lattice match, offering an strategy to confine electrons in qTP2 layer and holes in PbTe/SnTe layer. Such spatial separation is favourable to reduce electron-hole recombination and improve photocatalytic efficiency.

\subsubsection{Efficient carrier transport}

\begin{figure}[h]
    \centering
    \includegraphics[width=\linewidth]{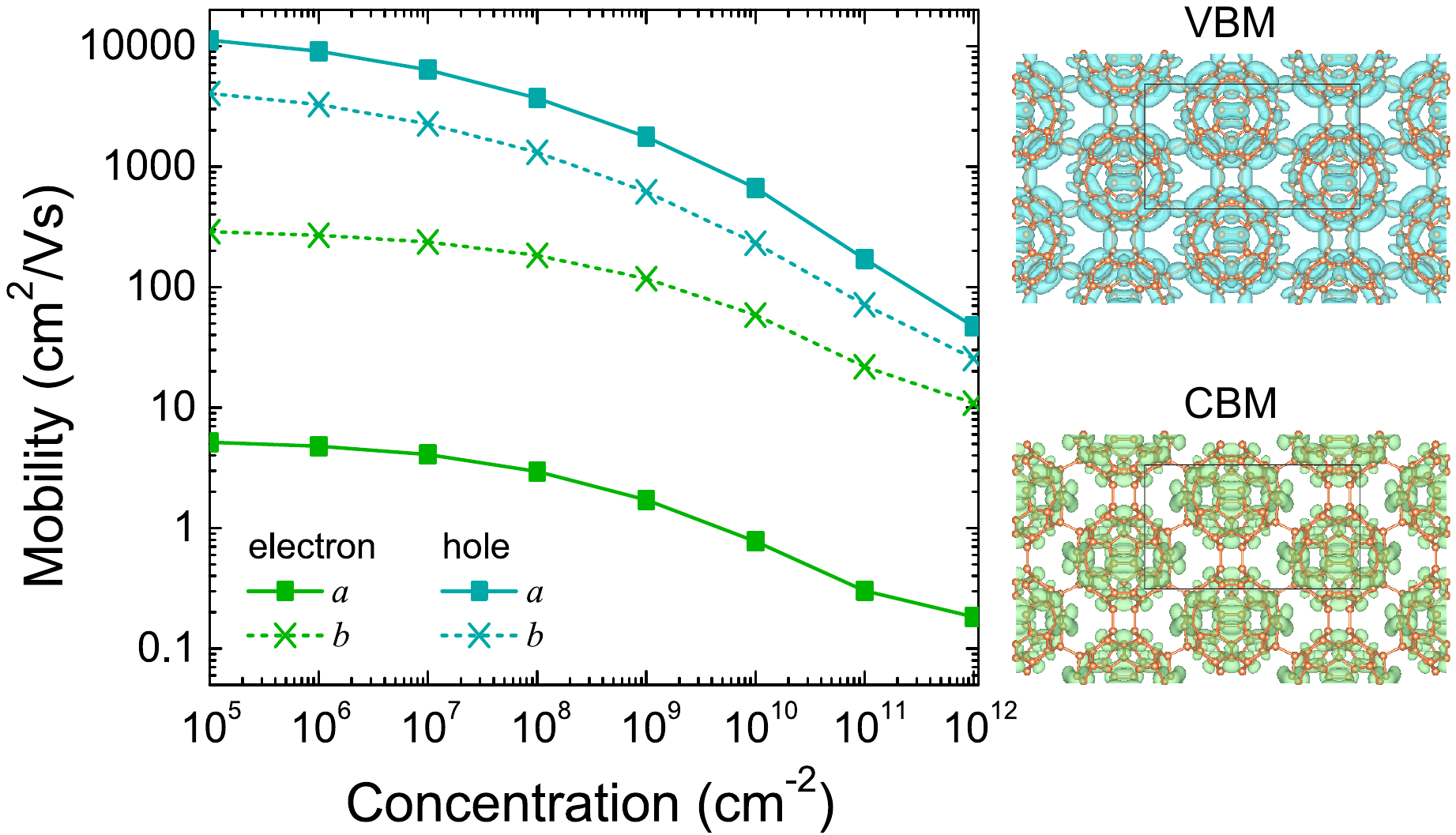}
    \caption{Mobility of monolayer qHP C$_{60}$ networks at 300\,K as a function of carrier concentration, as well as the corresponding partial charge density of the VBM and CBM states\,\cite{Peng2022c}. 
    }
    \label{fig:mobility}
\end{figure}

As shown in Fig.\,\ref{fig:mobility}, the VBM and CBM states with low effective masses facilitate high carrier mobility on the molecular surface for photocatalytic reactions. In qHP, the hole mobility is significantly enhanced due to the delocalised VBM across both directions, while the electron mobility is much lower than the hole mobility but still relatively high. The lowest electron mobility is $1.7-4.8$\,cm$^2$/(V\,s) along the crystallographic orientation $a$ at low carrier concentrations ($<10^9$\,cm$^{-2}$). The good agreement with the measured mobility\,\cite{Hou2022} suggests accurate descriptions of transport properties based on the Boltzmann transport equation under the momentum relaxation time approximation\,\cite{Rode1975,Faghaninia2015,Ganose2021}. The relatively high carrier mobility is attributed to the delocalised $\pi$ electrons. These results confirm efficient carrier transport across the monolayers, which is a critical prerequisite for photocatalysis.

\subsection{Reaction pathways}

\begin{figure}[t]
    \centering
    \includegraphics[width=\linewidth]{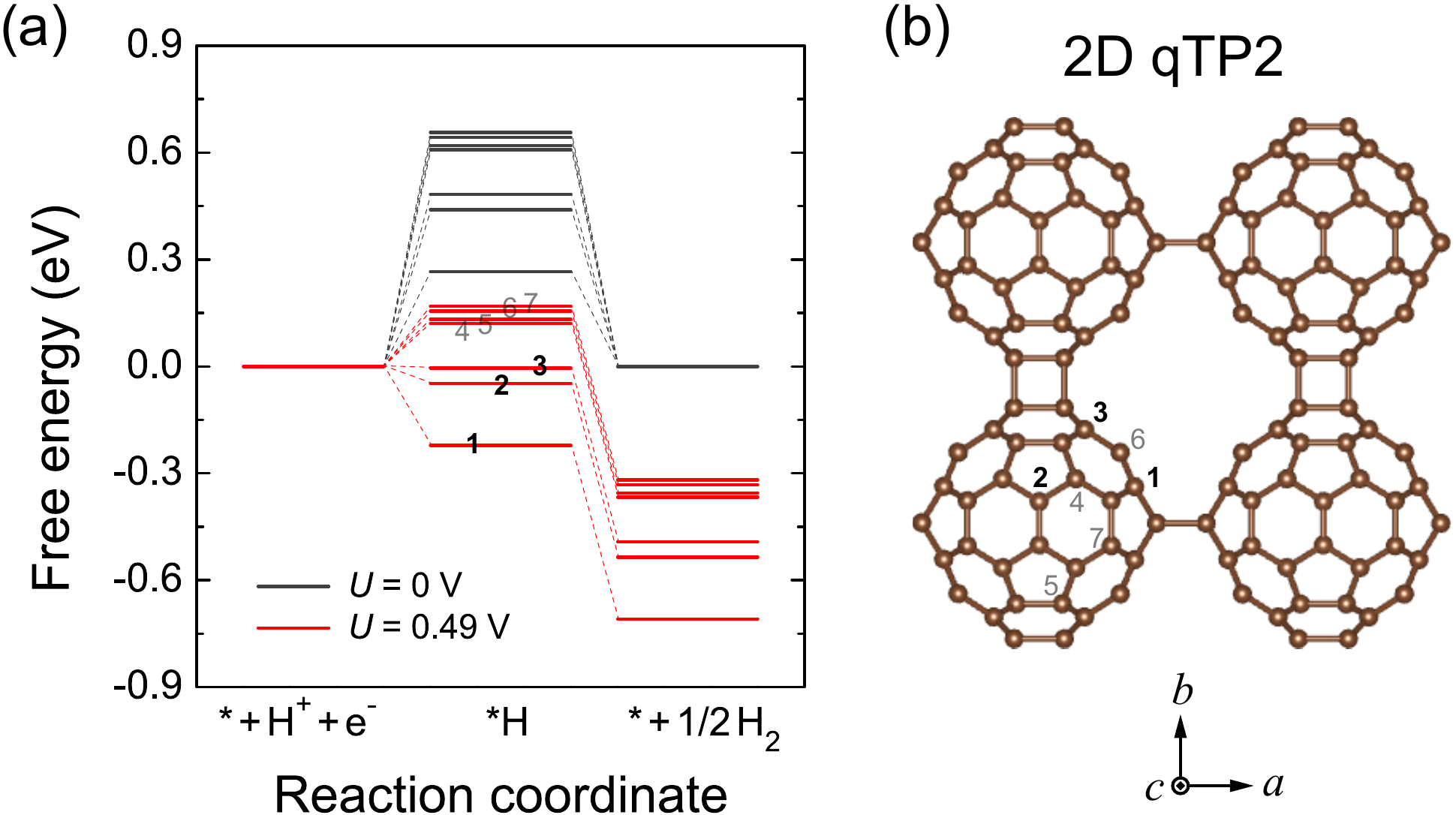}
    \caption{(a) Gibbs free energy changes associated with the HER in qTP2 C$_{60}$ networks at a pH of 0 and room temperature. (b) Adsorption sites for qTP2 C$_{60}$ networks, with lower to higher numbers corresponding to lower to higher free energies of the intermediates\,\cite{Jones2023}. 
    }
    \label{fig:HER}
\end{figure}

Thermodynamic analysis reveals favourable adsorption energies for water on the surface of monolayer fullerene networks\,\cite{Peng2022c}, which is the initial step of photocatalytic reaction. To thermodynamically drive the redox reactions, the next step is for the free-energy diagram to exhibit downhill reaction pathway. As shown in Fig.\,\ref{fig:HER}(a), the hydrogen evolution reaction has two steps. In the first step, monolayer fullerene networks (denoted as *) combine with a proton (H$^+$) and an electron (e$^-$) to form H* species. In the second step, H$_2$ molecules are formed from the H* species.

Without photoexcitation, there is always energy barrier posed by the intermediate adsorbate. Upon photoexcitation, the external potential generated by the photoexcited electrons in the CBM exceeds the energy barriers for hydrogen evolution for all three phases at the level of unscreened hybrid functional PBEsol0\,\cite{Peng2022c,Jones2023,Wu2025,Shearsby2025}. The presence of photoexcitation also creates a more favorable Gibbs free energy between the intermediates and the products, enabling spontaneous HER under acidic conditions at room temperature. 
The photocatalytic activity is further enhanced by the high surface area and abundant active sites provided by the spherical C$_{60}$ units, as shown in Fig.\,\ref{fig:HER}(b). In addition, the hydrogen storage capacity of various fullerene structures\,\cite{Zhao2005,Kim2006,Yoon2008,Pupysheva2008,Wang2009,Sun2009,Durbin2016,SankarDe2018,Ren2023} offers a unique dual function in facilitating hydrogen evolution and serving as hydrogen storage materials.

\subsection{Experimental verification}

Conventional fullerene-based heterostructures typically involve sparse and non-periodically distributed C$_{60}$ molecules\,\cite{Guan2018}. In comparison, monolayer polymeric C$_{60}$ networks offer atomically smooth surfaces with periodic C$_{60}$ arrangements. This enhanced structural order leads to improved crystallinity and higher C$_{60}$ content, thereby increasing the density of active sites and overall photocatalytic activity. 
Recent experimental studies\,\cite{Wang2023} have supported the theoretical predictions\,\cite{Peng2022c} regarding the photocatalytic capabilities of monolayer fullerene networks. In particular, the photocatalytic efficiency of 2D fullerene networks is much higher than those of 0D C$_{60}$ molecules or 3D C$_{60}$ crystals\,\cite{Wang2023}. These findings not only validate the previous theoretical predictions but also pave the way for further research into the optimisation and integration of 2D fullerene networks in photocatalytic systems. 

From a materials design perspective, monolayer fullerene networks represent a promising class of photocatalysts owing to their unique combination of molecular features and crystalline characteristics. The C$_{60}$ building blocks retain high surface area with delocalised $\pi$ electrons for abundant active sites, while their covalent 2D connectivity enables efficient charge transport across the monolayers. These structural features allow for strong optical absorption, high carrier mobility, and enhanced chemical reactivity. Compared to traditional metal oxides, monolayer fullerene networks offer a chemically pure, carbon-based alternative that combines structural flexibility and high efficiency, making them appealing candidates for next-generation photocatalysts.


\section{Are functionalities of fullerene networks tuneable?}\label{sect:tune}

The chemical functionalities of monolayer fullerene networks can be further modulated through structural design at multiple levels. Three key tuning strategies have emerged: variation in stacking configurations, control of molecular size, and dimensionality engineering. Vertical stacking of monolayers introduces interlayer degrees of freedom that alter optical absorption and exciton dynamics, offering a pathway to enhance light harvesting\,\cite{Shearsby2025}. At the molecular scale, substituting C$_{60}$ with smaller cage units such as C$_{24}$ modifies the electronic structure and increases the density of active sites, improving catalytic performance\,\cite{Wu2025}. Beyond monolayers, extending or reducing the lattice dimensionality yields rich behaviour: three-dimensional van der Waals crystals exhibit diverse structural phases\,\cite{Kayley2025}, one-dimensional chains enhance reactivity via increased surface active sites\,\cite{Jones2023}, and quasi-one-dimensional nanoribbons display strongly geometry- and edge-dependent band structures\,\cite{Peng2025}. Among these, fullerene nanoribbons are especially interesting due to their tuneable edge shapes, quantum confinement effects, and potential for directional charge transport, positioning them as promising candidates for next-generation molecular optoelectronics and catalysis.

\subsection{Stacking degree of freedom}

\begin{figure}[t]
    \centering
    \includegraphics[width=\linewidth]{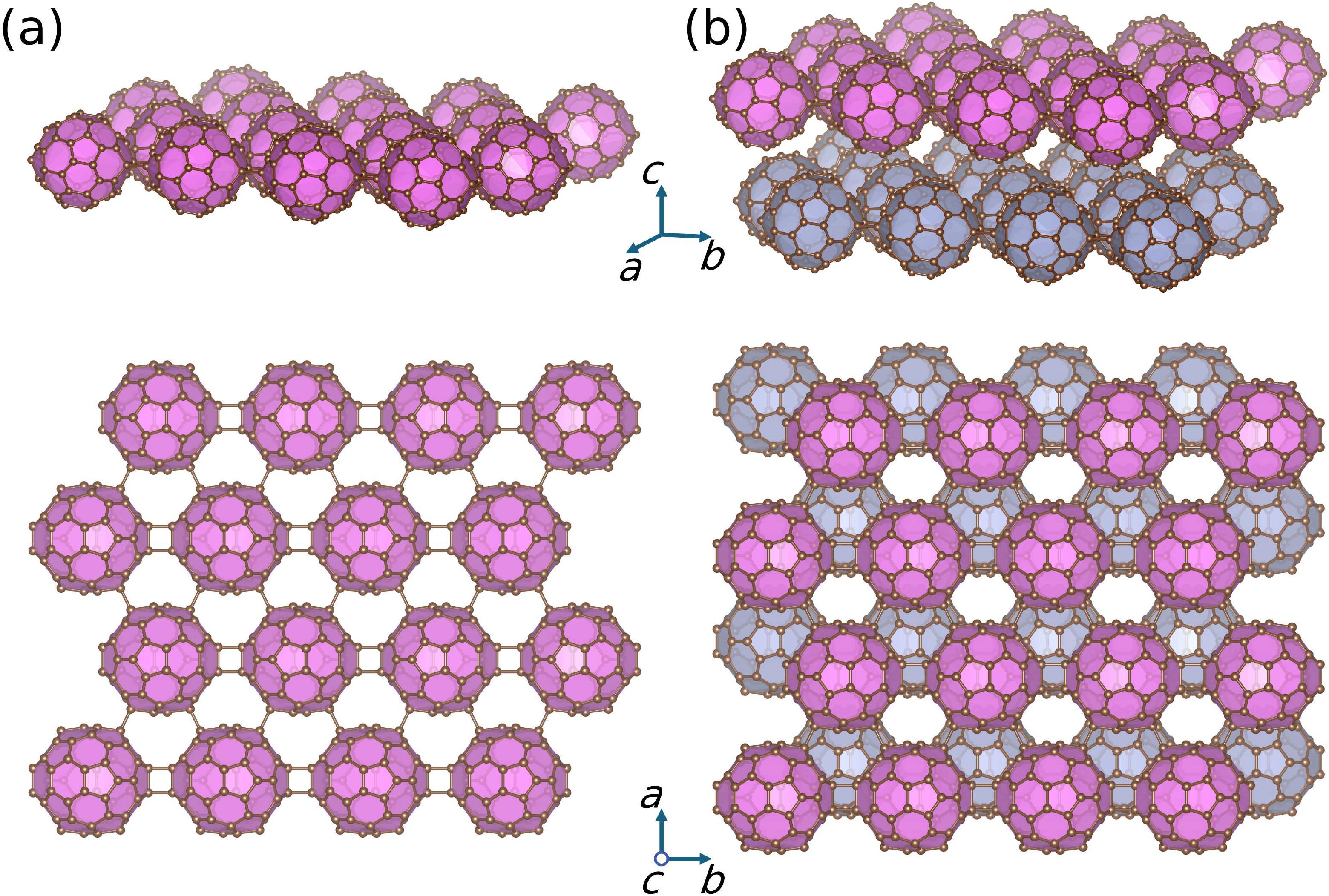}
    \caption{3D and top views of the crystal structures of (a) monolayer and (b) bilayer fullerene networks\,\cite{Shearsby2025}. 
    }
    \label{fig:bilayer}
\end{figure}

While monolayer polymeric qHP C$_{60}$ networks possess promising optoelectronic and photocatalytic properties, they suffer from limited thermodynamic stability, making experimental isolation challenging. As a result, most experimentally-realised qHP structures exist in the few-layer form\,\cite{Meirzadeh2023,Wang2023,Zhang2025}, which benefit from enhanced structural stability due to interlayer van der Waals interactions. These bilayer systems not only retain the desirable properties of monolayers such as appropriate band alignment for water splitting but also introduce new stacking degrees of freedom such as sliding, orientation, and twisting angles, which can be exploited to tune their functionalities. Taking AB-stacked bilayer fullerene networks as an example, it can be viewed as two van der Waals layers in a closely-packed stacking pattern for C$_{60}$ molecules with space group $P2/c$ (No.\,13), as shown in Fig.\,\ref{fig:bilayer}.

From an electronic structure perspective, the bilayer exhibits a slightly reduced direct band gap at $\Gamma$ (2.05\,eV) compared to the monolayer (2.08\,eV). Importantly, the band-edge positions of the bilayer still straddle the redox potentials of water at pH from 0 to 7, indicating that the bilayer retains its suitability for overall water splitting\,\cite{Shearsby2025}. The presence of two layers gives rise to nearly degenerate states at the VBM and CBM, with minor energy splittings between in-phase and out-of-phase combinations of the top and bottom layer states. The interlayer interaction slightly modifies the anisotropic effective mass, particularly in the conduction band, allowing for directional tuning of charge transport\,\cite{Shearsby2025}.

\begin{figure}[t]
    \centering    \includegraphics[width=\linewidth]{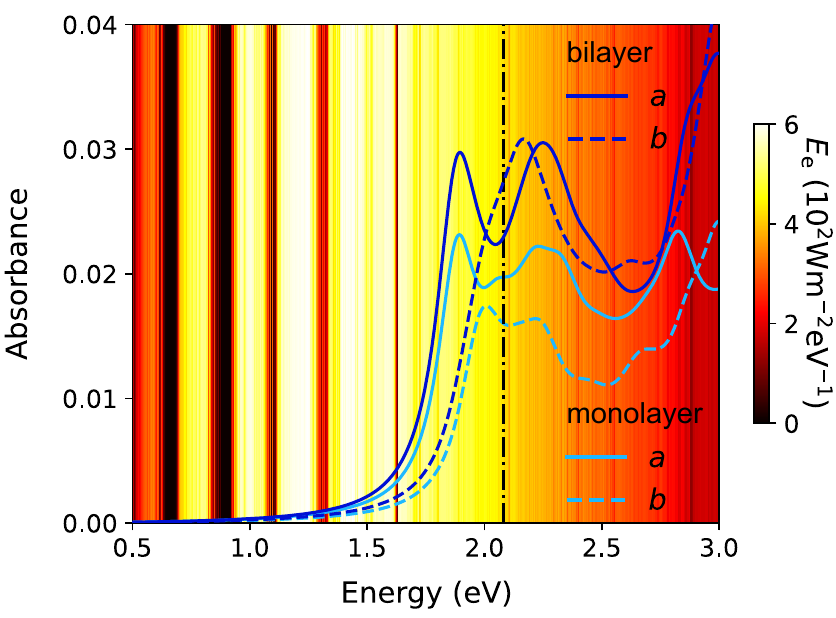}
    \caption{Optical absorbance of monolayer and bilayer fullerene networks for both $a$ and $b$ polarisations. The black vertical dash-dot lines represent the direct band gaps from independent-particle approximations in the absence of excitons. The background shows the global total spectral irradiance $E_{\rm e}$ from the Sun\,\cite{Shearsby2025}. 
    }
    \label{fig:optic}
\end{figure}

In terms of optical properties, the bilayer C$_{60}$ network demonstrates enhanced absorbance across the entire visible spectrum compared to its monolayer counterpart (Fig.\,\ref{fig:optic}). This enhancement arises from stronger excitonic transitions involving multiple degenerate states and is accompanied by increased anisotropy in polarisation-dependent absorbance. Notably, while the monolayer shows relatively isotropic exciton absorption, the bilayer exhibits more pronounced absorbance differences for polarisation along the $a$ and $b$ axes. Such anisotropy can be exploited in polarised light detectors and direction-sensitive photonic devices\,\cite{Shearsby2025}.

Overall, stacking monolayers into bilayer fullerene networks preserves photocatalytic performance while enhancing structural stability and optoelectronic tuneability. These stacking degrees of freedom also open avenues for emerging applications such as flexible displays, memory storage, and ferroelectric devices, where interlayer sliding and twist engineering may give rise to novel phenomena such as sliding ferroelectricity\,\cite{Wang2025} and moir{\'e} patterns\,\cite{Meirzadeh2023}.

\subsection{Molecular Size}

The choice of molecular building block provides further tuneablity in the stability and functionality of 2D fullerene networks. While C$_{60}$-based monolayers have attracted significant attention for their photocatalytic\,\cite{Peng2022c, Wang2023} and optoelectronic\,\cite{Zhang2025} properties, their relatively large molecular size limit the structural stability and density of accessible active sites. Recent studies have demonstrated that reducing the molecular size to the smallest stable conventional [5,6]fullerene unit, C$_{24}$, yields 2D networks with superior stability and enhanced photocatalytic performance\,\cite{Wu2025}. 

\subsubsection{Monolayer C$_{24}$ networks}

\begin{figure}[h]
    \centering
    \includegraphics[width=0.9\linewidth]{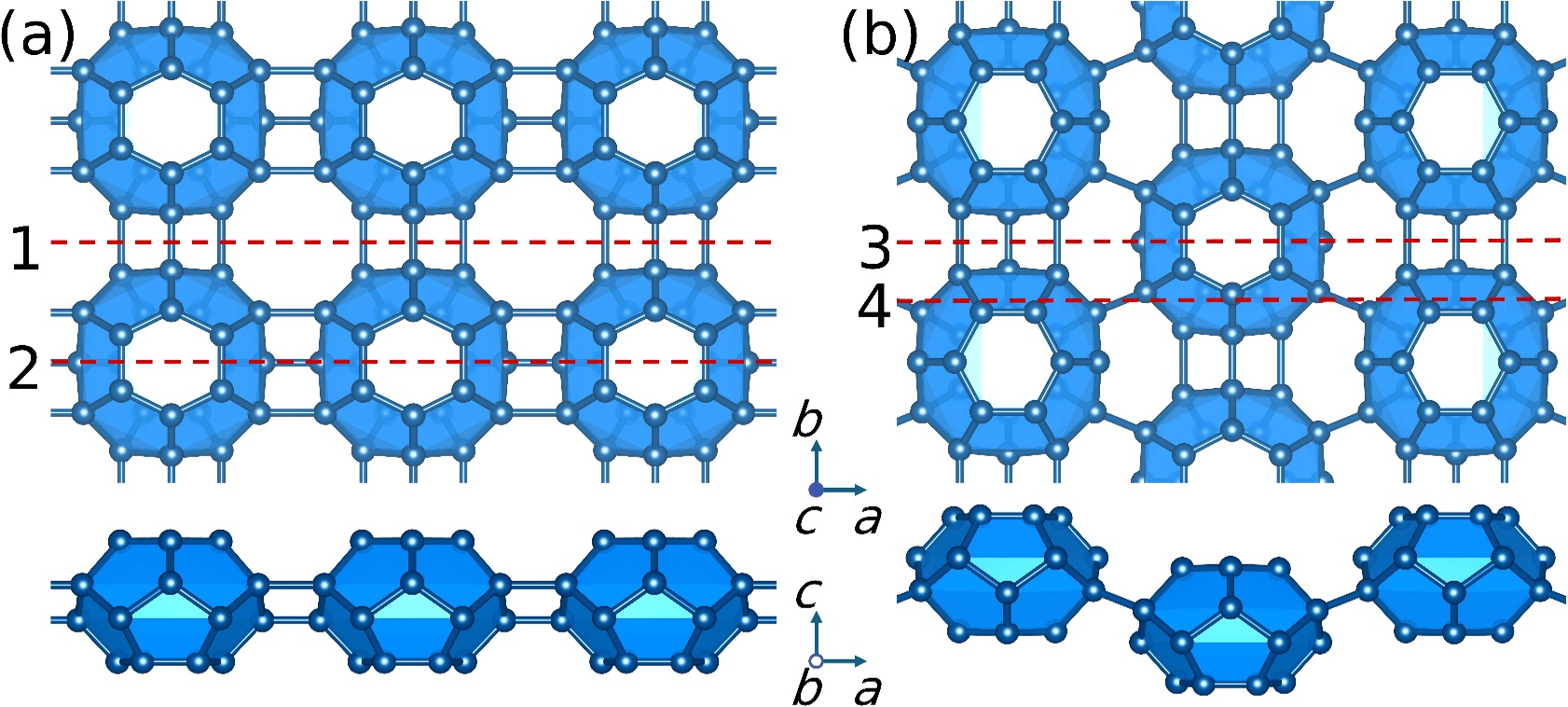}
    \caption{Top and side views of crystal structures of (a) qTP and (b) qHP C$_{24}$ monolayers\,\cite{Wu2025}. 
    }
    \label{fig:C24-crystal}
\end{figure}

Monolayer C$_{24}$ networks retain the key features of fullerene chemistry such as delocalised $\pi$ electrons and robust cage-like structures. As shown in Fig.\,\ref{fig:C24-crystal}, the crystal structures of monolayer C$_{24}$ networks are similar to monolayer polymeric C$_{60}$. The qTP C$_{24}$ structure has a nearly-square lattice with three noncoplanar intermolecular bonds between neighboring clusters. The qHP C$_{24}$ lattice can be viewed as periodically misaligned 1D chains along $b$ connected by the three noncoplanar intermolecular bonds, which are further joined through diagonal single bonds between neighboring chains. Monolayer qHP networks exhibit a buckled structure between neighbouring chains due to asymmetric interchain bonding positions.

Compared to monolayer polymeric C$_{60}$, C$_{24}$ monolayers exhibit significantly improved dynamic, thermodynamic, and mechanical stability\,\cite{Wu2025}. Cohesive energy calculations demonstrate that the formation of polymeric C$_{24}$ networks is energetically favourable, owing to the release of stereochemical strain through the formation of noncoplanar $sp^3$-like bonds. These networks are also dynamically stable, as confirmed by phonon dispersion calculations showing no imaginary modes. Mechanically, the C$_{24}$ lattices exhibit larger elastic and shear moduli than their C$_{60}$ counterparts, because of the smaller cage size and therefore higher density of covalent interfullerene bonds.

\begin{figure*}
    \centering
    \includegraphics[width=\linewidth]{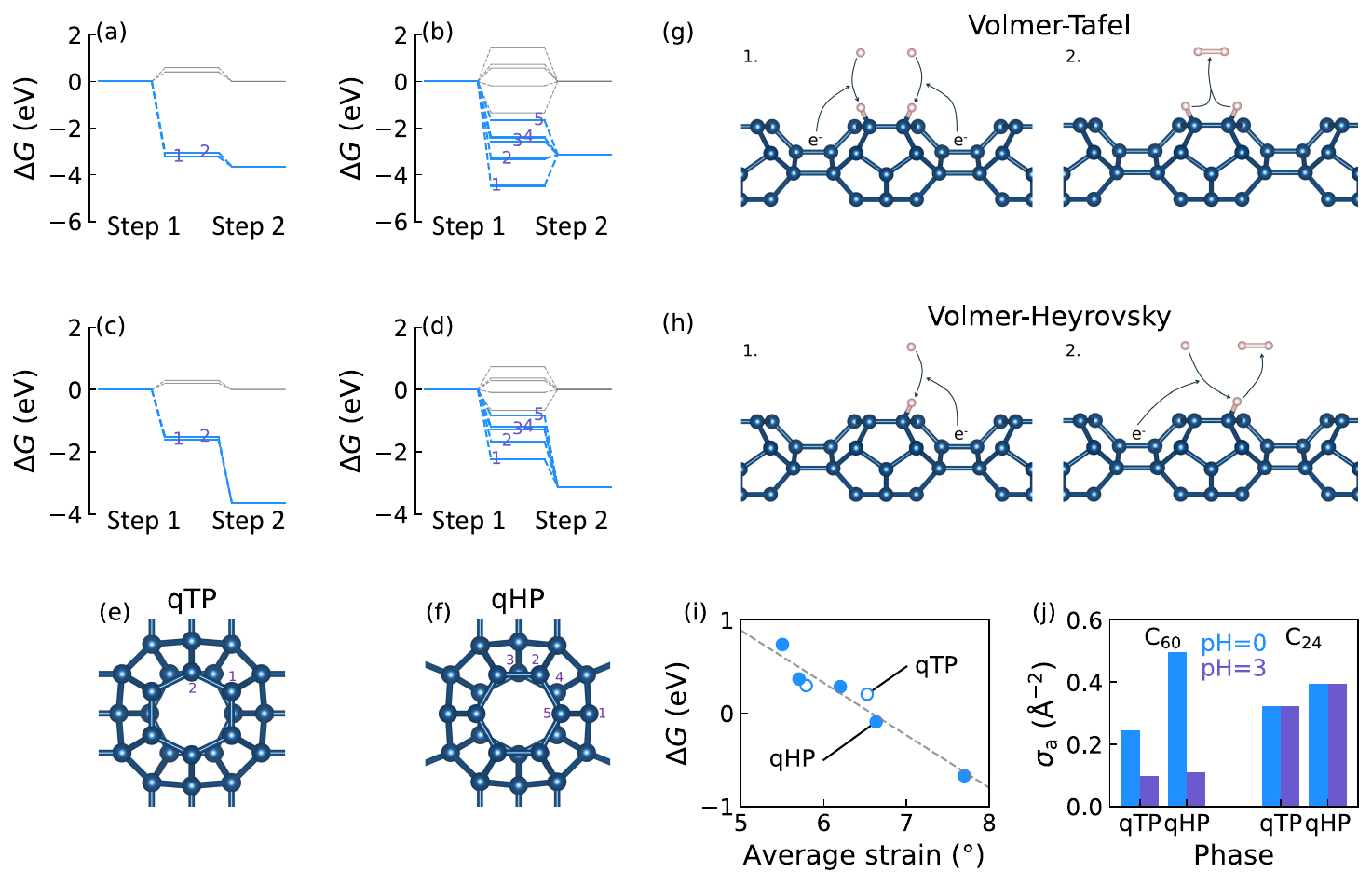}
    \caption{Free energy profiles of hydrogen evolution at different adsorption sites through the Volmer-Tafel mechanism for (a) qTP and (b) qHP C$_{24}$ and the Volmer-Heyrovsky mechanism for (c) qTP and (d) qHP C$_{24}$ at $\text{pH}=0$, with grey and blue lines representing the absence and presence of photoexcitation respectively. Symmetry-irreducible adsorption sites for (e) qTP and (f) qHP C$_{24}$. (g) Volmer-Tafel and (h) Volmer-Heyrovsky reaction mechanisms. (i) Correlation between adsorption free energy and average bond angle strain. (j) Area density $\sigma_\mathrm{a}$ of active sites for various phases of fullerene monolayers at changing pH\,\cite{Wu2025}. 
    }
    \label{fig:C24}
\end{figure*}

The reduction in molecular size also leads to decreased screening and hence increased band gaps. Unscreened hybrid functional calculations show that monolayer C$_{24}$ networks have wide band gaps (3.10–3.74\,eV depending on phase), which are comparable to those of TiO$_2$\,\cite{Deak2011,Pfeifer2013,Ju2014,Mi2015,Zhang2015n,Deak2016,Chiodo2010,Li2020a}, allowing for visible-to-UV light harvesting. Both qTP and qHP C$_{24}$ structures have band-edge positions that straddle the redox potentials of water across a broad pH range, satisfying the thermodynamic requirements for overall water splitting. Moreover, strongly bound bright excitons contribute to strong optical absorbance, particularly in the UV regimes, facilitating efficient generation of photoexcited carriers.

\subsubsection{Surface active sites}

To understand the thermodynamic driving force for photocatalytic water splitting in monolayer C$_{24}$ networks, we show the free-energy landscape of HER for both the Volmer-Tafel (V-T) and Volmer-Heyrovsky (V-H) mechanisms\,\cite{Danilovic2012} on C$_{24}$ surface in Fig.\,\ref{fig:C24}(a)-(d). Following full structural relaxation, hydrogen atoms preferentially adsorb on top sites, as enumerated in Fig.\,\ref{fig:C24}(e) and (f) for all symmetry-inequivalent adsorption sites ordered by increasing adsorption free energy. 

The two HER pathways proceed via ($i$) adsorption of two protons and then combination of two adsorbed protons into a H$_2$ molecule (V-T), and ($ii$) adsorption of a second proton on top of the previously adsorbed proton and then desorption into a H$_2$ molecule (V-H), as shown in Fig.\,\ref{fig:C24}(g) and (h) respectively. For both qTP and qHP C$_{24}$, the HER at $\text{pH}=0$ remains kinetically hindered without light, as all energy barriers exceed the thermal fluctuation threshold ($k_\mathrm{B}T \sim 26$\,meV). Under photoexcitation, however, all adsorption sites in qTP become catalytically active, enabling spontaneous HER via either V-T or V-H pathways. For qHP, although sites 1 and 2 are inactive under the V-T mechanism, the V-H pathway remains downhill for all adsorption sites. Remarkably, even at near-neutral pH conditions, at least one of the two mechanisms remains spontaneous for all sites in both qTP and qHP C$_{24}$ monolayers.

To further understand adsorption behaviour, we assess the correlation between hydrogen adsorption free energy and local bond angle strain. The bond angle strain is defined as the deviation of the average bond angle at a given site from the ideal $sp^2$ value of $120^\circ$, providing a geometric descriptor of local carbon environments. As depicted in Fig.\,\ref{fig:C24}(i), adsorption free energy exhibits a strong correlation with the bond angle strain, indicating that more curved C$_{24}$ cages enhance hydrogen adsorption. This suggests that the higher curvature of monolayer C$_{24}$ networks plays a significant role in increasing photocatalytic efficiency.

Finally, the area density $\sigma_{\rm a}$ of thermodynamically active sites at different pH values is summarised in Fig.\,\ref{fig:C24}(j). While qTP and qHP C$_{24}$ monolayers exhibit similar active site densities under acidic conditions ($\text{pH}=0$) to their C$_{60}$ counterparts, increasing pH to 3 significantly reduces the number of active sites on C$_{60}$ networks. On the other hand, C$_{24}$ monolayers retain full catalytic reactivity. As a result, the active site density in C$_{24}$ monolayers becomes approximately three times that of C$_{60}$ at moderate acidity, with this ratio increasing further under near-neutral conditions. This resilience under varying pH highlights the superior catalytic performance of C$_{24}$ networks for hydrogen evolution across a broader operational range.


\subsection{Lattice dimensionality}

Changes in lattice dimensionality offer a useful strategy to modulate the physical and chemical behaviour of fullerene networks. When increasing lattice dimension to 3D van der Waals crystals of C$_{60}$, molecular orientations and their corresponding crystalline symmetry govern electronic band gaps, excitonic effects, and phase stability, which are crucial for applications in optoelectronics and energy harvesting devices\,\cite{Kayley2025}. Conversely, reducing lattice dimension to 1D polymeric chains increases the band gap and the number of active sites, which are advantageous for photocatalytic HER\,\cite{Jones2023}. Quasi-1D C$_{60}$ nanoribbons, split from 2D monolayers, represent an intermediate geometry where quantum confinement and edge states give rise to tuneable band gaps and effective masses, with electronic properties highly sensitive to nanoribbon width and edge configuration\,\cite{Peng2025}. Together, these dimensional variants unlock even richer physical and chemical properties, offering design flexibility for targeted functionalities.

\subsubsection{3D van der Waals crystals}

Extending the dimensionality from 2D C$_{60}$ monolayers to 3D van der Waals crystals further modulates their structural and electronic properties. In van der Waals layered structures such as the orthorhombic ($Immm$) and trigonal ($R\bar{3}m$) phases, C$_{60}$ molecules are covalently bonded in quasi-2D sheets via [2\,+\,2] cycloaddition bonds, with adjacent layers stacked through van der Waals interactions. These van der Waals layered structures exhibit dynamic stability\,\cite{Kayley2025} and are viable for experimental synthesis under moderate pressure and temperature conditions\,\cite{Rao1993,Iwasa1994,Nunez-Regueiro1995,Eklund1995,Xu1995,Springborg1995,Marques1996,Giacalone2006,Murga2015}.

\begin{figure}[t]
    \centering
    \includegraphics[width=\linewidth]{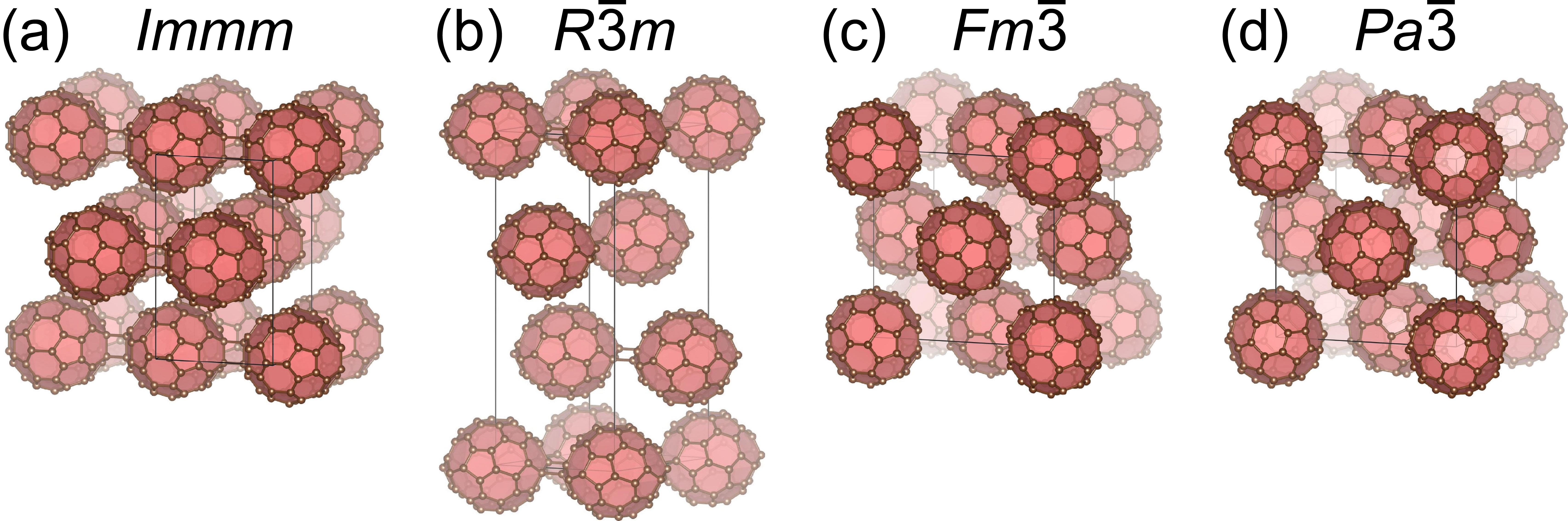}
    \caption{Crystal structures of solid C$_{60}$ in conventional unit cells with space group (a) $Immm$, (b) $R\bar{3}m$, (c) $Fm\bar{3}$, and (d) $Pa\bar{3}$\,\cite{Kayley2025}. 
    }
    \label{fig:3D}
\end{figure}

Instead of layered crystals, the cubic $Fm\bar{3}$ and $Pa\bar{3}$ phases remove all covalent intermolecular bonds in favour of purely non-covalent van der Waals interactions. In these structures, the C$_{60}$ units have extra rotational degrees of freedom with varied molecular orientations. The $Fm\bar{3}$ phase exhibits higher symmetry with a single C$_{60}$ molecule per primitive unit cell (i.e., all molecules have the same orientation), while the $Pa\bar{3}$ structure contains four inequivalent molecules with distinct orientations, leading to reduced symmetry but enhanced packing diversity. This rotationally-disordered phase can be realised at elevated temperatures as a result of the phase transitions observed near 255\,K\,\cite{Heiney1992}. These 3D van der Waals phases display large band gaps ($2.293-2.461$\,eV) and strong exciton binding energies ($>200$\,meV)\,\cite{Kayley2025}, which agree well with previous $GW$ calculations\,\cite{Shirley1993,Shirley1996} and (inverse-)photoemission spectra\,\cite{Weaver1991,Lof1992,Lof1995,Schwedhelm1998,Makarova2001}. 

The transition from 2D to 3D provides more than structural diversity. It introduces additional avenues for tuneable functionalities through control of crystalline symmetry and intermolecular spacing\,\cite{Kayley2025}. While the layered $Immm$ and $R\bar{3}m$ phases offer anisotropic optical absorption and exciton confinement suitable for photonic devices, the cubic $Fm\bar{3}$ and $Pa\bar{3}$ phases exhibit more isotropic absorption profiles comparable to halide perovskites\,\cite{Xie2020,Peng2022d}. These distinctions highlight the rich structural tuneability in fullerene-based networks and provide a versatile platform for energy harvesting and storage. Notably, fullerene molecules are highly stable in outer space and contribute to 1\% of the cosmic carbon in the interstellar medium\,\cite{Ehrenfreund1997,Cami2010,Woods2020}, which further extends their applications to astrochemistry.


\subsubsection{1D chain}

\begin{figure}[t]
    \centering
    \includegraphics[width=\linewidth]{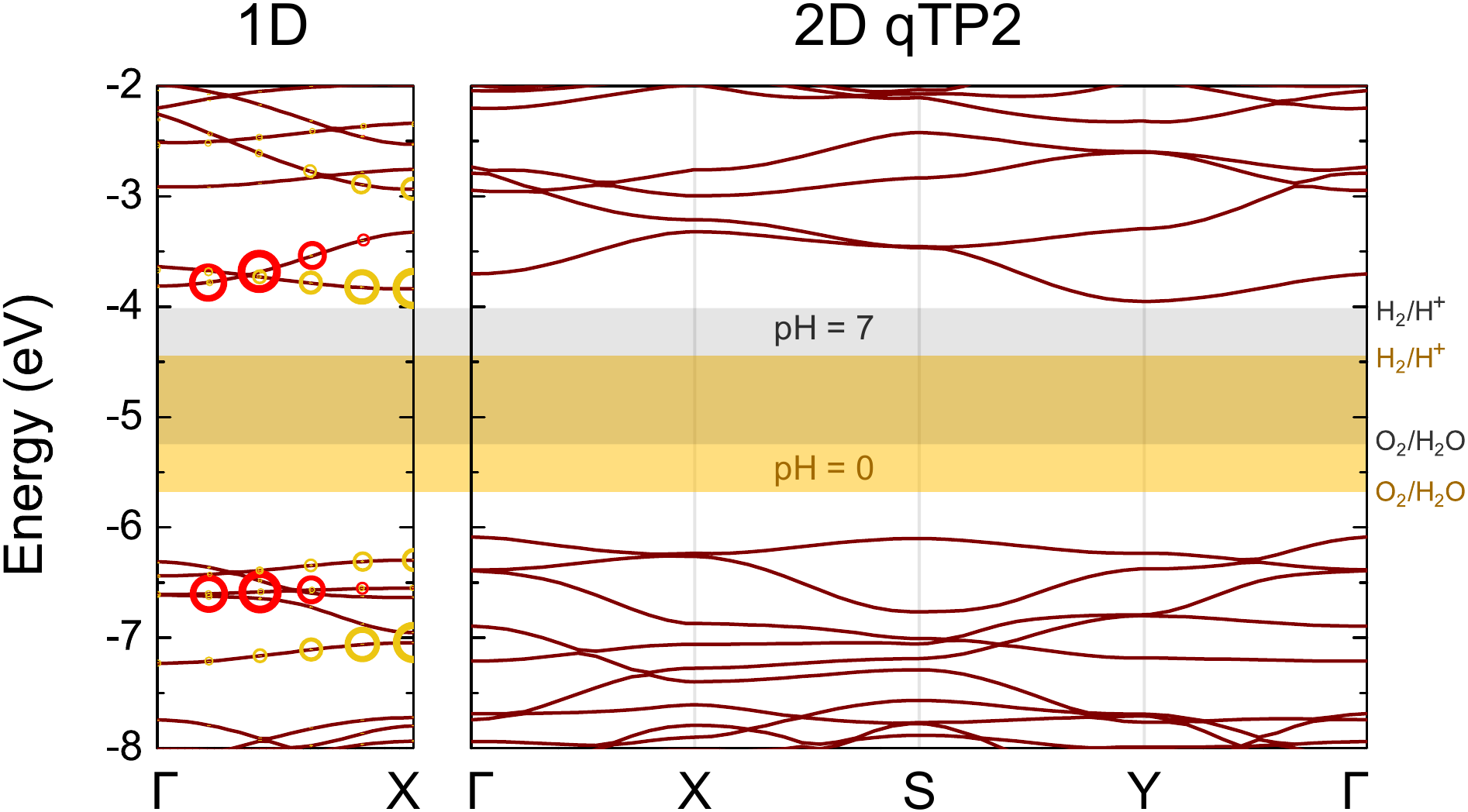}
    \caption{PBEsol0 band structures of the 1D C$_{60}$ chain and 2D C$_{60}$ networks. The red and yellow circles drawn on the 1D band structures indicate the contributions of the corresponding electron-hole pairs to the two brightest excitons near the band edges. The radii of these circles represent their corresponding oscillator strength\,\cite{Jones2023}. 
    }
    \label{fig:1D}
\end{figure}

Reducing the dimensionality of C$_{60}$ networks from 2D monolayers to 1D polymeric chains offers a unique opportunity to enhance photocatalytic performance. In these 1D structures, C$_{60}$ cages are covalently linked along [2\,+\,2] cycloaddition bonds, forming periodic chains. This geometry facilitates rapid charge extraction and increases the accessibility of catalytically active sites\,\cite{Jones2023}. Unlike monolayer fullerene networks, the linear confinement of 1D chains enforces directional charge transport, which can be harnessed in photocatalytic devices with nano architectures. Moreover, the reduced dielectric screening in 1D leads to larger band gaps compared to that of 2D networks (Fig.\,\ref{fig:1D}), indicating higher external potential for HER. 

To understand carrier separation in 1D chains, the electron-hole pairs for the two brightest excitons in 1D are indicated by red and yellow circles in Fig.\,\ref{fig:1D}, with larger radius representing higher oscillator strength. Because the holes are in much lower valence states than the VBM, their thermalisation to the VBM tends to dissociate the excitons, giving rise to effective electron-hole separation to their respective redox half reactions.

Benefiting from the favourable band-edge alignments for HER, 1D C$_{60}$ chains demonstrate higher catalytic potential than their 2D counterparts under a wider pH range. These chains also possess twice the number of surface active sites per C$_{60}$ unit compared to monolayer networks, leading to higher photocatalytic efficiency. Furthermore, the 1D chain is thermodynamically more stable than monolayer fullerene networks at room temperature, making it a more promising photocatalyst with higher efficiency and superior stability than the monolayers.

\subsubsection{Quasi-1D nanoribbons}

The fabrication of nanoribbons from monolayers has spurred advances in both fundamental science\,\cite{Nakada1996,Yazyev2013} and technological applications\,\cite{Chen2020b,Wang2021c}. Nanoribbons of graphene, for example, are quasi-one-dimensional strips of hexagonally bonded carbon atoms with edge-dependent properties\,\cite{Son2006} such as tuneable band gaps\,\cite{Chen2015b,Nguyen2017,Cernevics2020} and exotic phases including Dirac semimetallic\,\cite{Raza2008,Pizzochero2021b}, half-metallic\,\cite{Son2006a,Pizzochero2022}, magnetic\,\cite{Yazyev2010,Ma2025}, and topological\,\cite{Groning2018,Rizzo2018,Tepliakov2023} states. Using fullerene molecules as basic building blocks beyond carbon atoms can create even more complex edge geometry. Additionally, it has been predicted that many structural phases of monolayer fullerene networks tend to split into nanoribbons upon heating or under strain\,\cite{Peng2023,Ribeiro2022,Ying2023}. Thus understanding the impact of edges on fullerene nanoribbons\,\cite{Peng2025} is of particular relevance to a wide range of applications such as photocatalysis\,\cite{Peng2022c,Jones2023,Shearsby2025,Wu2025} and nanofiltration\,\cite{Tong2023,Tong2024}.

\begin{figure*}
    \centering
    \includegraphics[width=\linewidth]{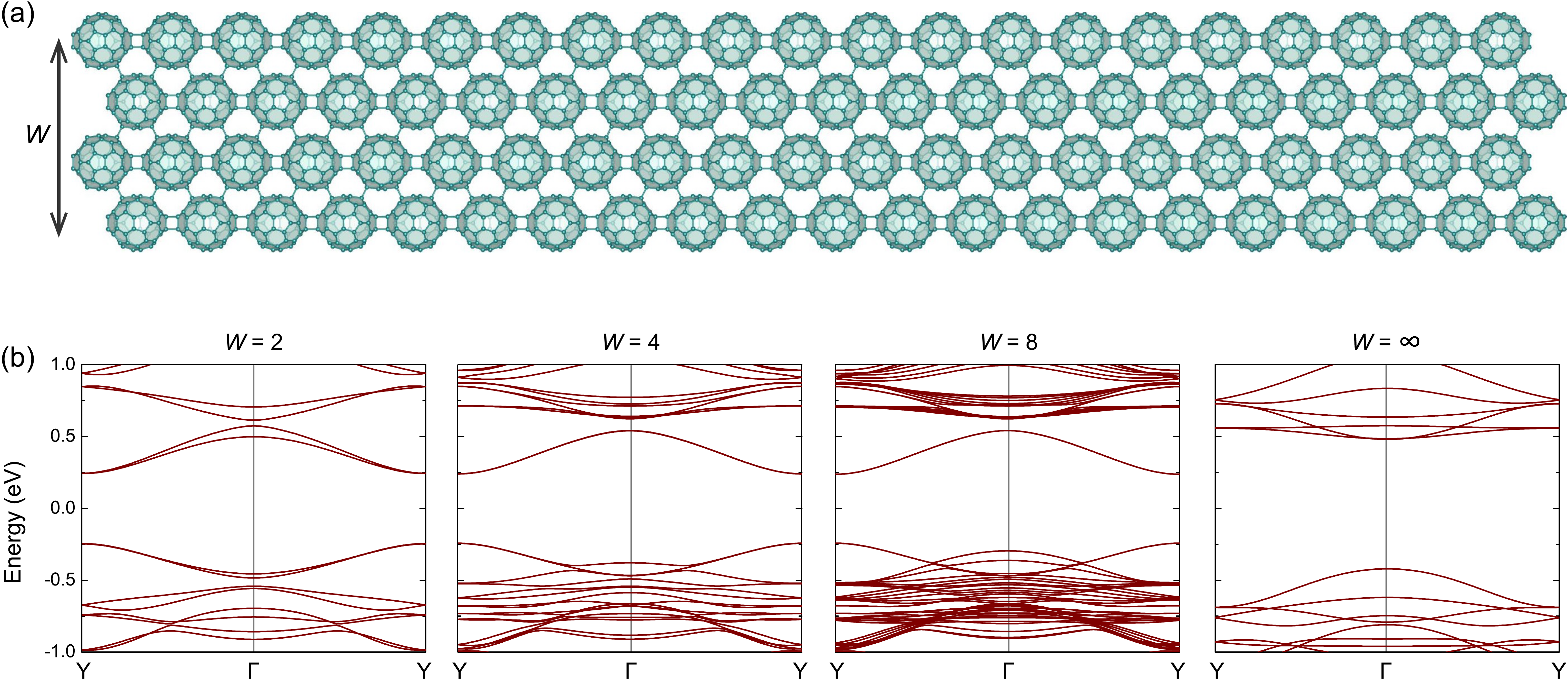}
    \caption{(a) Crystal structures of a representative qHP-AC nanoribbon, and (b) their corresponding band structures as a function of $W$\,\cite{Peng2025}. 
    }
    \label{fig:width}
\end{figure*}

In Fig.\,\ref{fig:width}(a), we show a representative qHP nanoribbon. Both edges terminate at the lattice site corresponding to intermolecular single bonds. This nanoribbon can be formed upon cleavage along the armchair crystallographic orientation of the monolayer. Therefore, we denote it as qHP-AC nanoribbons\,\cite{Peng2025}. The width of the nanoribbon $W$ is defined as the number of C$_{60}$ units spanning the non-periodic (transverse) direction. The qHP-AC nanoribbons crystallise in the $P222_1$ space group (No.\,17), with a lattice constant of 9.15\,\AA\ and a width of 30.75\,\AA\ for $W=4$.

The electronic band structure of qHP-AC nanoribbons shows additional in-gap states absent in the monolayer [Fig.\,\ref{fig:width}(b)]. These states are localised on the ribbon edges and remain fixed in number (two nearly-degenerate conduction bands and two nearly-degenerate valence bands) regardless of width $W$, whereas the bulk-like valence and conduction bands display replicas proportional to $W$. For narrow ribbons ($W=2$), the two in-gap valence states and the two in-gap conduction states appear within $\pm$0.5\,eV. As $W$ increases, interactions between the two edges diminish, leading to complete degeneracy of these in-gap states. As the CBMs at both $\Gamma$ and Y originate from the edge states, their effective masses quickly converge once $W > 2$. The same holds for the VBM at Y. On the other hand, the VBM at $\Gamma$ transits from edge-localised states to monolayer-like characters as $W$ increases, causing a sign reversal in the effective mass between between $W=3$ and $W=4$. Notably, the CBM at $\Gamma$ has negative effective mass, while the VBM at $\Gamma$ has positive effective mass with larger absolute value than that of the CBM. Therefore, the electron and hole at $\Gamma$ have positive total mass [$m({\rm e})+m({\rm h})$] but negative reduced mass [$1/m({\rm e})+1/m({\rm h})$]. In a classical picture, the electron-hole pairs at $\Gamma$ are expected to form excitons that behave differently from gravitational objects such as a binary star system\,\cite{Lin2021}.

In terms of band gaps, while the monolayer exhibits a direct band gap at $\Gamma$, the nanoribbons display their smallest direct band gap at Y. The presence of edge states leads to a 420\,meV reduction in the band gap relative to the monolayer. The band-gap reduction from edge states may account for the experimentally-observed discrepancy in electronic band gaps ($1.60-2.05$\,eV)\,\cite{Hou2022,Meirzadeh2023,Wang2023} and in optical band gaps ($1.10-1.55$\,eV)\,\cite{Hou2022,Zhang2025}, due to the finite size of the measured samples.

\begin{figure}[t]
    \centering
    \includegraphics[width=\linewidth]{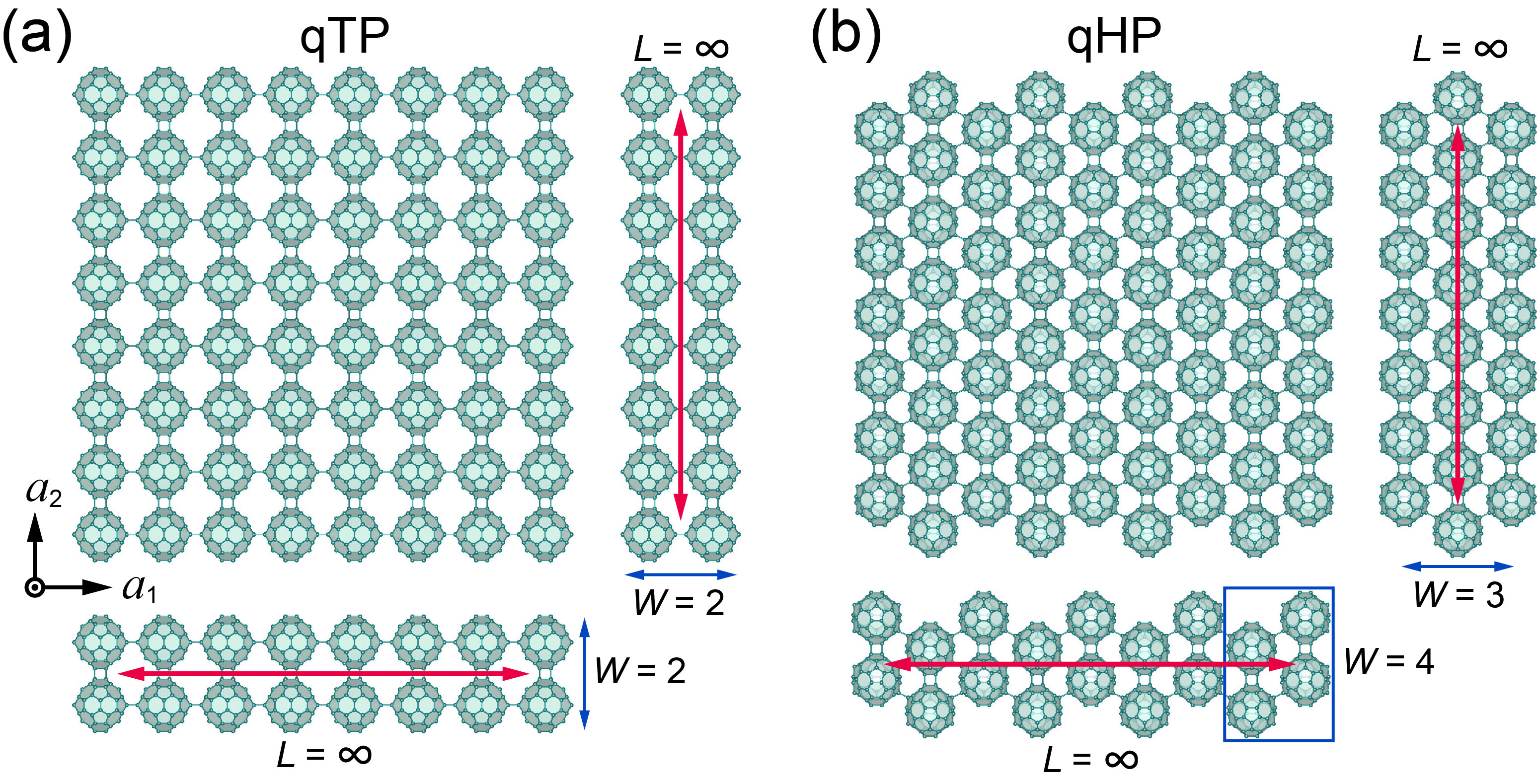}
    \caption{Crystal structures of monolayer (a) qTP and (b) qHP fullerene networks, as well as their corresponding nanoribbons\,\cite{Peng2025}. 
    }
    \label{fig:nanoribbons}
\end{figure}

Besides qHP-AC nanoribbons, a variety of other nanoribbons can be realised, depending on the crystalline direction and the resulting edge geometry, as shown in Fig.\,\ref{fig:nanoribbons}. For the qHP monolayers, nanoribbons cleaved along $a_1$ adopt a zigzag-like edge geometry, which is denoted as qHP-ZZ. The electronic structures of qHP-ZZ nanoribbons also exhibit in-gap edge states. Different from the dispersive in-gap states in the qHP-AC nanoribbons, the edge states in qHP-ZZ nanoribbons show flat-band features\,\cite{Peng2025}.  

For the qTP monolayers, the nanoribbons can be formed along either the vertical or the horizontal [2\,+\,2] cycloaddition bonding directions, which are denoted as qTP-V and qTP-H respectively. The two qTP nanoribbons also show a rich variety of electronic properties, including crossover from positive to negative effective mass with increased $W$, transitions from indirect to direct band gaps, and coexistence of both flat and dispersive bands\,\cite{Peng2025}.

Overall, fullerene nanoribbons, derived from experimentally-observed monolayer networks, offer a promising platform with tuneable electronic properties governed by edge geometry and width\,\cite{Peng2025}. First-principles calculations reveal diverse behaviours: depending on their crystallographic orientation, qTP nanoribbons exhibit either direct or indirect band gaps, while qHP nanoribbons host prominent in-gap states localised on the edges. These edge states lead to flat bands and unconventional effective masses, distinct from those of their parent monolayers. The evolution of band gaps, effective masses, and band widths with increasing nanoribbon width $W$ highlights the importance of finite-size effects. These findings position fullerene nanoribbons as versatile nanostructures for future nanoscale devices.

\section{Conclusion and outlook}\label{sect:conclusion}

In conclusion, monolayer fullerene networks represent an emerging family of molecular 2D materials with unique structural, electronic, and chemical properties distinct from atomically constructed monolayers. Their diversity in intermolecular bonding motifs enables the formation of rich lattice geometries with tuneable functionalities. Through first-principles calculations, we have provided a thorough understanding of the phase stability of the experimentally-observed structures and clarified the impact of intermolecular bonds on thermal expansion. Beyond their stability, monolayer polymeric C$_{60}$ fulfills all key criteria for photocatalysts: they possess suitable band-edge alignments, demonstrate strong optical absorption, and support efficient carrier separation and transport. Theoretical predictions of photocatalytic water splitting have been subsequently corroborated by experimental observations. Furthermore, the molecular nature of fullerene building blocks allows for functional tuning via interlayer stacking, molecular size, and dimensionality from 3D crystals to 1D chains and nanoribbons. These findings highlight a robust platform for the rational design of carbon-based nanomaterials.


Looking ahead, the development of controlled synthesis techniques and device integration strategies will be essential to fully harness the potential of 2D fullerene networks, while future theoretical studies will prove invaluable in complementing, guiding, and driving the experimental work.
Opportunities also lie in engineering moir{\'e} superlattices, heterostructures, and strain-tuneable systems for emergent quantum phenomena such as ferroelectricity and superconductivity. The intrinsic structural resilience as atom-like, stable building units, combined with the richness of their chemical degrees of freedom, positions monolayer fullerene networks as a platform to realise next-generation materials and devices for a wide range of applications at the nanoscale.

\section*{Conflicts of interest}
There are no conflicts to declare. 



\section*{Acknowledgements}
B.P. acknowledges support from Magdalene College Cambridge for a Nevile Research Fellowship. The calculations were performed using resources provided by the Cambridge Service for Data Driven Discovery (CSD3) operated by the University of Cambridge Research Computing Service (\url{www.csd3.cam.ac.uk}), provided by Dell EMC and Intel using Tier-2 funding from the Engineering and Physical Sciences Research Council (capital grant EP/T022159/1), and DiRAC funding from the Science and Technology Facilities Council (\url{http://www.dirac.ac.uk} ), as well as with computational support from the UK Materials and Molecular Modelling Hub, which is partially funded by EPSRC (EP/T022213/1, EP/W032260/1 and EP/P020194/1), for which access was obtained via the UKCP consortium and funded by EPSRC grant ref EP/P022561/1.


\begin{thebibliography}{243}%
\makeatletter
\providecommand \@ifxundefined [1]{%
 \@ifx{#1\undefined}
}%
\providecommand \@ifnum [1]{%
 \ifnum #1\expandafter \@firstoftwo
 \else \expandafter \@secondoftwo
 \fi
}%
\providecommand \@ifx [1]{%
 \ifx #1\expandafter \@firstoftwo
 \else \expandafter \@secondoftwo
 \fi
}%
\providecommand \natexlab [1]{#1}%
\providecommand \enquote  [1]{``#1''}%
\providecommand \bibnamefont  [1]{#1}%
\providecommand \bibfnamefont [1]{#1}%
\providecommand \citenamefont [1]{#1}%
\providecommand \href@noop [0]{\@secondoftwo}%
\providecommand \href [0]{\begingroup \@sanitize@url \@href}%
\providecommand \@href[1]{\@@startlink{#1}\@@href}%
\providecommand \@@href[1]{\endgroup#1\@@endlink}%
\providecommand \@sanitize@url [0]{\catcode `\\12\catcode `\$12\catcode
  `\&12\catcode `\#12\catcode `\^12\catcode `\_12\catcode `\%12\relax}%
\providecommand \@@startlink[1]{}%
\providecommand \@@endlink[0]{}%
\providecommand \url  [0]{\begingroup\@sanitize@url \@url }%
\providecommand \@url [1]{\endgroup\@href {#1}{\urlprefix }}%
\providecommand \urlprefix  [0]{URL }%
\providecommand \Eprint [0]{\href }%
\providecommand \doibase [0]{http://dx.doi.org/}%
\providecommand \selectlanguage [0]{\@gobble}%
\providecommand \bibinfo  [0]{\@secondoftwo}%
\providecommand \bibfield  [0]{\@secondoftwo}%
\providecommand \translation [1]{[#1]}%
\providecommand \BibitemOpen [0]{}%
\providecommand \bibitemStop [0]{}%
\providecommand \bibitemNoStop [0]{.\EOS\space}%
\providecommand \EOS [0]{\spacefactor3000\relax}%
\providecommand \BibitemShut  [1]{\csname bibitem#1\endcsname}%
\let\auto@bib@innerbib\@empty
\bibitem [{\citenamefont {Novoselov}\ \emph {et~al.}(2004)\citenamefont
  {Novoselov}, \citenamefont {Geim}, \citenamefont {Morozov}, \citenamefont
  {Jiang}, \citenamefont {Zhang}, \citenamefont {Dubonos}, \citenamefont
  {Grigorieva},\ and\ \citenamefont {Firsov}}]{Novoselov2004}%
  \BibitemOpen
  \bibfield  {author} {\bibinfo {author} {\bibfnamefont {K.~S.}\ \bibnamefont
  {Novoselov}}, \bibinfo {author} {\bibfnamefont {A.~K.}\ \bibnamefont {Geim}},
  \bibinfo {author} {\bibfnamefont {S.~V.}\ \bibnamefont {Morozov}}, \bibinfo
  {author} {\bibfnamefont {D.}~\bibnamefont {Jiang}}, \bibinfo {author}
  {\bibfnamefont {Y.}~\bibnamefont {Zhang}}, \bibinfo {author} {\bibfnamefont
  {S.~V.}\ \bibnamefont {Dubonos}}, \bibinfo {author} {\bibfnamefont {I.~V.}\
  \bibnamefont {Grigorieva}}, \ and\ \bibinfo {author} {\bibfnamefont {A.~A.}\
  \bibnamefont {Firsov}},\ }\href {\doibase 10.1126/science.1102896} {\bibfield
   {journal} {\bibinfo  {journal} {Science}\ }\textbf {\bibinfo {volume}
  {306}},\ \bibinfo {pages} {666} (\bibinfo {year} {2004})}\BibitemShut
  {NoStop}%
\bibitem [{\citenamefont {Novoselov}\ \emph {et~al.}(2005)\citenamefont
  {Novoselov}, \citenamefont {Geim}, \citenamefont {Morozov}, \citenamefont
  {Jiang}, \citenamefont {Katsnelson}, \citenamefont {Grigorieva},
  \citenamefont {Dubonos},\ and\ \citenamefont {Firsov}}]{Novoselov2005}%
  \BibitemOpen
  \bibfield  {author} {\bibinfo {author} {\bibfnamefont {K.~S.}\ \bibnamefont
  {Novoselov}}, \bibinfo {author} {\bibfnamefont {A.~K.}\ \bibnamefont {Geim}},
  \bibinfo {author} {\bibfnamefont {S.~V.}\ \bibnamefont {Morozov}}, \bibinfo
  {author} {\bibfnamefont {D.}~\bibnamefont {Jiang}}, \bibinfo {author}
  {\bibfnamefont {M.~I.}\ \bibnamefont {Katsnelson}}, \bibinfo {author}
  {\bibfnamefont {I.~V.}\ \bibnamefont {Grigorieva}}, \bibinfo {author}
  {\bibfnamefont {S.~V.}\ \bibnamefont {Dubonos}}, \ and\ \bibinfo {author}
  {\bibfnamefont {A.~A.}\ \bibnamefont {Firsov}},\ }\href {\doibase
  10.1038/nature04233} {\bibfield  {journal} {\bibinfo  {journal} {Nature}\
  }\textbf {\bibinfo {volume} {438}},\ \bibinfo {pages} {197} (\bibinfo {year}
  {2005})}\BibitemShut {NoStop}%
\bibitem [{\citenamefont {Zhang}\ \emph {et~al.}(2005)\citenamefont {Zhang},
  \citenamefont {Tan}, \citenamefont {Stormer},\ and\ \citenamefont
  {Kim}}]{Zhang2005}%
  \BibitemOpen
  \bibfield  {author} {\bibinfo {author} {\bibfnamefont {Y.}~\bibnamefont
  {Zhang}}, \bibinfo {author} {\bibfnamefont {Y.-W.}\ \bibnamefont {Tan}},
  \bibinfo {author} {\bibfnamefont {H.~L.}\ \bibnamefont {Stormer}}, \ and\
  \bibinfo {author} {\bibfnamefont {P.}~\bibnamefont {Kim}},\ }\href {\doibase
  10.1038/nature04235} {\bibfield  {journal} {\bibinfo  {journal} {Nature}\
  }\textbf {\bibinfo {volume} {438}},\ \bibinfo {pages} {201} (\bibinfo {year}
  {2005})}\BibitemShut {NoStop}%
\bibitem [{\citenamefont {Geim}(2009)}]{Geim2009}%
  \BibitemOpen
  \bibfield  {author} {\bibinfo {author} {\bibfnamefont {A.~K.}\ \bibnamefont
  {Geim}},\ }\href@noop {} {\bibfield  {journal} {\bibinfo  {journal}
  {Science}\ }\textbf {\bibinfo {volume} {324}},\ \bibinfo {pages} {1530}
  (\bibinfo {year} {2009})}\BibitemShut {NoStop}%
\bibitem [{\citenamefont {Novoselov}\ \emph {et~al.}(2012)\citenamefont
  {Novoselov}, \citenamefont {Fal'ko}, \citenamefont {Colombo}, \citenamefont
  {Gellert}, \citenamefont {Schwab},\ and\ \citenamefont
  {Kim}}]{Novoselov2012}%
  \BibitemOpen
  \bibfield  {author} {\bibinfo {author} {\bibfnamefont {K.~S.}\ \bibnamefont
  {Novoselov}}, \bibinfo {author} {\bibfnamefont {V.~I.}\ \bibnamefont
  {Fal'ko}}, \bibinfo {author} {\bibfnamefont {L.}~\bibnamefont {Colombo}},
  \bibinfo {author} {\bibfnamefont {P.~R.}\ \bibnamefont {Gellert}}, \bibinfo
  {author} {\bibfnamefont {M.~G.}\ \bibnamefont {Schwab}}, \ and\ \bibinfo
  {author} {\bibfnamefont {K.}~\bibnamefont {Kim}},\ }\href@noop {} {\bibfield
  {journal} {\bibinfo  {journal} {Nature}\ }\textbf {\bibinfo {volume} {490}},\
  \bibinfo {pages} {192} (\bibinfo {year} {2012})}\BibitemShut {NoStop}%
\bibitem [{\citenamefont {Gong}\ \emph {et~al.}(2017)\citenamefont {Gong},
  \citenamefont {Li}, \citenamefont {Li}, \citenamefont {Ji}, \citenamefont
  {Stern}, \citenamefont {Xia}, \citenamefont {Cao}, \citenamefont {Bao},
  \citenamefont {Wang}, \citenamefont {Wang}, \citenamefont {Qiu},
  \citenamefont {Cava}, \citenamefont {Louie}, \citenamefont {Xia},\ and\
  \citenamefont {Zhang}}]{Gong2017}%
  \BibitemOpen
  \bibfield  {author} {\bibinfo {author} {\bibfnamefont {C.}~\bibnamefont
  {Gong}}, \bibinfo {author} {\bibfnamefont {L.}~\bibnamefont {Li}}, \bibinfo
  {author} {\bibfnamefont {Z.}~\bibnamefont {Li}}, \bibinfo {author}
  {\bibfnamefont {H.}~\bibnamefont {Ji}}, \bibinfo {author} {\bibfnamefont
  {A.}~\bibnamefont {Stern}}, \bibinfo {author} {\bibfnamefont
  {Y.}~\bibnamefont {Xia}}, \bibinfo {author} {\bibfnamefont {T.}~\bibnamefont
  {Cao}}, \bibinfo {author} {\bibfnamefont {W.}~\bibnamefont {Bao}}, \bibinfo
  {author} {\bibfnamefont {C.}~\bibnamefont {Wang}}, \bibinfo {author}
  {\bibfnamefont {Y.}~\bibnamefont {Wang}}, \bibinfo {author} {\bibfnamefont
  {Z.~Q.}\ \bibnamefont {Qiu}}, \bibinfo {author} {\bibfnamefont {R.~J.}\
  \bibnamefont {Cava}}, \bibinfo {author} {\bibfnamefont {S.~G.}\ \bibnamefont
  {Louie}}, \bibinfo {author} {\bibfnamefont {J.}~\bibnamefont {Xia}}, \ and\
  \bibinfo {author} {\bibfnamefont {X.}~\bibnamefont {Zhang}},\ }\href
  {\doibase 10.1038/nature22060} {\bibfield  {journal} {\bibinfo  {journal}
  {Nature}\ }\textbf {\bibinfo {volume} {546}},\ \bibinfo {pages} {265}
  (\bibinfo {year} {2017})}\BibitemShut {NoStop}%
\bibitem [{\citenamefont {Huang}\ \emph {et~al.}(2017)\citenamefont {Huang},
  \citenamefont {Clark}, \citenamefont {Navarro-Moratalla}, \citenamefont
  {Klein}, \citenamefont {Cheng}, \citenamefont {Seyler}, \citenamefont
  {Zhong}, \citenamefont {Schmidgall}, \citenamefont {McGuire}, \citenamefont
  {Cobden}, \citenamefont {Yao}, \citenamefont {Xiao}, \citenamefont
  {Jarillo-Herrero},\ and\ \citenamefont {Xu}}]{Huang2017}%
  \BibitemOpen
  \bibfield  {author} {\bibinfo {author} {\bibfnamefont {B.}~\bibnamefont
  {Huang}}, \bibinfo {author} {\bibfnamefont {G.}~\bibnamefont {Clark}},
  \bibinfo {author} {\bibfnamefont {E.}~\bibnamefont {Navarro-Moratalla}},
  \bibinfo {author} {\bibfnamefont {D.~R.}\ \bibnamefont {Klein}}, \bibinfo
  {author} {\bibfnamefont {R.}~\bibnamefont {Cheng}}, \bibinfo {author}
  {\bibfnamefont {K.~L.}\ \bibnamefont {Seyler}}, \bibinfo {author}
  {\bibfnamefont {D.}~\bibnamefont {Zhong}}, \bibinfo {author} {\bibfnamefont
  {E.}~\bibnamefont {Schmidgall}}, \bibinfo {author} {\bibfnamefont {M.~A.}\
  \bibnamefont {McGuire}}, \bibinfo {author} {\bibfnamefont {D.~H.}\
  \bibnamefont {Cobden}}, \bibinfo {author} {\bibfnamefont {W.}~\bibnamefont
  {Yao}}, \bibinfo {author} {\bibfnamefont {D.}~\bibnamefont {Xiao}}, \bibinfo
  {author} {\bibfnamefont {P.}~\bibnamefont {Jarillo-Herrero}}, \ and\ \bibinfo
  {author} {\bibfnamefont {X.}~\bibnamefont {Xu}},\ }\href {\doibase
  10.1038/nature22391} {\bibfield  {journal} {\bibinfo  {journal} {Nature}\
  }\textbf {\bibinfo {volume} {546}},\ \bibinfo {pages} {270} (\bibinfo {year}
  {2017})}\BibitemShut {NoStop}%
\bibitem [{\citenamefont {Deng}\ \emph {et~al.}(2018)\citenamefont {Deng},
  \citenamefont {Yu}, \citenamefont {Song}, \citenamefont {Zhang},
  \citenamefont {Wang}, \citenamefont {Sun}, \citenamefont {Yi}, \citenamefont
  {Wu}, \citenamefont {Wu}, \citenamefont {Zhu}, \citenamefont {Wang},
  \citenamefont {Chen},\ and\ \citenamefont {Zhang}}]{Deng2018}%
  \BibitemOpen
  \bibfield  {author} {\bibinfo {author} {\bibfnamefont {Y.}~\bibnamefont
  {Deng}}, \bibinfo {author} {\bibfnamefont {Y.}~\bibnamefont {Yu}}, \bibinfo
  {author} {\bibfnamefont {Y.}~\bibnamefont {Song}}, \bibinfo {author}
  {\bibfnamefont {J.}~\bibnamefont {Zhang}}, \bibinfo {author} {\bibfnamefont
  {N.~Z.}\ \bibnamefont {Wang}}, \bibinfo {author} {\bibfnamefont
  {Z.}~\bibnamefont {Sun}}, \bibinfo {author} {\bibfnamefont {Y.}~\bibnamefont
  {Yi}}, \bibinfo {author} {\bibfnamefont {Y.~Z.}\ \bibnamefont {Wu}}, \bibinfo
  {author} {\bibfnamefont {S.}~\bibnamefont {Wu}}, \bibinfo {author}
  {\bibfnamefont {J.}~\bibnamefont {Zhu}}, \bibinfo {author} {\bibfnamefont
  {J.}~\bibnamefont {Wang}}, \bibinfo {author} {\bibfnamefont {X.~H.}\
  \bibnamefont {Chen}}, \ and\ \bibinfo {author} {\bibfnamefont
  {Y.}~\bibnamefont {Zhang}},\ }\href {\doibase 10.1038/s41586-018-0626-9}
  {\bibfield  {journal} {\bibinfo  {journal} {Nature}\ }\textbf {\bibinfo
  {volume} {563}},\ \bibinfo {pages} {94} (\bibinfo {year} {2018})}\BibitemShut
  {NoStop}%
\bibitem [{\citenamefont {Pizzochero}\ \emph {et~al.}(2020)\citenamefont
  {Pizzochero}, \citenamefont {Yadav},\ and\ \citenamefont
  {Yazyev}}]{Pizzochero2020}%
  \BibitemOpen
  \bibfield  {author} {\bibinfo {author} {\bibfnamefont {M.}~\bibnamefont
  {Pizzochero}}, \bibinfo {author} {\bibfnamefont {R.}~\bibnamefont {Yadav}}, \
  and\ \bibinfo {author} {\bibfnamefont {O.~V.}\ \bibnamefont {Yazyev}},\
  }\href {\doibase 10.1088/2053-1583/ab7cab} {\bibfield  {journal} {\bibinfo
  {journal} {2D Mater.}\ }\textbf {\bibinfo {volume} {7}},\ \bibinfo {pages}
  {035005} (\bibinfo {year} {2020})}\BibitemShut {NoStop}%
\bibitem [{\citenamefont {Pizzochero}\ and\ \citenamefont
  {Yazyev}(2020)}]{Pizzochero2020a}%
  \BibitemOpen
  \bibfield  {author} {\bibinfo {author} {\bibfnamefont {M.}~\bibnamefont
  {Pizzochero}}\ and\ \bibinfo {author} {\bibfnamefont {O.~V.}\ \bibnamefont
  {Yazyev}},\ }\href {\doibase 10.1021/acs.jpcc.0c01873} {\bibfield  {journal}
  {\bibinfo  {journal} {J. Phys. Chem. C}\ }\textbf {\bibinfo {volume} {124}},\
  \bibinfo {pages} {7585} (\bibinfo {year} {2020})}\BibitemShut {NoStop}%
\bibitem [{\citenamefont {Zhang}\ \emph
  {et~al.}(2015{\natexlab{a}})\citenamefont {Zhang}, \citenamefont {Yan},
  \citenamefont {Li}, \citenamefont {Chen},\ and\ \citenamefont
  {Zeng}}]{Zhang2015e}%
  \BibitemOpen
  \bibfield  {author} {\bibinfo {author} {\bibfnamefont {S.}~\bibnamefont
  {Zhang}}, \bibinfo {author} {\bibfnamefont {Z.}~\bibnamefont {Yan}}, \bibinfo
  {author} {\bibfnamefont {Y.}~\bibnamefont {Li}}, \bibinfo {author}
  {\bibfnamefont {Z.}~\bibnamefont {Chen}}, \ and\ \bibinfo {author}
  {\bibfnamefont {H.}~\bibnamefont {Zeng}},\ }\href {\doibase
  10.1002/anie.201411246} {\bibfield  {journal} {\bibinfo  {journal} {Angew.
  Chem. Int. Ed.}\ }\textbf {\bibinfo {volume} {54}},\ \bibinfo {pages} {3112}
  (\bibinfo {year} {2015}{\natexlab{a}})}\BibitemShut {NoStop}%
\bibitem [{\citenamefont {Peng}\ \emph
  {et~al.}(2016{\natexlab{a}})\citenamefont {Peng}, \citenamefont {Zhang},
  \citenamefont {Shao}, \citenamefont {Xu}, \citenamefont {Zhang},\ and\
  \citenamefont {Zhu}}]{Peng2016d}%
  \BibitemOpen
  \bibfield  {author} {\bibinfo {author} {\bibfnamefont {B.}~\bibnamefont
  {Peng}}, \bibinfo {author} {\bibfnamefont {H.}~\bibnamefont {Zhang}},
  \bibinfo {author} {\bibfnamefont {H.}~\bibnamefont {Shao}}, \bibinfo {author}
  {\bibfnamefont {Y.}~\bibnamefont {Xu}}, \bibinfo {author} {\bibfnamefont
  {R.}~\bibnamefont {Zhang}}, \ and\ \bibinfo {author} {\bibfnamefont
  {H.}~\bibnamefont {Zhu}},\ }\href {\doibase 10.1039/c6tc00115g} {\bibfield
  {journal} {\bibinfo  {journal} {J. Mater. Chem. C}\ }\textbf {\bibinfo
  {volume} {4}},\ \bibinfo {pages} {3592} (\bibinfo {year}
  {2016}{\natexlab{a}})}\BibitemShut {NoStop}%
\bibitem [{\citenamefont {Peng}\ \emph
  {et~al.}(2018{\natexlab{a}})\citenamefont {Peng}, \citenamefont {Zhang},
  \citenamefont {Shao}, \citenamefont {Xu}, \citenamefont {Ni}, \citenamefont
  {Wu}, \citenamefont {Li}, \citenamefont {Lu}, \citenamefont {Jin},\ and\
  \citenamefont {Zhu}}]{Peng2018c}%
  \BibitemOpen
  \bibfield  {author} {\bibinfo {author} {\bibfnamefont {B.}~\bibnamefont
  {Peng}}, \bibinfo {author} {\bibfnamefont {H.}~\bibnamefont {Zhang}},
  \bibinfo {author} {\bibfnamefont {H.}~\bibnamefont {Shao}}, \bibinfo {author}
  {\bibfnamefont {K.}~\bibnamefont {Xu}}, \bibinfo {author} {\bibfnamefont
  {G.}~\bibnamefont {Ni}}, \bibinfo {author} {\bibfnamefont {L.}~\bibnamefont
  {Wu}}, \bibinfo {author} {\bibfnamefont {J.}~\bibnamefont {Li}}, \bibinfo
  {author} {\bibfnamefont {H.}~\bibnamefont {Lu}}, \bibinfo {author}
  {\bibfnamefont {Q.}~\bibnamefont {Jin}}, \ and\ \bibinfo {author}
  {\bibfnamefont {H.}~\bibnamefont {Zhu}},\ }\href {\doibase
  10.1021/acsphotonics.8b00757} {\bibfield  {journal} {\bibinfo  {journal} {ACS
  Photonics}\ }\textbf {\bibinfo {volume} {5}},\ \bibinfo {pages} {4081}
  (\bibinfo {year} {2018}{\natexlab{a}})}\BibitemShut {NoStop}%
\bibitem [{\citenamefont {Kane}\ and\ \citenamefont {Mele}(2005)}]{Kane2005}%
  \BibitemOpen
  \bibfield  {author} {\bibinfo {author} {\bibfnamefont {C.~L.}\ \bibnamefont
  {Kane}}\ and\ \bibinfo {author} {\bibfnamefont {E.~J.}\ \bibnamefont
  {Mele}},\ }\href {\doibase 10.1103/PhysRevLett.95.226801} {\bibfield
  {journal} {\bibinfo  {journal} {Phys. Rev. Lett.}\ }\textbf {\bibinfo
  {volume} {95}},\ \bibinfo {pages} {226801} (\bibinfo {year}
  {2005})}\BibitemShut {NoStop}%
\bibitem [{\citenamefont {Wada}\ \emph {et~al.}(2011)\citenamefont {Wada},
  \citenamefont {Murakami}, \citenamefont {Freimuth},\ and\ \citenamefont
  {Bihlmayer}}]{Wada2011}%
  \BibitemOpen
  \bibfield  {author} {\bibinfo {author} {\bibfnamefont {M.}~\bibnamefont
  {Wada}}, \bibinfo {author} {\bibfnamefont {S.}~\bibnamefont {Murakami}},
  \bibinfo {author} {\bibfnamefont {F.}~\bibnamefont {Freimuth}}, \ and\
  \bibinfo {author} {\bibfnamefont {G.}~\bibnamefont {Bihlmayer}},\ }\href
  {\doibase 10.1103/PhysRevB.83.121310} {\bibfield  {journal} {\bibinfo
  {journal} {Phys. Rev. B}\ }\textbf {\bibinfo {volume} {83}},\ \bibinfo
  {pages} {121310} (\bibinfo {year} {2011})}\BibitemShut {NoStop}%
\bibitem [{\citenamefont {Liu}\ \emph {et~al.}(2011)\citenamefont {Liu},
  \citenamefont {Jiang},\ and\ \citenamefont {Yao}}]{Liu2011}%
  \BibitemOpen
  \bibfield  {author} {\bibinfo {author} {\bibfnamefont {C.-C.}\ \bibnamefont
  {Liu}}, \bibinfo {author} {\bibfnamefont {H.}~\bibnamefont {Jiang}}, \ and\
  \bibinfo {author} {\bibfnamefont {Y.}~\bibnamefont {Yao}},\ }\href {\doibase
  10.1103/PhysRevB.84.195430} {\bibfield  {journal} {\bibinfo  {journal} {Phys.
  Rev. B}\ }\textbf {\bibinfo {volume} {84}},\ \bibinfo {pages} {195430}
  (\bibinfo {year} {2011})}\BibitemShut {NoStop}%
\bibitem [{\citenamefont {Cai}\ \emph {et~al.}(2015)\citenamefont {Cai},
  \citenamefont {Zhang}, \citenamefont {Hu}, \citenamefont {Hu}, \citenamefont
  {Zou},\ and\ \citenamefont {Zeng}}]{Cai2015}%
  \BibitemOpen
  \bibfield  {author} {\bibinfo {author} {\bibfnamefont {B.}~\bibnamefont
  {Cai}}, \bibinfo {author} {\bibfnamefont {S.}~\bibnamefont {Zhang}}, \bibinfo
  {author} {\bibfnamefont {Z.}~\bibnamefont {Hu}}, \bibinfo {author}
  {\bibfnamefont {Y.}~\bibnamefont {Hu}}, \bibinfo {author} {\bibfnamefont
  {Y.}~\bibnamefont {Zou}}, \ and\ \bibinfo {author} {\bibfnamefont
  {H.}~\bibnamefont {Zeng}},\ }\href {\doibase 10.1039/C5CP00563A} {\bibfield
  {journal} {\bibinfo  {journal} {Phys. Chem. Chem. Phys.}\ }\textbf {\bibinfo
  {volume} {17}},\ \bibinfo {pages} {12634} (\bibinfo {year}
  {2015})}\BibitemShut {NoStop}%
\bibitem [{\citenamefont {Zhu}\ \emph {et~al.}(2015)\citenamefont {Zhu},
  \citenamefont {Chen}, \citenamefont {Xu}, \citenamefont {Gao}, \citenamefont
  {Guan}, \citenamefont {Liu}, \citenamefont {Qian}, \citenamefont {Zhang},\
  and\ \citenamefont {Jia}}]{Zhu2015}%
  \BibitemOpen
  \bibfield  {author} {\bibinfo {author} {\bibfnamefont {F.-f.}\ \bibnamefont
  {Zhu}}, \bibinfo {author} {\bibfnamefont {W.-j.}\ \bibnamefont {Chen}},
  \bibinfo {author} {\bibfnamefont {Y.}~\bibnamefont {Xu}}, \bibinfo {author}
  {\bibfnamefont {C.-l.}\ \bibnamefont {Gao}}, \bibinfo {author} {\bibfnamefont
  {D.-d.}\ \bibnamefont {Guan}}, \bibinfo {author} {\bibfnamefont {C.-h.}\
  \bibnamefont {Liu}}, \bibinfo {author} {\bibfnamefont {D.}~\bibnamefont
  {Qian}}, \bibinfo {author} {\bibfnamefont {S.-C.}\ \bibnamefont {Zhang}}, \
  and\ \bibinfo {author} {\bibfnamefont {J.-f.}\ \bibnamefont {Jia}},\ }\href
  {\doibase 10.1038/nmat4384} {\bibfield  {journal} {\bibinfo  {journal} {Nat
  Mater}\ }\textbf {\bibinfo {volume} {14}},\ \bibinfo {pages} {1020} (\bibinfo
  {year} {2015})}\BibitemShut {NoStop}%
\bibitem [{\citenamefont {Balandin}\ \emph {et~al.}(2008)\citenamefont
  {Balandin}, \citenamefont {Ghosh}, \citenamefont {Bao}, \citenamefont
  {Calizo}, \citenamefont {Teweldebrhan}, \citenamefont {Miao},\ and\
  \citenamefont {Lau}}]{Balandin2008}%
  \BibitemOpen
  \bibfield  {author} {\bibinfo {author} {\bibfnamefont {A.~A.}\ \bibnamefont
  {Balandin}}, \bibinfo {author} {\bibfnamefont {S.}~\bibnamefont {Ghosh}},
  \bibinfo {author} {\bibfnamefont {W.}~\bibnamefont {Bao}}, \bibinfo {author}
  {\bibfnamefont {I.}~\bibnamefont {Calizo}}, \bibinfo {author} {\bibfnamefont
  {D.}~\bibnamefont {Teweldebrhan}}, \bibinfo {author} {\bibfnamefont
  {F.}~\bibnamefont {Miao}}, \ and\ \bibinfo {author} {\bibfnamefont {C.~N.}\
  \bibnamefont {Lau}},\ }\href {\doibase 10.1021/nl0731872} {\bibfield
  {journal} {\bibinfo  {journal} {Nano Lett.}\ }\textbf {\bibinfo {volume}
  {8}},\ \bibinfo {pages} {902} (\bibinfo {year} {2008})}\BibitemShut {NoStop}%
\bibitem [{\citenamefont {Seol}\ \emph {et~al.}(2010)\citenamefont {Seol},
  \citenamefont {Jo}, \citenamefont {Moore}, \citenamefont {Lindsay},
  \citenamefont {Aitken}, \citenamefont {Pettes}, \citenamefont {Li},
  \citenamefont {Yao}, \citenamefont {Huang}, \citenamefont {Broido},
  \citenamefont {Mingo}, \citenamefont {Ruoff},\ and\ \citenamefont
  {Shi}}]{Seol2010}%
  \BibitemOpen
  \bibfield  {author} {\bibinfo {author} {\bibfnamefont {J.~H.}\ \bibnamefont
  {Seol}}, \bibinfo {author} {\bibfnamefont {I.}~\bibnamefont {Jo}}, \bibinfo
  {author} {\bibfnamefont {A.~L.}\ \bibnamefont {Moore}}, \bibinfo {author}
  {\bibfnamefont {L.}~\bibnamefont {Lindsay}}, \bibinfo {author} {\bibfnamefont
  {Z.~H.}\ \bibnamefont {Aitken}}, \bibinfo {author} {\bibfnamefont {M.~T.}\
  \bibnamefont {Pettes}}, \bibinfo {author} {\bibfnamefont {X.}~\bibnamefont
  {Li}}, \bibinfo {author} {\bibfnamefont {Z.}~\bibnamefont {Yao}}, \bibinfo
  {author} {\bibfnamefont {R.}~\bibnamefont {Huang}}, \bibinfo {author}
  {\bibfnamefont {D.}~\bibnamefont {Broido}}, \bibinfo {author} {\bibfnamefont
  {N.}~\bibnamefont {Mingo}}, \bibinfo {author} {\bibfnamefont {R.~S.}\
  \bibnamefont {Ruoff}}, \ and\ \bibinfo {author} {\bibfnamefont
  {L.}~\bibnamefont {Shi}},\ }\href {\doibase 10.1126/science.1184014}
  {\bibfield  {journal} {\bibinfo  {journal} {Science}\ }\textbf {\bibinfo
  {volume} {328}},\ \bibinfo {pages} {213} (\bibinfo {year}
  {2010})}\BibitemShut {NoStop}%
\bibitem [{\citenamefont {Peng}\ \emph
  {et~al.}(2016{\natexlab{b}})\citenamefont {Peng}, \citenamefont {Zhang},
  \citenamefont {Shao}, \citenamefont {Xu}, \citenamefont {Zhang},\ and\
  \citenamefont {Zhu}}]{Peng2016b}%
  \BibitemOpen
  \bibfield  {author} {\bibinfo {author} {\bibfnamefont {B.}~\bibnamefont
  {Peng}}, \bibinfo {author} {\bibfnamefont {H.}~\bibnamefont {Zhang}},
  \bibinfo {author} {\bibfnamefont {H.}~\bibnamefont {Shao}}, \bibinfo {author}
  {\bibfnamefont {Y.}~\bibnamefont {Xu}}, \bibinfo {author} {\bibfnamefont
  {X.}~\bibnamefont {Zhang}}, \ and\ \bibinfo {author} {\bibfnamefont
  {H.}~\bibnamefont {Zhu}},\ }\href {\doibase 10.1038/srep20225} {\bibfield
  {journal} {\bibinfo  {journal} {Sci. Rep.}\ }\textbf {\bibinfo {volume}
  {6}},\ \bibinfo {pages} {20225} (\bibinfo {year}
  {2016}{\natexlab{b}})}\BibitemShut {NoStop}%
\bibitem [{\citenamefont {Peng}\ \emph
  {et~al.}(2016{\natexlab{c}})\citenamefont {Peng}, \citenamefont {Zhang},
  \citenamefont {Shao}, \citenamefont {Xu}, \citenamefont {Zhang},\ and\
  \citenamefont {Zhu}}]{Peng2016c}%
  \BibitemOpen
  \bibfield  {author} {\bibinfo {author} {\bibfnamefont {B.}~\bibnamefont
  {Peng}}, \bibinfo {author} {\bibfnamefont {H.}~\bibnamefont {Zhang}},
  \bibinfo {author} {\bibfnamefont {H.}~\bibnamefont {Shao}}, \bibinfo {author}
  {\bibfnamefont {Y.}~\bibnamefont {Xu}}, \bibinfo {author} {\bibfnamefont
  {X.}~\bibnamefont {Zhang}}, \ and\ \bibinfo {author} {\bibfnamefont
  {H.}~\bibnamefont {Zhu}},\ }\href {\doibase 10.1002/andp.201500354}
  {\bibfield  {journal} {\bibinfo  {journal} {Annalen der Physik}\ }\textbf
  {\bibinfo {volume} {528}},\ \bibinfo {pages} {504} (\bibinfo {year}
  {2016}{\natexlab{c}})}\BibitemShut {NoStop}%
\bibitem [{\citenamefont {Peng}\ \emph
  {et~al.}(2016{\natexlab{d}})\citenamefont {Peng}, \citenamefont {Zhang},
  \citenamefont {Shao}, \citenamefont {Xu}, \citenamefont {Zhang},
  \citenamefont {Lu}, \citenamefont {Zhang},\ and\ \citenamefont
  {Zhu}}]{Peng2016e}%
  \BibitemOpen
  \bibfield  {author} {\bibinfo {author} {\bibfnamefont {B.}~\bibnamefont
  {Peng}}, \bibinfo {author} {\bibfnamefont {H.}~\bibnamefont {Zhang}},
  \bibinfo {author} {\bibfnamefont {H.}~\bibnamefont {Shao}}, \bibinfo {author}
  {\bibfnamefont {Y.}~\bibnamefont {Xu}}, \bibinfo {author} {\bibfnamefont
  {R.}~\bibnamefont {Zhang}}, \bibinfo {author} {\bibfnamefont
  {H.}~\bibnamefont {Lu}}, \bibinfo {author} {\bibfnamefont {D.~W.}\
  \bibnamefont {Zhang}}, \ and\ \bibinfo {author} {\bibfnamefont
  {H.}~\bibnamefont {Zhu}},\ }\href {\doibase 10.1021/acsami.6b04211}
  {\bibfield  {journal} {\bibinfo  {journal} {ACS Appl. Mater. Interfaces}\
  }\textbf {\bibinfo {volume} {8}},\ \bibinfo {pages} {20977} (\bibinfo {year}
  {2016}{\natexlab{d}})}\BibitemShut {NoStop}%
\bibitem [{\citenamefont {Chen}\ \emph {et~al.}(2019)\citenamefont {Chen},
  \citenamefont {Peng}, \citenamefont {Cong}, \citenamefont {Shang},
  \citenamefont {Wu}, \citenamefont {Yang}, \citenamefont {Zhou}, \citenamefont
  {Yu}, \citenamefont {Zhang}, \citenamefont {Wang}, \citenamefont {Zou},
  \citenamefont {Zhang}, \citenamefont {Liu}, \citenamefont {Xiong},
  \citenamefont {Shao}, \citenamefont {Liu}, \citenamefont {Zhang},
  \citenamefont {Huang},\ and\ \citenamefont {Yu}}]{Chen2019}%
  \BibitemOpen
  \bibfield  {author} {\bibinfo {author} {\bibfnamefont {Y.}~\bibnamefont
  {Chen}}, \bibinfo {author} {\bibfnamefont {B.}~\bibnamefont {Peng}}, \bibinfo
  {author} {\bibfnamefont {C.}~\bibnamefont {Cong}}, \bibinfo {author}
  {\bibfnamefont {J.}~\bibnamefont {Shang}}, \bibinfo {author} {\bibfnamefont
  {L.}~\bibnamefont {Wu}}, \bibinfo {author} {\bibfnamefont {W.}~\bibnamefont
  {Yang}}, \bibinfo {author} {\bibfnamefont {J.}~\bibnamefont {Zhou}}, \bibinfo
  {author} {\bibfnamefont {P.}~\bibnamefont {Yu}}, \bibinfo {author}
  {\bibfnamefont {H.}~\bibnamefont {Zhang}}, \bibinfo {author} {\bibfnamefont
  {Y.}~\bibnamefont {Wang}}, \bibinfo {author} {\bibfnamefont {C.}~\bibnamefont
  {Zou}}, \bibinfo {author} {\bibfnamefont {J.}~\bibnamefont {Zhang}}, \bibinfo
  {author} {\bibfnamefont {S.}~\bibnamefont {Liu}}, \bibinfo {author}
  {\bibfnamefont {Q.}~\bibnamefont {Xiong}}, \bibinfo {author} {\bibfnamefont
  {H.}~\bibnamefont {Shao}}, \bibinfo {author} {\bibfnamefont {Z.}~\bibnamefont
  {Liu}}, \bibinfo {author} {\bibfnamefont {H.}~\bibnamefont {Zhang}}, \bibinfo
  {author} {\bibfnamefont {W.}~\bibnamefont {Huang}}, \ and\ \bibinfo {author}
  {\bibfnamefont {T.}~\bibnamefont {Yu}},\ }\href {\doibase
  10.1002/adma.201804979} {\bibfield  {journal} {\bibinfo  {journal} {Adv.
  Mater.}\ }\textbf {\bibinfo {volume} {31}},\ \bibinfo {pages} {1804979}
  (\bibinfo {year} {2019})}\BibitemShut {NoStop}%
\bibitem [{\citenamefont {\c{S}ahin}\ \emph {et~al.}(2009)\citenamefont
  {\c{S}ahin}, \citenamefont {Cahangirov}, \citenamefont {Topsakal},
  \citenamefont {Bekaroglu}, \citenamefont {Akturk}, \citenamefont {Senger},\
  and\ \citenamefont {Ciraci}}]{Sahin2009}%
  \BibitemOpen
  \bibfield  {author} {\bibinfo {author} {\bibfnamefont {H.}~\bibnamefont
  {\c{S}ahin}}, \bibinfo {author} {\bibfnamefont {S.}~\bibnamefont
  {Cahangirov}}, \bibinfo {author} {\bibfnamefont {M.}~\bibnamefont
  {Topsakal}}, \bibinfo {author} {\bibfnamefont {E.}~\bibnamefont {Bekaroglu}},
  \bibinfo {author} {\bibfnamefont {E.}~\bibnamefont {Akturk}}, \bibinfo
  {author} {\bibfnamefont {R.~T.}\ \bibnamefont {Senger}}, \ and\ \bibinfo
  {author} {\bibfnamefont {S.}~\bibnamefont {Ciraci}},\ }\href {\doibase
  10.1103/PhysRevB.80.155453} {\bibfield  {journal} {\bibinfo  {journal} {Phys.
  Rev. B}\ }\textbf {\bibinfo {volume} {80}},\ \bibinfo {pages} {155453}
  (\bibinfo {year} {2009})}\BibitemShut {NoStop}%
\bibitem [{\citenamefont {Cahangirov}\ \emph {et~al.}(2009)\citenamefont
  {Cahangirov}, \citenamefont {Topsakal}, \citenamefont {Akt\"urk},
  \citenamefont {\ifmmode~\mbox{\c{S}}\else \c{S}\fi{}ahin},\ and\
  \citenamefont {Ciraci}}]{Cahangirov2009}%
  \BibitemOpen
  \bibfield  {author} {\bibinfo {author} {\bibfnamefont {S.}~\bibnamefont
  {Cahangirov}}, \bibinfo {author} {\bibfnamefont {M.}~\bibnamefont
  {Topsakal}}, \bibinfo {author} {\bibfnamefont {E.}~\bibnamefont {Akt\"urk}},
  \bibinfo {author} {\bibfnamefont {H.}~\bibnamefont
  {\ifmmode~\mbox{\c{S}}\else \c{S}\fi{}ahin}}, \ and\ \bibinfo {author}
  {\bibfnamefont {S.}~\bibnamefont {Ciraci}},\ }\href {\doibase
  10.1103/PhysRevLett.102.236804} {\bibfield  {journal} {\bibinfo  {journal}
  {Phys. Rev. Lett.}\ }\textbf {\bibinfo {volume} {102}},\ \bibinfo {pages}
  {236804} (\bibinfo {year} {2009})}\BibitemShut {NoStop}%
\bibitem [{\citenamefont {O'Hare}\ \emph {et~al.}(2012)\citenamefont {O'Hare},
  \citenamefont {Kusmartsev},\ and\ \citenamefont {Kugel}}]{OHare2012}%
  \BibitemOpen
  \bibfield  {author} {\bibinfo {author} {\bibfnamefont {A.}~\bibnamefont
  {O'Hare}}, \bibinfo {author} {\bibfnamefont {F.~V.}\ \bibnamefont
  {Kusmartsev}}, \ and\ \bibinfo {author} {\bibfnamefont {K.~I.}\ \bibnamefont
  {Kugel}},\ }\href {\doibase 10.1021/nl204283q} {\bibfield  {journal}
  {\bibinfo  {journal} {Nano Lett.}\ }\textbf {\bibinfo {volume} {12}},\
  \bibinfo {pages} {1045} (\bibinfo {year} {2012})}\BibitemShut {NoStop}%
\bibitem [{\citenamefont {Roome}\ and\ \citenamefont
  {Carey}(2014)}]{Roome2014}%
  \BibitemOpen
  \bibfield  {author} {\bibinfo {author} {\bibfnamefont {N.~J.}\ \bibnamefont
  {Roome}}\ and\ \bibinfo {author} {\bibfnamefont {J.~D.}\ \bibnamefont
  {Carey}},\ }\href {\doibase 10.1021/am501022x} {\bibfield  {journal}
  {\bibinfo  {journal} {ACS Appl. Mater. Interfaces}\ }\textbf {\bibinfo
  {volume} {6}},\ \bibinfo {pages} {7743} (\bibinfo {year} {2014})}\BibitemShut
  {NoStop}%
\bibitem [{\citenamefont {Peng}\ \emph
  {et~al.}(2016{\natexlab{e}})\citenamefont {Peng}, \citenamefont {Zhang},
  \citenamefont {Shao}, \citenamefont {Xu}, \citenamefont {Zhang},\ and\
  \citenamefont {Zhu}}]{Peng2016a}%
  \BibitemOpen
  \bibfield  {author} {\bibinfo {author} {\bibfnamefont {B.}~\bibnamefont
  {Peng}}, \bibinfo {author} {\bibfnamefont {H.}~\bibnamefont {Zhang}},
  \bibinfo {author} {\bibfnamefont {H.}~\bibnamefont {Shao}}, \bibinfo {author}
  {\bibfnamefont {Y.}~\bibnamefont {Xu}}, \bibinfo {author} {\bibfnamefont
  {X.}~\bibnamefont {Zhang}}, \ and\ \bibinfo {author} {\bibfnamefont
  {H.}~\bibnamefont {Zhu}},\ }\href {\doibase 10.1039/C5RA19747C} {\bibfield
  {journal} {\bibinfo  {journal} {RSC Adv.}\ }\textbf {\bibinfo {volume} {6}},\
  \bibinfo {pages} {5767} (\bibinfo {year} {2016}{\natexlab{e}})}\BibitemShut
  {NoStop}%
\bibitem [{\citenamefont {Peng}\ \emph
  {et~al.}(2016{\natexlab{f}})\citenamefont {Peng}, \citenamefont {Zhang},
  \citenamefont {Shao}, \citenamefont {Xu}, \citenamefont {Ni}, \citenamefont
  {Zhang},\ and\ \citenamefont {Zhu}}]{Peng2016f}%
  \BibitemOpen
  \bibfield  {author} {\bibinfo {author} {\bibfnamefont {B.}~\bibnamefont
  {Peng}}, \bibinfo {author} {\bibfnamefont {H.}~\bibnamefont {Zhang}},
  \bibinfo {author} {\bibfnamefont {H.}~\bibnamefont {Shao}}, \bibinfo {author}
  {\bibfnamefont {Y.}~\bibnamefont {Xu}}, \bibinfo {author} {\bibfnamefont
  {G.}~\bibnamefont {Ni}}, \bibinfo {author} {\bibfnamefont {R.}~\bibnamefont
  {Zhang}}, \ and\ \bibinfo {author} {\bibfnamefont {H.}~\bibnamefont {Zhu}},\
  }\href {\doibase 10.1103/PhysRevB.94.245420} {\bibfield  {journal} {\bibinfo
  {journal} {Phys. Rev. B}\ }\textbf {\bibinfo {volume} {94}},\ \bibinfo
  {pages} {245420} (\bibinfo {year} {2016}{\natexlab{f}})}\BibitemShut
  {NoStop}%
\bibitem [{\citenamefont {Peng}\ \emph
  {et~al.}(2017{\natexlab{a}})\citenamefont {Peng}, \citenamefont {Zhang},
  \citenamefont {Zhang}, \citenamefont {Shao}, \citenamefont {Ni},
  \citenamefont {Zhu},\ and\ \citenamefont {Zhu}}]{Peng2017}%
  \BibitemOpen
  \bibfield  {author} {\bibinfo {author} {\bibfnamefont {B.}~\bibnamefont
  {Peng}}, \bibinfo {author} {\bibfnamefont {D.}~\bibnamefont {Zhang}},
  \bibinfo {author} {\bibfnamefont {H.}~\bibnamefont {Zhang}}, \bibinfo
  {author} {\bibfnamefont {H.}~\bibnamefont {Shao}}, \bibinfo {author}
  {\bibfnamefont {G.}~\bibnamefont {Ni}}, \bibinfo {author} {\bibfnamefont
  {Y.}~\bibnamefont {Zhu}}, \ and\ \bibinfo {author} {\bibfnamefont
  {H.}~\bibnamefont {Zhu}},\ }\href {\doibase 10.1039/C7NR00838D} {\bibfield
  {journal} {\bibinfo  {journal} {Nanoscale}\ }\textbf {\bibinfo {volume}
  {9}},\ \bibinfo {pages} {7397} (\bibinfo {year}
  {2017}{\natexlab{a}})}\BibitemShut {NoStop}%
\bibitem [{\citenamefont {Peng}\ \emph
  {et~al.}(2018{\natexlab{b}})\citenamefont {Peng}, \citenamefont {Mortazavi},
  \citenamefont {Zhang}, \citenamefont {Shao}, \citenamefont {Xu},
  \citenamefont {Li}, \citenamefont {Ni}, \citenamefont {Rabczuk},\ and\
  \citenamefont {Zhu}}]{Peng2018b}%
  \BibitemOpen
  \bibfield  {author} {\bibinfo {author} {\bibfnamefont {B.}~\bibnamefont
  {Peng}}, \bibinfo {author} {\bibfnamefont {B.}~\bibnamefont {Mortazavi}},
  \bibinfo {author} {\bibfnamefont {H.}~\bibnamefont {Zhang}}, \bibinfo
  {author} {\bibfnamefont {H.}~\bibnamefont {Shao}}, \bibinfo {author}
  {\bibfnamefont {K.}~\bibnamefont {Xu}}, \bibinfo {author} {\bibfnamefont
  {J.}~\bibnamefont {Li}}, \bibinfo {author} {\bibfnamefont {G.}~\bibnamefont
  {Ni}}, \bibinfo {author} {\bibfnamefont {T.}~\bibnamefont {Rabczuk}}, \ and\
  \bibinfo {author} {\bibfnamefont {H.}~\bibnamefont {Zhu}},\ }\href {\doibase
  10.1103/PhysRevApplied.10.034046} {\bibfield  {journal} {\bibinfo  {journal}
  {Phys. Rev. Applied}\ }\textbf {\bibinfo {volume} {10}},\ \bibinfo {pages}
  {034046} (\bibinfo {year} {2018}{\natexlab{b}})}\BibitemShut {NoStop}%
\bibitem [{\citenamefont {Peng}\ \emph
  {et~al.}(2018{\natexlab{c}})\citenamefont {Peng}, \citenamefont {Zhang},
  \citenamefont {Shao}, \citenamefont {Xu}, \citenamefont {Ni}, \citenamefont
  {Li}, \citenamefont {Zhu},\ and\ \citenamefont {Soukoulis}}]{Peng2018a}%
  \BibitemOpen
  \bibfield  {author} {\bibinfo {author} {\bibfnamefont {B.}~\bibnamefont
  {Peng}}, \bibinfo {author} {\bibfnamefont {H.}~\bibnamefont {Zhang}},
  \bibinfo {author} {\bibfnamefont {H.}~\bibnamefont {Shao}}, \bibinfo {author}
  {\bibfnamefont {K.}~\bibnamefont {Xu}}, \bibinfo {author} {\bibfnamefont
  {G.}~\bibnamefont {Ni}}, \bibinfo {author} {\bibfnamefont {J.}~\bibnamefont
  {Li}}, \bibinfo {author} {\bibfnamefont {H.}~\bibnamefont {Zhu}}, \ and\
  \bibinfo {author} {\bibfnamefont {C.~M.}\ \bibnamefont {Soukoulis}},\ }\href
  {\doibase 10.1039/C7TA09480A} {\bibfield  {journal} {\bibinfo  {journal} {J.
  Mater. Chem. A}\ }\textbf {\bibinfo {volume} {6}},\ \bibinfo {pages} {2018}
  (\bibinfo {year} {2018}{\natexlab{c}})}\BibitemShut {NoStop}%
\bibitem [{\citenamefont {Peng}\ \emph
  {et~al.}(2019{\natexlab{a}})\citenamefont {Peng}, \citenamefont {Mei},
  \citenamefont {Zhang}, \citenamefont {Shao}, \citenamefont {Xu},
  \citenamefont {Ni}, \citenamefont {Jin}, \citenamefont {Soukoulis},\ and\
  \citenamefont {Zhu}}]{Peng2019a}%
  \BibitemOpen
  \bibfield  {author} {\bibinfo {author} {\bibfnamefont {B.}~\bibnamefont
  {Peng}}, \bibinfo {author} {\bibfnamefont {H.}~\bibnamefont {Mei}}, \bibinfo
  {author} {\bibfnamefont {H.}~\bibnamefont {Zhang}}, \bibinfo {author}
  {\bibfnamefont {H.}~\bibnamefont {Shao}}, \bibinfo {author} {\bibfnamefont
  {K.}~\bibnamefont {Xu}}, \bibinfo {author} {\bibfnamefont {G.}~\bibnamefont
  {Ni}}, \bibinfo {author} {\bibfnamefont {Q.}~\bibnamefont {Jin}}, \bibinfo
  {author} {\bibfnamefont {C.~M.}\ \bibnamefont {Soukoulis}}, \ and\ \bibinfo
  {author} {\bibfnamefont {H.}~\bibnamefont {Zhu}},\ }\href {\doibase
  10.1039/C8QI01297K} {\bibfield  {journal} {\bibinfo  {journal} {Inorg. Chem.
  Front.}\ }\textbf {\bibinfo {volume} {6}},\ \bibinfo {pages} {920} (\bibinfo
  {year} {2019}{\natexlab{a}})}\BibitemShut {NoStop}%
\bibitem [{\citenamefont {Peng}\ \emph
  {et~al.}(2022{\natexlab{a}})\citenamefont {Peng}, \citenamefont {Bouhon},
  \citenamefont {Monserrat},\ and\ \citenamefont {Slager}}]{Peng2022}%
  \BibitemOpen
  \bibfield  {author} {\bibinfo {author} {\bibfnamefont {B.}~\bibnamefont
  {Peng}}, \bibinfo {author} {\bibfnamefont {A.}~\bibnamefont {Bouhon}},
  \bibinfo {author} {\bibfnamefont {B.}~\bibnamefont {Monserrat}}, \ and\
  \bibinfo {author} {\bibfnamefont {R.-J.}\ \bibnamefont {Slager}},\ }\href
  {\doibase 10.1038/s41467-022-28046-9} {\bibfield  {journal} {\bibinfo
  {journal} {Nature Communications}\ }\textbf {\bibinfo {volume} {13}},\
  \bibinfo {pages} {423} (\bibinfo {year} {2022}{\natexlab{a}})}\BibitemShut
  {NoStop}%
\bibitem [{\citenamefont {Peng}\ \emph
  {et~al.}(2022{\natexlab{b}})\citenamefont {Peng}, \citenamefont {Bouhon},
  \citenamefont {Slager},\ and\ \citenamefont {Monserrat}}]{Peng2022a}%
  \BibitemOpen
  \bibfield  {author} {\bibinfo {author} {\bibfnamefont {B.}~\bibnamefont
  {Peng}}, \bibinfo {author} {\bibfnamefont {A.}~\bibnamefont {Bouhon}},
  \bibinfo {author} {\bibfnamefont {R.-J.}\ \bibnamefont {Slager}}, \ and\
  \bibinfo {author} {\bibfnamefont {B.}~\bibnamefont {Monserrat}},\ }\href
  {\doibase 10.1103/PhysRevB.105.085115} {\bibfield  {journal} {\bibinfo
  {journal} {Phys. Rev. B}\ }\textbf {\bibinfo {volume} {105}},\ \bibinfo
  {pages} {085115} (\bibinfo {year} {2022}{\natexlab{b}})}\BibitemShut
  {NoStop}%
\bibitem [{\citenamefont {Cho}\ \emph {et~al.}(2015)\citenamefont {Cho},
  \citenamefont {Kim}, \citenamefont {Kim}, \citenamefont {Zhao}, \citenamefont
  {Seok}, \citenamefont {Keum}, \citenamefont {Baik}, \citenamefont {Choe},
  \citenamefont {Chang}, \citenamefont {Suenaga}, \citenamefont {Kim},
  \citenamefont {Lee},\ and\ \citenamefont {Yang}}]{Cho2015a}%
  \BibitemOpen
  \bibfield  {author} {\bibinfo {author} {\bibfnamefont {S.}~\bibnamefont
  {Cho}}, \bibinfo {author} {\bibfnamefont {S.}~\bibnamefont {Kim}}, \bibinfo
  {author} {\bibfnamefont {J.~H.}\ \bibnamefont {Kim}}, \bibinfo {author}
  {\bibfnamefont {J.}~\bibnamefont {Zhao}}, \bibinfo {author} {\bibfnamefont
  {J.}~\bibnamefont {Seok}}, \bibinfo {author} {\bibfnamefont {D.~H.}\
  \bibnamefont {Keum}}, \bibinfo {author} {\bibfnamefont {J.}~\bibnamefont
  {Baik}}, \bibinfo {author} {\bibfnamefont {D.-H.}\ \bibnamefont {Choe}},
  \bibinfo {author} {\bibfnamefont {K.~J.}\ \bibnamefont {Chang}}, \bibinfo
  {author} {\bibfnamefont {K.}~\bibnamefont {Suenaga}}, \bibinfo {author}
  {\bibfnamefont {S.~W.}\ \bibnamefont {Kim}}, \bibinfo {author} {\bibfnamefont
  {Y.~H.}\ \bibnamefont {Lee}}, \ and\ \bibinfo {author} {\bibfnamefont
  {H.}~\bibnamefont {Yang}},\ }\href {\doibase 10.1126/science.aab3175}
  {\bibfield  {journal} {\bibinfo  {journal} {Science}\ }\textbf {\bibinfo
  {volume} {349}},\ \bibinfo {pages} {625} (\bibinfo {year}
  {2015})}\BibitemShut {NoStop}%
\bibitem [{\citenamefont {Peng}\ \emph {et~al.}(2020)\citenamefont {Peng},
  \citenamefont {Zhang}, \citenamefont {Chen}, \citenamefont {Hou},
  \citenamefont {Qiu}, \citenamefont {Shao}, \citenamefont {Zhu}, \citenamefont
  {Monserrat}, \citenamefont {Fu}, \citenamefont {Weng},\ and\ \citenamefont
  {Soukoulis}}]{Peng2020}%
  \BibitemOpen
  \bibfield  {author} {\bibinfo {author} {\bibfnamefont {B.}~\bibnamefont
  {Peng}}, \bibinfo {author} {\bibfnamefont {H.}~\bibnamefont {Zhang}},
  \bibinfo {author} {\bibfnamefont {W.}~\bibnamefont {Chen}}, \bibinfo {author}
  {\bibfnamefont {B.}~\bibnamefont {Hou}}, \bibinfo {author} {\bibfnamefont
  {Z.-J.}\ \bibnamefont {Qiu}}, \bibinfo {author} {\bibfnamefont
  {H.}~\bibnamefont {Shao}}, \bibinfo {author} {\bibfnamefont {H.}~\bibnamefont
  {Zhu}}, \bibinfo {author} {\bibfnamefont {B.}~\bibnamefont {Monserrat}},
  \bibinfo {author} {\bibfnamefont {D.}~\bibnamefont {Fu}}, \bibinfo {author}
  {\bibfnamefont {H.}~\bibnamefont {Weng}}, \ and\ \bibinfo {author}
  {\bibfnamefont {C.~M.}\ \bibnamefont {Soukoulis}},\ }\href {\doibase
  10.1038/s41699-020-0147-x} {\bibfield  {journal} {\bibinfo  {journal} {npj 2D
  Materials and Applications}\ }\textbf {\bibinfo {volume} {4}},\ \bibinfo
  {pages} {14} (\bibinfo {year} {2020})}\BibitemShut {NoStop}%
\bibitem [{\citenamefont {Gou}\ \emph {et~al.}(2023)\citenamefont {Gou},
  \citenamefont {Bai}, \citenamefont {Zhang}, \citenamefont {Huang},
  \citenamefont {Duan}, \citenamefont {Ariando}, \citenamefont {Yang},
  \citenamefont {Chen}, \citenamefont {Lu},\ and\ \citenamefont
  {Wee}}]{Gou2023}%
  \BibitemOpen
  \bibfield  {author} {\bibinfo {author} {\bibfnamefont {J.}~\bibnamefont
  {Gou}}, \bibinfo {author} {\bibfnamefont {H.}~\bibnamefont {Bai}}, \bibinfo
  {author} {\bibfnamefont {X.}~\bibnamefont {Zhang}}, \bibinfo {author}
  {\bibfnamefont {Y.~L.}\ \bibnamefont {Huang}}, \bibinfo {author}
  {\bibfnamefont {S.}~\bibnamefont {Duan}}, \bibinfo {author} {\bibfnamefont
  {A.}~\bibnamefont {Ariando}}, \bibinfo {author} {\bibfnamefont {S.~A.}\
  \bibnamefont {Yang}}, \bibinfo {author} {\bibfnamefont {L.}~\bibnamefont
  {Chen}}, \bibinfo {author} {\bibfnamefont {Y.}~\bibnamefont {Lu}}, \ and\
  \bibinfo {author} {\bibfnamefont {A.~T.~S.}\ \bibnamefont {Wee}},\ }\href
  {\doibase 10.1038/s41586-023-05848-5} {\bibfield  {journal} {\bibinfo
  {journal} {Nature}\ }\textbf {\bibinfo {volume} {617}},\ \bibinfo {pages}
  {67} (\bibinfo {year} {2023})}\BibitemShut {NoStop}%
\bibitem [{\citenamefont {Peng}\ \emph {et~al.}(2024)\citenamefont {Peng},
  \citenamefont {Lange}, \citenamefont {Bennett}, \citenamefont {Wang},
  \citenamefont {Slager},\ and\ \citenamefont {Monserrat}}]{Peng2024}%
  \BibitemOpen
  \bibfield  {author} {\bibinfo {author} {\bibfnamefont {B.}~\bibnamefont
  {Peng}}, \bibinfo {author} {\bibfnamefont {G.~F.}\ \bibnamefont {Lange}},
  \bibinfo {author} {\bibfnamefont {D.}~\bibnamefont {Bennett}}, \bibinfo
  {author} {\bibfnamefont {K.}~\bibnamefont {Wang}}, \bibinfo {author}
  {\bibfnamefont {R.-J.}\ \bibnamefont {Slager}}, \ and\ \bibinfo {author}
  {\bibfnamefont {B.}~\bibnamefont {Monserrat}},\ }\href {\doibase
  10.1103/PhysRevLett.132.116601} {\bibfield  {journal} {\bibinfo  {journal}
  {Phys. Rev. Lett.}\ }\textbf {\bibinfo {volume} {132}},\ \bibinfo {pages}
  {116601} (\bibinfo {year} {2024})}\BibitemShut {NoStop}%
\bibitem [{\citenamefont {Qiu}\ \emph {et~al.}(2013)\citenamefont {Qiu},
  \citenamefont {Xu}, \citenamefont {Wang}, \citenamefont {Ren}, \citenamefont
  {Nan}, \citenamefont {Ni}, \citenamefont {Chen}, \citenamefont {Yuan},
  \citenamefont {Miao}, \citenamefont {Song}, \citenamefont {Long},
  \citenamefont {Shi}, \citenamefont {Sun}, \citenamefont {Wang},\ and\
  \citenamefont {Wang}}]{Qiu2013}%
  \BibitemOpen
  \bibfield  {author} {\bibinfo {author} {\bibfnamefont {H.}~\bibnamefont
  {Qiu}}, \bibinfo {author} {\bibfnamefont {T.}~\bibnamefont {Xu}}, \bibinfo
  {author} {\bibfnamefont {Z.}~\bibnamefont {Wang}}, \bibinfo {author}
  {\bibfnamefont {W.}~\bibnamefont {Ren}}, \bibinfo {author} {\bibfnamefont
  {H.}~\bibnamefont {Nan}}, \bibinfo {author} {\bibfnamefont {Z.}~\bibnamefont
  {Ni}}, \bibinfo {author} {\bibfnamefont {Q.}~\bibnamefont {Chen}}, \bibinfo
  {author} {\bibfnamefont {S.}~\bibnamefont {Yuan}}, \bibinfo {author}
  {\bibfnamefont {F.}~\bibnamefont {Miao}}, \bibinfo {author} {\bibfnamefont
  {F.}~\bibnamefont {Song}}, \bibinfo {author} {\bibfnamefont {G.}~\bibnamefont
  {Long}}, \bibinfo {author} {\bibfnamefont {Y.}~\bibnamefont {Shi}}, \bibinfo
  {author} {\bibfnamefont {L.}~\bibnamefont {Sun}}, \bibinfo {author}
  {\bibfnamefont {J.}~\bibnamefont {Wang}}, \ and\ \bibinfo {author}
  {\bibfnamefont {X.}~\bibnamefont {Wang}},\ }\href {\doibase
  10.1038/ncomms3642} {\bibfield  {journal} {\bibinfo  {journal} {Nat Commun}\
  }\textbf {\bibinfo {volume} {4}},\ \bibinfo {pages} {2642} (\bibinfo {year}
  {2013})}\BibitemShut {NoStop}%
\bibitem [{\citenamefont {Peng}\ \emph
  {et~al.}(2016{\natexlab{g}})\citenamefont {Peng}, \citenamefont {Ning},
  \citenamefont {Zhang}, \citenamefont {Shao}, \citenamefont {Xu},
  \citenamefont {Ni},\ and\ \citenamefont {Zhu}}]{Peng2016g}%
  \BibitemOpen
  \bibfield  {author} {\bibinfo {author} {\bibfnamefont {B.}~\bibnamefont
  {Peng}}, \bibinfo {author} {\bibfnamefont {Z.}~\bibnamefont {Ning}}, \bibinfo
  {author} {\bibfnamefont {H.}~\bibnamefont {Zhang}}, \bibinfo {author}
  {\bibfnamefont {H.}~\bibnamefont {Shao}}, \bibinfo {author} {\bibfnamefont
  {Y.}~\bibnamefont {Xu}}, \bibinfo {author} {\bibfnamefont {G.}~\bibnamefont
  {Ni}}, \ and\ \bibinfo {author} {\bibfnamefont {H.}~\bibnamefont {Zhu}},\
  }\href {\doibase 10.1021/acs.jpcc.6b10812} {\bibfield  {journal} {\bibinfo
  {journal} {J. Phys. Chem. C}\ }\textbf {\bibinfo {volume} {120}},\ \bibinfo
  {pages} {29324} (\bibinfo {year} {2016}{\natexlab{g}})}\BibitemShut {NoStop}%
\bibitem [{\citenamefont {Pizzochero}\ and\ \citenamefont
  {Yazyev}(2018)}]{Pizzochero2018}%
  \BibitemOpen
  \bibfield  {author} {\bibinfo {author} {\bibfnamefont {M.}~\bibnamefont
  {Pizzochero}}\ and\ \bibinfo {author} {\bibfnamefont {O.~V.}\ \bibnamefont
  {Yazyev}},\ }\href {\doibase 10.1088/2053-1583/aab279} {\bibfield  {journal}
  {\bibinfo  {journal} {2D Mater.}\ }\textbf {\bibinfo {volume} {5}},\ \bibinfo
  {pages} {025022} (\bibinfo {year} {2018})}\BibitemShut {NoStop}%
\bibitem [{\citenamefont {Pizzochero}(2020)}]{Pizzochero2020b}%
  \BibitemOpen
  \bibfield  {author} {\bibinfo {author} {\bibfnamefont {M.}~\bibnamefont
  {Pizzochero}},\ }\href {\doibase 10.1088/1361-6463/ab7ca3} {\bibfield
  {journal} {\bibinfo  {journal} {J. Phys. D: Appl. Phys.}\ }\textbf {\bibinfo
  {volume} {53}},\ \bibinfo {pages} {244003} (\bibinfo {year}
  {2020})}\BibitemShut {NoStop}%
\bibitem [{\citenamefont {Li}\ and\ \citenamefont {Gali}(2022)}]{Li2022b}%
  \BibitemOpen
  \bibfield  {author} {\bibinfo {author} {\bibfnamefont {S.}~\bibnamefont
  {Li}}\ and\ \bibinfo {author} {\bibfnamefont {A.}~\bibnamefont {Gali}},\
  }\href {\doibase 10.1021/acs.jpclett.2c02687} {\bibfield  {journal} {\bibinfo
   {journal} {J. Phys. Chem. Lett.}\ }\textbf {\bibinfo {volume} {13}},\
  \bibinfo {pages} {9544} (\bibinfo {year} {2022})}\BibitemShut {NoStop}%
\bibitem [{\citenamefont {Trainer}\ \emph {et~al.}(2022)\citenamefont
  {Trainer}, \citenamefont {Nieminen}, \citenamefont {Bobba}, \citenamefont
  {Wang}, \citenamefont {Xi}, \citenamefont {Bansil},\ and\ \citenamefont
  {Iavarone}}]{Trainer2022}%
  \BibitemOpen
  \bibfield  {author} {\bibinfo {author} {\bibfnamefont {D.~J.}\ \bibnamefont
  {Trainer}}, \bibinfo {author} {\bibfnamefont {J.}~\bibnamefont {Nieminen}},
  \bibinfo {author} {\bibfnamefont {F.}~\bibnamefont {Bobba}}, \bibinfo
  {author} {\bibfnamefont {B.}~\bibnamefont {Wang}}, \bibinfo {author}
  {\bibfnamefont {X.}~\bibnamefont {Xi}}, \bibinfo {author} {\bibfnamefont
  {A.}~\bibnamefont {Bansil}}, \ and\ \bibinfo {author} {\bibfnamefont
  {M.}~\bibnamefont {Iavarone}},\ }\href {\doibase 10.1038/s41699-022-00286-9}
  {\bibfield  {journal} {\bibinfo  {journal} {npj 2D Materials and
  Applications}\ }\textbf {\bibinfo {volume} {6}},\ \bibinfo {pages} {13}
  (\bibinfo {year} {2022})}\BibitemShut {NoStop}%
\bibitem [{\citenamefont {Li}\ \emph {et~al.}(2025)\citenamefont {Li},
  \citenamefont {Li},\ and\ \citenamefont {Gali}}]{Li2025}%
  \BibitemOpen
  \bibfield  {author} {\bibinfo {author} {\bibfnamefont {S.}~\bibnamefont
  {Li}}, \bibinfo {author} {\bibfnamefont {P.}~\bibnamefont {Li}}, \ and\
  \bibinfo {author} {\bibfnamefont {A.}~\bibnamefont {Gali}},\ }\href {\doibase
  10.1063/5.0248897} {\bibfield  {journal} {\bibinfo  {journal} {Appl. Phys.
  Lett.}\ }\textbf {\bibinfo {volume} {126}},\ \bibinfo {pages} {062104}
  (\bibinfo {year} {2025})}\BibitemShut {NoStop}%
\bibitem [{\citenamefont {Peng}\ \emph
  {et~al.}(2017{\natexlab{b}})\citenamefont {Peng}, \citenamefont {Zhang},
  \citenamefont {Shao}, \citenamefont {Ning}, \citenamefont {Xu}, \citenamefont
  {Ni}, \citenamefont {Lu}, \citenamefont {Zhang},\ and\ \citenamefont
  {Zhu}}]{Peng2017a}%
  \BibitemOpen
  \bibfield  {author} {\bibinfo {author} {\bibfnamefont {B.}~\bibnamefont
  {Peng}}, \bibinfo {author} {\bibfnamefont {H.}~\bibnamefont {Zhang}},
  \bibinfo {author} {\bibfnamefont {H.}~\bibnamefont {Shao}}, \bibinfo {author}
  {\bibfnamefont {Z.}~\bibnamefont {Ning}}, \bibinfo {author} {\bibfnamefont
  {Y.}~\bibnamefont {Xu}}, \bibinfo {author} {\bibfnamefont {G.}~\bibnamefont
  {Ni}}, \bibinfo {author} {\bibfnamefont {H.}~\bibnamefont {Lu}}, \bibinfo
  {author} {\bibfnamefont {D.~W.}\ \bibnamefont {Zhang}}, \ and\ \bibinfo
  {author} {\bibfnamefont {H.}~\bibnamefont {Zhu}},\ }\href {\doibase
  10.1080/21663831.2017.1298539} {\bibfield  {journal} {\bibinfo  {journal}
  {Materials Research Letters}\ }\textbf {\bibinfo {volume} {5}},\ \bibinfo
  {pages} {399} (\bibinfo {year} {2017}{\natexlab{b}})}\BibitemShut {NoStop}%
\bibitem [{\citenamefont {Mannix}\ \emph {et~al.}(2017)\citenamefont {Mannix},
  \citenamefont {Kiraly}, \citenamefont {Hersam},\ and\ \citenamefont
  {Guisinger}}]{Mannix2017}%
  \BibitemOpen
  \bibfield  {author} {\bibinfo {author} {\bibfnamefont {A.~J.}\ \bibnamefont
  {Mannix}}, \bibinfo {author} {\bibfnamefont {B.}~\bibnamefont {Kiraly}},
  \bibinfo {author} {\bibfnamefont {M.~C.}\ \bibnamefont {Hersam}}, \ and\
  \bibinfo {author} {\bibfnamefont {N.~P.}\ \bibnamefont {Guisinger}},\ }\href
  {\doibase 10.1038/s41570-016-0014} {\bibfield  {journal} {\bibinfo  {journal}
  {Nature Reviews Chemistry}\ }\textbf {\bibinfo {volume} {1}},\ \bibinfo
  {pages} {0014} (\bibinfo {year} {2017})}\BibitemShut {NoStop}%
\bibitem [{\citenamefont {Niu}\ \emph {et~al.}(2023{\natexlab{a}})\citenamefont
  {Niu}, \citenamefont {Sopp}, \citenamefont {Lodi}, \citenamefont {Gee},
  \citenamefont {Kong}, \citenamefont {Pei}, \citenamefont {Gehring},
  \citenamefont {N{\"a}gele}, \citenamefont {Lau}, \citenamefont {Ma},
  \citenamefont {Liu}, \citenamefont {Narita}, \citenamefont {Mol},
  \citenamefont {Burghard}, \citenamefont {M{\"u}llen}, \citenamefont {Mai},
  \citenamefont {Feng},\ and\ \citenamefont {Bogani}}]{Niu2023}%
  \BibitemOpen
  \bibfield  {author} {\bibinfo {author} {\bibfnamefont {W.}~\bibnamefont
  {Niu}}, \bibinfo {author} {\bibfnamefont {S.}~\bibnamefont {Sopp}}, \bibinfo
  {author} {\bibfnamefont {A.}~\bibnamefont {Lodi}}, \bibinfo {author}
  {\bibfnamefont {A.}~\bibnamefont {Gee}}, \bibinfo {author} {\bibfnamefont
  {F.}~\bibnamefont {Kong}}, \bibinfo {author} {\bibfnamefont {T.}~\bibnamefont
  {Pei}}, \bibinfo {author} {\bibfnamefont {P.}~\bibnamefont {Gehring}},
  \bibinfo {author} {\bibfnamefont {J.}~\bibnamefont {N{\"a}gele}}, \bibinfo
  {author} {\bibfnamefont {C.~S.}\ \bibnamefont {Lau}}, \bibinfo {author}
  {\bibfnamefont {J.}~\bibnamefont {Ma}}, \bibinfo {author} {\bibfnamefont
  {J.}~\bibnamefont {Liu}}, \bibinfo {author} {\bibfnamefont {A.}~\bibnamefont
  {Narita}}, \bibinfo {author} {\bibfnamefont {J.}~\bibnamefont {Mol}},
  \bibinfo {author} {\bibfnamefont {M.}~\bibnamefont {Burghard}}, \bibinfo
  {author} {\bibfnamefont {K.}~\bibnamefont {M{\"u}llen}}, \bibinfo {author}
  {\bibfnamefont {Y.}~\bibnamefont {Mai}}, \bibinfo {author} {\bibfnamefont
  {X.}~\bibnamefont {Feng}}, \ and\ \bibinfo {author} {\bibfnamefont
  {L.}~\bibnamefont {Bogani}},\ }\href {\doibase 10.1038/s41563-022-01460-6}
  {\bibfield  {journal} {\bibinfo  {journal} {Nature Materials}\ }\textbf
  {\bibinfo {volume} {22}},\ \bibinfo {pages} {180} (\bibinfo {year}
  {2023}{\natexlab{a}})}\BibitemShut {NoStop}%
\bibitem [{\citenamefont {Qie}\ \emph {et~al.}(2024)\citenamefont {Qie},
  \citenamefont {Wang}, \citenamefont {Jiang}, \citenamefont {Zhang},
  \citenamefont {Jacobse}, \citenamefont {Lu}, \citenamefont {Li},
  \citenamefont {Liu}, \citenamefont {Alexandrova}, \citenamefont {Louie},
  \citenamefont {Crommie},\ and\ \citenamefont {Fischer}}]{Qie2024}%
  \BibitemOpen
  \bibfield  {author} {\bibinfo {author} {\bibfnamefont {B.}~\bibnamefont
  {Qie}}, \bibinfo {author} {\bibfnamefont {Z.}~\bibnamefont {Wang}}, \bibinfo
  {author} {\bibfnamefont {J.}~\bibnamefont {Jiang}}, \bibinfo {author}
  {\bibfnamefont {Z.}~\bibnamefont {Zhang}}, \bibinfo {author} {\bibfnamefont
  {P.~H.}\ \bibnamefont {Jacobse}}, \bibinfo {author} {\bibfnamefont
  {J.}~\bibnamefont {Lu}}, \bibinfo {author} {\bibfnamefont {X.}~\bibnamefont
  {Li}}, \bibinfo {author} {\bibfnamefont {F.}~\bibnamefont {Liu}}, \bibinfo
  {author} {\bibfnamefont {A.~N.}\ \bibnamefont {Alexandrova}}, \bibinfo
  {author} {\bibfnamefont {S.~G.}\ \bibnamefont {Louie}}, \bibinfo {author}
  {\bibfnamefont {M.~F.}\ \bibnamefont {Crommie}}, \ and\ \bibinfo {author}
  {\bibfnamefont {F.~R.}\ \bibnamefont {Fischer}},\ }\href {\doibase
  10.1126/science.adm9814} {\bibfield  {journal} {\bibinfo  {journal}
  {Science}\ }\textbf {\bibinfo {volume} {384}},\ \bibinfo {pages} {895}
  (\bibinfo {year} {2024})}\BibitemShut {NoStop}%
\bibitem [{\citenamefont {Dubey}\ \emph {et~al.}(2021)\citenamefont {Dubey},
  \citenamefont {Melle-Franco},\ and\ \citenamefont
  {Mateo-Alonso}}]{Dubey2021}%
  \BibitemOpen
  \bibfield  {author} {\bibinfo {author} {\bibfnamefont {R.~K.}\ \bibnamefont
  {Dubey}}, \bibinfo {author} {\bibfnamefont {M.}~\bibnamefont {Melle-Franco}},
  \ and\ \bibinfo {author} {\bibfnamefont {A.}~\bibnamefont {Mateo-Alonso}},\
  }\href {\doibase 10.1021/jacs.1c01849} {\bibfield  {journal} {\bibinfo
  {journal} {J. Am. Chem. Soc.}\ }\textbf {\bibinfo {volume} {143}},\ \bibinfo
  {pages} {6593} (\bibinfo {year} {2021})}\BibitemShut {NoStop}%
\bibitem [{\citenamefont {Niu}\ \emph {et~al.}(2023{\natexlab{b}})\citenamefont
  {Niu}, \citenamefont {Fu}, \citenamefont {Serra}, \citenamefont {Liu},
  \citenamefont {Droste}, \citenamefont {Lee}, \citenamefont {Ling},
  \citenamefont {Xu}, \citenamefont {Cojal~Gonz{\'a}lez}, \citenamefont
  {Lucotti}, \citenamefont {Rabe}, \citenamefont {Ryan~Hansen}, \citenamefont
  {Pisula}, \citenamefont {Blom}, \citenamefont {Palma}, \citenamefont
  {Tommasini}, \citenamefont {Mai}, \citenamefont {Ma},\ and\ \citenamefont
  {Feng}}]{Niu2023a}%
  \BibitemOpen
  \bibfield  {author} {\bibinfo {author} {\bibfnamefont {W.}~\bibnamefont
  {Niu}}, \bibinfo {author} {\bibfnamefont {Y.}~\bibnamefont {Fu}}, \bibinfo
  {author} {\bibfnamefont {G.}~\bibnamefont {Serra}}, \bibinfo {author}
  {\bibfnamefont {K.}~\bibnamefont {Liu}}, \bibinfo {author} {\bibfnamefont
  {J.}~\bibnamefont {Droste}}, \bibinfo {author} {\bibfnamefont
  {Y.}~\bibnamefont {Lee}}, \bibinfo {author} {\bibfnamefont {Z.}~\bibnamefont
  {Ling}}, \bibinfo {author} {\bibfnamefont {F.}~\bibnamefont {Xu}}, \bibinfo
  {author} {\bibfnamefont {J.~D.}\ \bibnamefont {Cojal~Gonz{\'a}lez}}, \bibinfo
  {author} {\bibfnamefont {A.}~\bibnamefont {Lucotti}}, \bibinfo {author}
  {\bibfnamefont {J.~P.}\ \bibnamefont {Rabe}}, \bibinfo {author}
  {\bibfnamefont {M.}~\bibnamefont {Ryan~Hansen}}, \bibinfo {author}
  {\bibfnamefont {W.}~\bibnamefont {Pisula}}, \bibinfo {author} {\bibfnamefont
  {P.~W.~M.}\ \bibnamefont {Blom}}, \bibinfo {author} {\bibfnamefont {C.-A.}\
  \bibnamefont {Palma}}, \bibinfo {author} {\bibfnamefont {M.}~\bibnamefont
  {Tommasini}}, \bibinfo {author} {\bibfnamefont {Y.}~\bibnamefont {Mai}},
  \bibinfo {author} {\bibfnamefont {J.}~\bibnamefont {Ma}}, \ and\ \bibinfo
  {author} {\bibfnamefont {X.}~\bibnamefont {Feng}},\ }\href {\doibase
  10.1002/anie.202305737} {\bibfield  {journal} {\bibinfo  {journal} {Angew.
  Chem. Int. Ed.}\ }\textbf {\bibinfo {volume} {62}},\ \bibinfo {pages}
  {e202305737} (\bibinfo {year} {2023}{\natexlab{b}})}\BibitemShut {NoStop}%
\bibitem [{\citenamefont {Su}\ \emph {et~al.}(2025)\citenamefont {Su},
  \citenamefont {Ding}, \citenamefont {Hong}, \citenamefont {Ke}, \citenamefont
  {Yan}, \citenamefont {Li}, \citenamefont {Jiang},\ and\ \citenamefont
  {Yu}}]{Su2025}%
  \BibitemOpen
  \bibfield  {author} {\bibinfo {author} {\bibfnamefont {X.}~\bibnamefont
  {Su}}, \bibinfo {author} {\bibfnamefont {Z.}~\bibnamefont {Ding}}, \bibinfo
  {author} {\bibfnamefont {Y.}~\bibnamefont {Hong}}, \bibinfo {author}
  {\bibfnamefont {N.}~\bibnamefont {Ke}}, \bibinfo {author} {\bibfnamefont
  {K.}~\bibnamefont {Yan}}, \bibinfo {author} {\bibfnamefont {C.}~\bibnamefont
  {Li}}, \bibinfo {author} {\bibfnamefont {Y.-F.}\ \bibnamefont {Jiang}}, \
  and\ \bibinfo {author} {\bibfnamefont {P.}~\bibnamefont {Yu}},\ }\href
  {\doibase 10.1038/s44160-025-00744-4} {\bibfield  {journal} {\bibinfo
  {journal} {Nature Synthesis}\ ,\ } (\bibinfo {year} {2025})}\BibitemShut
  {NoStop}%
\bibitem [{\citenamefont {Fu}\ \emph {et~al.}(2025)\citenamefont {Fu},
  \citenamefont {Huang}, \citenamefont {Liu}, \citenamefont {Henriques},
  \citenamefont {Gao}, \citenamefont {Han}, \citenamefont {Chen}, \citenamefont
  {Wang}, \citenamefont {Palma}, \citenamefont {Cheng}, \citenamefont {Lin},
  \citenamefont {Du}, \citenamefont {Ma}, \citenamefont
  {Fern{\'{a}}ndez-Rossier}, \citenamefont {Feng},\ and\ \citenamefont
  {Gao}}]{Fu2025}%
  \BibitemOpen
  \bibfield  {author} {\bibinfo {author} {\bibfnamefont {X.}~\bibnamefont
  {Fu}}, \bibinfo {author} {\bibfnamefont {L.}~\bibnamefont {Huang}}, \bibinfo
  {author} {\bibfnamefont {K.}~\bibnamefont {Liu}}, \bibinfo {author}
  {\bibfnamefont {J.~C.~G.}\ \bibnamefont {Henriques}}, \bibinfo {author}
  {\bibfnamefont {Y.}~\bibnamefont {Gao}}, \bibinfo {author} {\bibfnamefont
  {X.}~\bibnamefont {Han}}, \bibinfo {author} {\bibfnamefont {H.}~\bibnamefont
  {Chen}}, \bibinfo {author} {\bibfnamefont {Y.}~\bibnamefont {Wang}}, \bibinfo
  {author} {\bibfnamefont {C.-A.}\ \bibnamefont {Palma}}, \bibinfo {author}
  {\bibfnamefont {Z.}~\bibnamefont {Cheng}}, \bibinfo {author} {\bibfnamefont
  {X.}~\bibnamefont {Lin}}, \bibinfo {author} {\bibfnamefont {S.}~\bibnamefont
  {Du}}, \bibinfo {author} {\bibfnamefont {J.}~\bibnamefont {Ma}}, \bibinfo
  {author} {\bibfnamefont {J.}~\bibnamefont {Fern{\'{a}}ndez-Rossier}},
  \bibinfo {author} {\bibfnamefont {X.}~\bibnamefont {Feng}}, \ and\ \bibinfo
  {author} {\bibfnamefont {H.-J.}\ \bibnamefont {Gao}},\ }\href {\doibase
  10.1038/s44160-025-00743-5} {\bibfield  {journal} {\bibinfo  {journal}
  {Nature Synthesis}\ ,\ } (\bibinfo {year} {2025})}\BibitemShut {NoStop}%
\bibitem [{\citenamefont {Kroto}\ \emph {et~al.}(1985)\citenamefont {Kroto},
  \citenamefont {Heath}, \citenamefont {O'Brien}, \citenamefont {Curl},\ and\
  \citenamefont {Smalley}}]{Kroto1985}%
  \BibitemOpen
  \bibfield  {author} {\bibinfo {author} {\bibfnamefont {H.~W.}\ \bibnamefont
  {Kroto}}, \bibinfo {author} {\bibfnamefont {J.~R.}\ \bibnamefont {Heath}},
  \bibinfo {author} {\bibfnamefont {S.~C.}\ \bibnamefont {O'Brien}}, \bibinfo
  {author} {\bibfnamefont {R.~F.}\ \bibnamefont {Curl}}, \ and\ \bibinfo
  {author} {\bibfnamefont {R.~E.}\ \bibnamefont {Smalley}},\ }\href {\doibase
  10.1038/318162a0} {\bibfield  {journal} {\bibinfo  {journal} {Nature}\
  }\textbf {\bibinfo {volume} {318}},\ \bibinfo {pages} {162} (\bibinfo {year}
  {1985})}\BibitemShut {NoStop}%
\bibitem [{\citenamefont {Kroto}(1988)}]{Kroto1988}%
  \BibitemOpen
  \bibfield  {author} {\bibinfo {author} {\bibfnamefont {H.}~\bibnamefont
  {Kroto}},\ }\href {\doibase 10.1126/science.242.4882.1139} {\bibfield
  {journal} {\bibinfo  {journal} {Science}\ }\textbf {\bibinfo {volume}
  {242}},\ \bibinfo {pages} {1139} (\bibinfo {year} {1988})}\BibitemShut
  {NoStop}%
\bibitem [{\citenamefont {Kroto}(1990)}]{Kroto1990}%
  \BibitemOpen
  \bibfield  {author} {\bibinfo {author} {\bibfnamefont {H.}~\bibnamefont
  {Kroto}},\ }\href {\doibase 10.1351/pac199062030407} {\bibfield  {journal}
  {\bibinfo  {journal} {Pure and Applied Chemistry}\ }\textbf {\bibinfo
  {volume} {62}},\ \bibinfo {pages} {407} (\bibinfo {year} {1990})}\BibitemShut
  {NoStop}%
\bibitem [{\citenamefont {Lamb}\ and\ \citenamefont
  {Huffman}(1993)}]{Lamb1993}%
  \BibitemOpen
  \bibfield  {author} {\bibinfo {author} {\bibfnamefont {L.~D.}\ \bibnamefont
  {Lamb}}\ and\ \bibinfo {author} {\bibfnamefont {D.~R.}\ \bibnamefont
  {Huffman}},\ }\href {\doibase 10.1016/0022-3697(93)90277-X} {\bibfield
  {journal} {\bibinfo  {journal} {Journal of Physics and Chemistry of Solids}\
  }\textbf {\bibinfo {volume} {54}},\ \bibinfo {pages} {1635} (\bibinfo {year}
  {1993})}\BibitemShut {NoStop}%
\bibitem [{\citenamefont {Kr\"atschmer}\ \emph {et~al.}(1990)\citenamefont
  {Kr\"atschmer}, \citenamefont {Lamb}, \citenamefont {Fostiropoulos},\ and\
  \citenamefont {Huffman}}]{Kratschmer1990}%
  \BibitemOpen
  \bibfield  {author} {\bibinfo {author} {\bibfnamefont {W.}~\bibnamefont
  {Kr\"atschmer}}, \bibinfo {author} {\bibfnamefont {L.~D.}\ \bibnamefont
  {Lamb}}, \bibinfo {author} {\bibfnamefont {K.}~\bibnamefont {Fostiropoulos}},
  \ and\ \bibinfo {author} {\bibfnamefont {D.~R.}\ \bibnamefont {Huffman}},\
  }\href {\doibase 10.1038/347354a0} {\bibfield  {journal} {\bibinfo  {journal}
  {Nature}\ }\textbf {\bibinfo {volume} {347}},\ \bibinfo {pages} {354}
  (\bibinfo {year} {1990})}\BibitemShut {NoStop}%
\bibitem [{\citenamefont {Heiney}\ \emph {et~al.}(1991)\citenamefont {Heiney},
  \citenamefont {Fischer}, \citenamefont {McGhie}, \citenamefont {Romanow},
  \citenamefont {Denenstein}, \citenamefont {McCauley~Jr.}, \citenamefont
  {Smith},\ and\ \citenamefont {Cox}}]{Heiney1991}%
  \BibitemOpen
  \bibfield  {author} {\bibinfo {author} {\bibfnamefont {P.~A.}\ \bibnamefont
  {Heiney}}, \bibinfo {author} {\bibfnamefont {J.~E.}\ \bibnamefont {Fischer}},
  \bibinfo {author} {\bibfnamefont {A.~R.}\ \bibnamefont {McGhie}}, \bibinfo
  {author} {\bibfnamefont {W.~J.}\ \bibnamefont {Romanow}}, \bibinfo {author}
  {\bibfnamefont {A.~M.}\ \bibnamefont {Denenstein}}, \bibinfo {author}
  {\bibfnamefont {J.~P.}\ \bibnamefont {McCauley~Jr.}}, \bibinfo {author}
  {\bibfnamefont {A.~B.}\ \bibnamefont {Smith}}, \ and\ \bibinfo {author}
  {\bibfnamefont {D.~E.}\ \bibnamefont {Cox}},\ }\href {\doibase
  10.1103/PhysRevLett.66.2911} {\bibfield  {journal} {\bibinfo  {journal}
  {Phys. Rev. Lett.}\ }\textbf {\bibinfo {volume} {66}},\ \bibinfo {pages}
  {2911} (\bibinfo {year} {1991})}\BibitemShut {NoStop}%
\bibitem [{\citenamefont {Fischer}\ \emph {et~al.}(1991)\citenamefont
  {Fischer}, \citenamefont {Heiney}, \citenamefont {McGhie}, \citenamefont
  {Romanow}, \citenamefont {Denenstein}, \citenamefont {McCauley},\ and\
  \citenamefont {Smith}}]{Fischer1991}%
  \BibitemOpen
  \bibfield  {author} {\bibinfo {author} {\bibfnamefont {J.~E.}\ \bibnamefont
  {Fischer}}, \bibinfo {author} {\bibfnamefont {P.~A.}\ \bibnamefont {Heiney}},
  \bibinfo {author} {\bibfnamefont {A.~R.}\ \bibnamefont {McGhie}}, \bibinfo
  {author} {\bibfnamefont {W.~J.}\ \bibnamefont {Romanow}}, \bibinfo {author}
  {\bibfnamefont {A.~M.}\ \bibnamefont {Denenstein}}, \bibinfo {author}
  {\bibfnamefont {J.~P.}\ \bibnamefont {McCauley}}, \ and\ \bibinfo {author}
  {\bibfnamefont {A.~B.}\ \bibnamefont {Smith}},\ }\href {\doibase
  10.1126/science.252.5010.1288} {\bibfield  {journal} {\bibinfo  {journal}
  {Science}\ }\textbf {\bibinfo {volume} {252}},\ \bibinfo {pages} {1288}
  (\bibinfo {year} {1991})}\BibitemShut {NoStop}%
\bibitem [{\citenamefont {David}\ \emph {et~al.}(1991)\citenamefont {David},
  \citenamefont {Ibberson}, \citenamefont {Matthewman}, \citenamefont
  {Prassides}, \citenamefont {Dennis}, \citenamefont {Hare}, \citenamefont
  {Kroto}, \citenamefont {Taylor},\ and\ \citenamefont {Walton}}]{David1991}%
  \BibitemOpen
  \bibfield  {author} {\bibinfo {author} {\bibfnamefont {W.~I.~F.}\
  \bibnamefont {David}}, \bibinfo {author} {\bibfnamefont {R.~M.}\ \bibnamefont
  {Ibberson}}, \bibinfo {author} {\bibfnamefont {J.~C.}\ \bibnamefont
  {Matthewman}}, \bibinfo {author} {\bibfnamefont {K.}~\bibnamefont
  {Prassides}}, \bibinfo {author} {\bibfnamefont {T.~J.~S.}\ \bibnamefont
  {Dennis}}, \bibinfo {author} {\bibfnamefont {J.~P.}\ \bibnamefont {Hare}},
  \bibinfo {author} {\bibfnamefont {H.~W.}\ \bibnamefont {Kroto}}, \bibinfo
  {author} {\bibfnamefont {R.}~\bibnamefont {Taylor}}, \ and\ \bibinfo {author}
  {\bibfnamefont {D.~R.~M.}\ \bibnamefont {Walton}},\ }\href {\doibase
  10.1038/353147a0} {\bibfield  {journal} {\bibinfo  {journal} {Nature}\
  }\textbf {\bibinfo {volume} {353}},\ \bibinfo {pages} {147} (\bibinfo {year}
  {1991})}\BibitemShut {NoStop}%
\bibitem [{\citenamefont {Heiney}(1992)}]{Heiney1992}%
  \BibitemOpen
  \bibfield  {author} {\bibinfo {author} {\bibfnamefont {P.~A.}\ \bibnamefont
  {Heiney}},\ }\href {\doibase 10.1016/0022-3697(92)90231-2} {\bibfield
  {journal} {\bibinfo  {journal} {Journal of Physics and Chemistry of Solids}\
  }\textbf {\bibinfo {volume} {53}},\ \bibinfo {pages} {1333} (\bibinfo {year}
  {1992})}\BibitemShut {NoStop}%
\bibitem [{\citenamefont {Rao}\ \emph {et~al.}(1993)\citenamefont {Rao},
  \citenamefont {Zhou}, \citenamefont {Wang}, \citenamefont {Hager},
  \citenamefont {Holden}, \citenamefont {Wang}, \citenamefont {Lee},
  \citenamefont {Bi}, \citenamefont {Eklund}, \citenamefont {Cornett},
  \citenamefont {Duncan},\ and\ \citenamefont {Amster}}]{Rao1993}%
  \BibitemOpen
  \bibfield  {author} {\bibinfo {author} {\bibfnamefont {A.~M.}\ \bibnamefont
  {Rao}}, \bibinfo {author} {\bibfnamefont {P.}~\bibnamefont {Zhou}}, \bibinfo
  {author} {\bibfnamefont {K.-A.}\ \bibnamefont {Wang}}, \bibinfo {author}
  {\bibfnamefont {G.~T.}\ \bibnamefont {Hager}}, \bibinfo {author}
  {\bibfnamefont {J.~M.}\ \bibnamefont {Holden}}, \bibinfo {author}
  {\bibfnamefont {Y.}~\bibnamefont {Wang}}, \bibinfo {author} {\bibfnamefont
  {W.-T.}\ \bibnamefont {Lee}}, \bibinfo {author} {\bibfnamefont {X.-X.}\
  \bibnamefont {Bi}}, \bibinfo {author} {\bibfnamefont {P.~C.}\ \bibnamefont
  {Eklund}}, \bibinfo {author} {\bibfnamefont {D.~S.}\ \bibnamefont {Cornett}},
  \bibinfo {author} {\bibfnamefont {M.~A.}\ \bibnamefont {Duncan}}, \ and\
  \bibinfo {author} {\bibfnamefont {I.~J.}\ \bibnamefont {Amster}},\ }\href
  {\doibase 10.1126/science.259.5097.955} {\bibfield  {journal} {\bibinfo
  {journal} {Science}\ }\textbf {\bibinfo {volume} {259}},\ \bibinfo {pages}
  {955} (\bibinfo {year} {1993})}\BibitemShut {NoStop}%
\bibitem [{\citenamefont {Iwasa}\ \emph {et~al.}(1994)\citenamefont {Iwasa},
  \citenamefont {Arima}, \citenamefont {Fleming}, \citenamefont {Siegrist},
  \citenamefont {Zhou}, \citenamefont {Haddon}, \citenamefont {Rothberg},
  \citenamefont {Lyons}, \citenamefont {Carter}, \citenamefont {Hebard},
  \citenamefont {Tycko}, \citenamefont {Dabbagh}, \citenamefont {Krajewski},
  \citenamefont {Thomas},\ and\ \citenamefont {Yagi}}]{Iwasa1994}%
  \BibitemOpen
  \bibfield  {author} {\bibinfo {author} {\bibfnamefont {Y.}~\bibnamefont
  {Iwasa}}, \bibinfo {author} {\bibfnamefont {T.}~\bibnamefont {Arima}},
  \bibinfo {author} {\bibfnamefont {R.~M.}\ \bibnamefont {Fleming}}, \bibinfo
  {author} {\bibfnamefont {T.}~\bibnamefont {Siegrist}}, \bibinfo {author}
  {\bibfnamefont {O.}~\bibnamefont {Zhou}}, \bibinfo {author} {\bibfnamefont
  {R.~C.}\ \bibnamefont {Haddon}}, \bibinfo {author} {\bibfnamefont {L.~J.}\
  \bibnamefont {Rothberg}}, \bibinfo {author} {\bibfnamefont {K.~B.}\
  \bibnamefont {Lyons}}, \bibinfo {author} {\bibfnamefont {H.~L.}\ \bibnamefont
  {Carter}}, \bibinfo {author} {\bibfnamefont {A.~F.}\ \bibnamefont {Hebard}},
  \bibinfo {author} {\bibfnamefont {R.}~\bibnamefont {Tycko}}, \bibinfo
  {author} {\bibfnamefont {G.}~\bibnamefont {Dabbagh}}, \bibinfo {author}
  {\bibfnamefont {J.~J.}\ \bibnamefont {Krajewski}}, \bibinfo {author}
  {\bibfnamefont {G.~A.}\ \bibnamefont {Thomas}}, \ and\ \bibinfo {author}
  {\bibfnamefont {T.}~\bibnamefont {Yagi}},\ }\href {\doibase
  10.1126/science.264.5165.1570} {\bibfield  {journal} {\bibinfo  {journal}
  {Science}\ }\textbf {\bibinfo {volume} {264}},\ \bibinfo {pages} {1570}
  (\bibinfo {year} {1994})}\BibitemShut {NoStop}%
\bibitem [{\citenamefont {N{\'u}{\~n}ez-Regueiro}\ \emph
  {et~al.}(1995)\citenamefont {N{\'u}{\~n}ez-Regueiro}, \citenamefont
  {Marques}, \citenamefont {Hodeau}, \citenamefont {B\'ethoux},\ and\
  \citenamefont {Perroux}}]{Nunez-Regueiro1995}%
  \BibitemOpen
  \bibfield  {author} {\bibinfo {author} {\bibfnamefont {M.}~\bibnamefont
  {N{\'u}{\~n}ez-Regueiro}}, \bibinfo {author} {\bibfnamefont {L.}~\bibnamefont
  {Marques}}, \bibinfo {author} {\bibfnamefont {J.~L.}\ \bibnamefont {Hodeau}},
  \bibinfo {author} {\bibfnamefont {O.}~\bibnamefont {B\'ethoux}}, \ and\
  \bibinfo {author} {\bibfnamefont {M.}~\bibnamefont {Perroux}},\ }\href
  {\doibase 10.1103/PhysRevLett.74.278} {\bibfield  {journal} {\bibinfo
  {journal} {Phys. Rev. Lett.}\ }\textbf {\bibinfo {volume} {74}},\ \bibinfo
  {pages} {278} (\bibinfo {year} {1995})}\BibitemShut {NoStop}%
\bibitem [{\citenamefont {Eklund}\ \emph {et~al.}(1995)\citenamefont {Eklund},
  \citenamefont {Rao}, \citenamefont {Zhou}, \citenamefont {Wang},\ and\
  \citenamefont {Holden}}]{Eklund1995}%
  \BibitemOpen
  \bibfield  {author} {\bibinfo {author} {\bibfnamefont {P.}~\bibnamefont
  {Eklund}}, \bibinfo {author} {\bibfnamefont {A.}~\bibnamefont {Rao}},
  \bibinfo {author} {\bibfnamefont {P.}~\bibnamefont {Zhou}}, \bibinfo {author}
  {\bibfnamefont {Y.}~\bibnamefont {Wang}}, \ and\ \bibinfo {author}
  {\bibfnamefont {J.}~\bibnamefont {Holden}},\ }\href {\doibase
  10.1016/0040-6090(94)05704-4} {\bibfield  {journal} {\bibinfo  {journal}
  {Thin Solid Films}\ }\textbf {\bibinfo {volume} {257}},\ \bibinfo {pages}
  {185} (\bibinfo {year} {1995})}\BibitemShut {NoStop}%
\bibitem [{\citenamefont {Xu}\ and\ \citenamefont {Scuseria}(1995)}]{Xu1995}%
  \BibitemOpen
  \bibfield  {author} {\bibinfo {author} {\bibfnamefont {C.~H.}\ \bibnamefont
  {Xu}}\ and\ \bibinfo {author} {\bibfnamefont {G.~E.}\ \bibnamefont
  {Scuseria}},\ }\href {\doibase 10.1103/PhysRevLett.74.274} {\bibfield
  {journal} {\bibinfo  {journal} {Phys. Rev. Lett.}\ }\textbf {\bibinfo
  {volume} {74}},\ \bibinfo {pages} {274} (\bibinfo {year} {1995})}\BibitemShut
  {NoStop}%
\bibitem [{\citenamefont {Springborg}(1995)}]{Springborg1995}%
  \BibitemOpen
  \bibfield  {author} {\bibinfo {author} {\bibfnamefont {M.}~\bibnamefont
  {Springborg}},\ }\href {\doibase 10.1103/PhysRevB.52.2935} {\bibfield
  {journal} {\bibinfo  {journal} {Phys. Rev. B}\ }\textbf {\bibinfo {volume}
  {52}},\ \bibinfo {pages} {2935} (\bibinfo {year} {1995})}\BibitemShut
  {NoStop}%
\bibitem [{\citenamefont {Marques}\ \emph {et~al.}(1996)\citenamefont
  {Marques}, \citenamefont {Hodeau}, \citenamefont {N\'u\~nez Regueiro},\ and\
  \citenamefont {Perroux}}]{Marques1996}%
  \BibitemOpen
  \bibfield  {author} {\bibinfo {author} {\bibfnamefont {L.}~\bibnamefont
  {Marques}}, \bibinfo {author} {\bibfnamefont {J.~L.}\ \bibnamefont {Hodeau}},
  \bibinfo {author} {\bibfnamefont {M.}~\bibnamefont {N\'u\~nez Regueiro}}, \
  and\ \bibinfo {author} {\bibfnamefont {M.}~\bibnamefont {Perroux}},\ }\href
  {\doibase 10.1103/PhysRevB.54.R12633} {\bibfield  {journal} {\bibinfo
  {journal} {Phys. Rev. B}\ }\textbf {\bibinfo {volume} {54}},\ \bibinfo
  {pages} {R12633} (\bibinfo {year} {1996})}\BibitemShut {NoStop}%
\bibitem [{\citenamefont {Giacalone}\ and\ \citenamefont
  {Mart\'in}(2006)}]{Giacalone2006}%
  \BibitemOpen
  \bibfield  {author} {\bibinfo {author} {\bibfnamefont {F.}~\bibnamefont
  {Giacalone}}\ and\ \bibinfo {author} {\bibfnamefont {N.}~\bibnamefont
  {Mart\'in}},\ }\href {\doibase 10.1021/cr068389h} {\bibfield  {journal}
  {\bibinfo  {journal} {Chem. Rev.}\ }\textbf {\bibinfo {volume} {106}},\
  \bibinfo {pages} {5136} (\bibinfo {year} {2006})}\BibitemShut {NoStop}%
\bibitem [{\citenamefont {{\'A}lvarez~Murga}\ and\ \citenamefont
  {Hodeau}(2015)}]{Murga2015}%
  \BibitemOpen
  \bibfield  {author} {\bibinfo {author} {\bibfnamefont {M.}~\bibnamefont
  {{\'A}lvarez~Murga}}\ and\ \bibinfo {author} {\bibfnamefont {J.}~\bibnamefont
  {Hodeau}},\ }\href {\doibase 10.1016/j.carbon.2014.10.083} {\bibfield
  {journal} {\bibinfo  {journal} {Carbon}\ }\textbf {\bibinfo {volume} {82}},\
  \bibinfo {pages} {381} (\bibinfo {year} {2015})}\BibitemShut {NoStop}%
\bibitem [{\citenamefont {Ruoff}\ and\ \citenamefont
  {Ruoff}(1991)}]{Ruoff1991}%
  \BibitemOpen
  \bibfield  {author} {\bibinfo {author} {\bibfnamefont {R.~S.}\ \bibnamefont
  {Ruoff}}\ and\ \bibinfo {author} {\bibfnamefont {A.~L.}\ \bibnamefont
  {Ruoff}},\ }\href {\doibase 10.1038/350663b0} {\bibfield  {journal} {\bibinfo
   {journal} {Nature}\ }\textbf {\bibinfo {volume} {350}},\ \bibinfo {pages}
  {663} (\bibinfo {year} {1991})}\BibitemShut {NoStop}%
\bibitem [{\citenamefont {Venkateswaran}\ \emph {et~al.}(1996)\citenamefont
  {Venkateswaran}, \citenamefont {Sanzi}, \citenamefont {Krishnappa},
  \citenamefont {Marques}, \citenamefont {Hodeau}, \citenamefont
  {N{\'u}{\~n}ez-Regueiro}, \citenamefont {Rao},\ and\ \citenamefont
  {Eklund}}]{Venkateswaran1996}%
  \BibitemOpen
  \bibfield  {author} {\bibinfo {author} {\bibfnamefont {U.~D.}\ \bibnamefont
  {Venkateswaran}}, \bibinfo {author} {\bibfnamefont {D.}~\bibnamefont
  {Sanzi}}, \bibinfo {author} {\bibfnamefont {J.}~\bibnamefont {Krishnappa}},
  \bibinfo {author} {\bibfnamefont {L.}~\bibnamefont {Marques}}, \bibinfo
  {author} {\bibfnamefont {J.-L.}\ \bibnamefont {Hodeau}}, \bibinfo {author}
  {\bibfnamefont {M.}~\bibnamefont {N{\'u}{\~n}ez-Regueiro}}, \bibinfo {author}
  {\bibfnamefont {A.~M.}\ \bibnamefont {Rao}}, \ and\ \bibinfo {author}
  {\bibfnamefont {P.~C.}\ \bibnamefont {Eklund}},\ }\href {\doibase
  10.1002/pssb.2221980173} {\bibfield  {journal} {\bibinfo  {journal} {Phys.
  Stat. Sol. (b)}\ }\textbf {\bibinfo {volume} {198}},\ \bibinfo {pages} {545}
  (\bibinfo {year} {1996})}\BibitemShut {NoStop}%
\bibitem [{\citenamefont {Rao}\ \emph {et~al.}(1997)\citenamefont {Rao},
  \citenamefont {Eklund}, \citenamefont {Hodeau}, \citenamefont {Marques},\
  and\ \citenamefont {Nunez-Regueiro}}]{Rao1997}%
  \BibitemOpen
  \bibfield  {author} {\bibinfo {author} {\bibfnamefont {A.~M.}\ \bibnamefont
  {Rao}}, \bibinfo {author} {\bibfnamefont {P.~C.}\ \bibnamefont {Eklund}},
  \bibinfo {author} {\bibfnamefont {J.-L.}\ \bibnamefont {Hodeau}}, \bibinfo
  {author} {\bibfnamefont {L.}~\bibnamefont {Marques}}, \ and\ \bibinfo
  {author} {\bibfnamefont {M.}~\bibnamefont {Nunez-Regueiro}},\ }\href
  {\doibase 10.1103/PhysRevB.55.4766} {\bibfield  {journal} {\bibinfo
  {journal} {Phys. Rev. B}\ }\textbf {\bibinfo {volume} {55}},\ \bibinfo
  {pages} {4766} (\bibinfo {year} {1997})}\BibitemShut {NoStop}%
\bibitem [{\citenamefont {Forr\'o}\ and\ \citenamefont
  {Mih\'aly}(2001)}]{Forro2001}%
  \BibitemOpen
  \bibfield  {author} {\bibinfo {author} {\bibfnamefont {L.}~\bibnamefont
  {Forr\'o}}\ and\ \bibinfo {author} {\bibfnamefont {L.}~\bibnamefont
  {Mih\'aly}},\ }\href {\doibase 10.1088/0034-4885/64/5/202} {\bibfield
  {journal} {\bibinfo  {journal} {Rep. Prog. Phys.}\ }\textbf {\bibinfo
  {volume} {64}},\ \bibinfo {pages} {649} (\bibinfo {year} {2001})}\BibitemShut
  {NoStop}%
\bibitem [{\citenamefont {Makarova}(2001)}]{Makarova2001}%
  \BibitemOpen
  \bibfield  {author} {\bibinfo {author} {\bibfnamefont {T.~L.}\ \bibnamefont
  {Makarova}},\ }\href {\doibase 10.1134/1.1356145} {\bibfield  {journal}
  {\bibinfo  {journal} {Semiconductors}\ }\textbf {\bibinfo {volume} {35}},\
  \bibinfo {pages} {243} (\bibinfo {year} {2001})}\BibitemShut {NoStop}%
\bibitem [{\citenamefont {Sun}\ \emph {et~al.}(2005)\citenamefont {Sun},
  \citenamefont {Liang}, \citenamefont {Yang},\ and\ \citenamefont
  {Gao}}]{Sun2005}%
  \BibitemOpen
  \bibfield  {author} {\bibinfo {author} {\bibfnamefont {J.}~\bibnamefont
  {Sun}}, \bibinfo {author} {\bibfnamefont {W.}~\bibnamefont {Liang}}, \bibinfo
  {author} {\bibfnamefont {J.}~\bibnamefont {Yang}}, \ and\ \bibinfo {author}
  {\bibfnamefont {J.}~\bibnamefont {Gao}},\ }\href {\doibase
  10.1016/j.theochem.2005.07.030} {\bibfield  {journal} {\bibinfo  {journal}
  {Journal of Molecular Structure: THEOCHEM}\ }\textbf {\bibinfo {volume}
  {755}},\ \bibinfo {pages} {105} (\bibinfo {year} {2005})}\BibitemShut
  {NoStop}%
\bibitem [{\citenamefont {Belosludov}\ \emph {et~al.}(2003)\citenamefont
  {Belosludov}, \citenamefont {Inerbaev}, \citenamefont {Belosludov},\ and\
  \citenamefont {Kawazoe}}]{Belosludov2003}%
  \BibitemOpen
  \bibfield  {author} {\bibinfo {author} {\bibfnamefont {V.~R.}\ \bibnamefont
  {Belosludov}}, \bibinfo {author} {\bibfnamefont {T.~M.}\ \bibnamefont
  {Inerbaev}}, \bibinfo {author} {\bibfnamefont {R.~V.}\ \bibnamefont
  {Belosludov}}, \ and\ \bibinfo {author} {\bibfnamefont {Y.}~\bibnamefont
  {Kawazoe}},\ }\href {\doibase 10.1103/PhysRevB.67.155410} {\bibfield
  {journal} {\bibinfo  {journal} {Phys. Rev. B}\ }\textbf {\bibinfo {volume}
  {67}},\ \bibinfo {pages} {155410} (\bibinfo {year} {2003})}\BibitemShut
  {NoStop}%
\bibitem [{\citenamefont {Belosludov}\ \emph {et~al.}(2006)\citenamefont
  {Belosludov}, \citenamefont {Inerbaev}, \citenamefont {Belosludov},
  \citenamefont {Kawazoe},\ and\ \citenamefont {Kudoh}}]{Belosludov2006}%
  \BibitemOpen
  \bibfield  {author} {\bibinfo {author} {\bibfnamefont {V.}~\bibnamefont
  {Belosludov}}, \bibinfo {author} {\bibfnamefont {T.}~\bibnamefont
  {Inerbaev}}, \bibinfo {author} {\bibfnamefont {R.}~\bibnamefont
  {Belosludov}}, \bibinfo {author} {\bibfnamefont {Y.}~\bibnamefont {Kawazoe}},
  \ and\ \bibinfo {author} {\bibfnamefont {J.}~\bibnamefont {Kudoh}},\ }\href
  {\doibase 10.1016/j.commatsci.2005.01.014} {\bibfield  {journal} {\bibinfo
  {journal} {Computational Materials Science}\ }\textbf {\bibinfo {volume}
  {36}},\ \bibinfo {pages} {17} (\bibinfo {year} {2006})}\BibitemShut {NoStop}%
\bibitem [{\citenamefont {Kennedy}\ \emph {et~al.}(2008)\citenamefont
  {Kennedy}, \citenamefont {Ayzner}, \citenamefont {Wanger}, \citenamefont
  {Day}, \citenamefont {Halim}, \citenamefont {Khan}, \citenamefont {Tolbert},
  \citenamefont {Schwartz},\ and\ \citenamefont {Rubin}}]{Kennedy2008}%
  \BibitemOpen
  \bibfield  {author} {\bibinfo {author} {\bibfnamefont {R.~D.}\ \bibnamefont
  {Kennedy}}, \bibinfo {author} {\bibfnamefont {A.~L.}\ \bibnamefont {Ayzner}},
  \bibinfo {author} {\bibfnamefont {D.~D.}\ \bibnamefont {Wanger}}, \bibinfo
  {author} {\bibfnamefont {C.~T.}\ \bibnamefont {Day}}, \bibinfo {author}
  {\bibfnamefont {M.}~\bibnamefont {Halim}}, \bibinfo {author} {\bibfnamefont
  {S.~I.}\ \bibnamefont {Khan}}, \bibinfo {author} {\bibfnamefont {S.~H.}\
  \bibnamefont {Tolbert}}, \bibinfo {author} {\bibfnamefont {B.~J.}\
  \bibnamefont {Schwartz}}, \ and\ \bibinfo {author} {\bibfnamefont
  {Y.}~\bibnamefont {Rubin}},\ }\href {\doibase 10.1021/ja807627u} {\bibfield
  {journal} {\bibinfo  {journal} {J. Am. Chem. Soc.}\ }\textbf {\bibinfo
  {volume} {130}},\ \bibinfo {pages} {17290} (\bibinfo {year}
  {2008})}\BibitemShut {NoStop}%
\bibitem [{\citenamefont {Dennler}\ \emph {et~al.}(2009)\citenamefont
  {Dennler}, \citenamefont {Scharber},\ and\ \citenamefont
  {Brabec}}]{Dennler2009}%
  \BibitemOpen
  \bibfield  {author} {\bibinfo {author} {\bibfnamefont {G.}~\bibnamefont
  {Dennler}}, \bibinfo {author} {\bibfnamefont {M.~C.}\ \bibnamefont
  {Scharber}}, \ and\ \bibinfo {author} {\bibfnamefont {C.~J.}\ \bibnamefont
  {Brabec}},\ }\href {\doibase 10.1002/adma.200801283} {\bibfield  {journal}
  {\bibinfo  {journal} {Adv. Mater.}\ }\textbf {\bibinfo {volume} {21}},\
  \bibinfo {pages} {1323} (\bibinfo {year} {2009})}\BibitemShut {NoStop}%
\bibitem [{\citenamefont {Stephens}\ \emph {et~al.}(1991)\citenamefont
  {Stephens}, \citenamefont {Mihaly}, \citenamefont {Lee}, \citenamefont
  {Whetten}, \citenamefont {Huang}, \citenamefont {Kaner}, \citenamefont
  {Deiderich},\ and\ \citenamefont {Holczer}}]{Stephens1991}%
  \BibitemOpen
  \bibfield  {author} {\bibinfo {author} {\bibfnamefont {P.~W.}\ \bibnamefont
  {Stephens}}, \bibinfo {author} {\bibfnamefont {L.}~\bibnamefont {Mihaly}},
  \bibinfo {author} {\bibfnamefont {P.~L.}\ \bibnamefont {Lee}}, \bibinfo
  {author} {\bibfnamefont {R.~L.}\ \bibnamefont {Whetten}}, \bibinfo {author}
  {\bibfnamefont {S.-M.}\ \bibnamefont {Huang}}, \bibinfo {author}
  {\bibfnamefont {R.}~\bibnamefont {Kaner}}, \bibinfo {author} {\bibfnamefont
  {F.}~\bibnamefont {Deiderich}}, \ and\ \bibinfo {author} {\bibfnamefont
  {K.}~\bibnamefont {Holczer}},\ }\href {\doibase 10.1038/351632a0} {\bibfield
  {journal} {\bibinfo  {journal} {Nature}\ }\textbf {\bibinfo {volume} {351}},\
  \bibinfo {pages} {632} (\bibinfo {year} {1991})}\BibitemShut {NoStop}%
\bibitem [{\citenamefont {Chauvet}\ \emph {et~al.}(1994)\citenamefont
  {Chauvet}, \citenamefont {Oszl\`anyi}, \citenamefont {Forro}, \citenamefont
  {Stephens}, \citenamefont {Tegze}, \citenamefont {Faigel},\ and\
  \citenamefont {J\`anossy}}]{Chauvet1994}%
  \BibitemOpen
  \bibfield  {author} {\bibinfo {author} {\bibfnamefont {O.}~\bibnamefont
  {Chauvet}}, \bibinfo {author} {\bibfnamefont {G.}~\bibnamefont {Oszl\`anyi}},
  \bibinfo {author} {\bibfnamefont {L.}~\bibnamefont {Forro}}, \bibinfo
  {author} {\bibfnamefont {P.~W.}\ \bibnamefont {Stephens}}, \bibinfo {author}
  {\bibfnamefont {M.}~\bibnamefont {Tegze}}, \bibinfo {author} {\bibfnamefont
  {G.}~\bibnamefont {Faigel}}, \ and\ \bibinfo {author} {\bibfnamefont
  {A.}~\bibnamefont {J\`anossy}},\ }\href {\doibase
  10.1103/PhysRevLett.72.2721} {\bibfield  {journal} {\bibinfo  {journal}
  {Phys. Rev. Lett.}\ }\textbf {\bibinfo {volume} {72}},\ \bibinfo {pages}
  {2721} (\bibinfo {year} {1994})}\BibitemShut {NoStop}%
\bibitem [{\citenamefont {Stephens}\ \emph {et~al.}(1994)\citenamefont
  {Stephens}, \citenamefont {Bortel}, \citenamefont {Faigel}, \citenamefont
  {Tegze}, \citenamefont {J\'anossy}, \citenamefont {Pekker}, \citenamefont
  {Oszlanyi},\ and\ \citenamefont {Forr\'o}}]{Stephens1994}%
  \BibitemOpen
  \bibfield  {author} {\bibinfo {author} {\bibfnamefont {P.~W.}\ \bibnamefont
  {Stephens}}, \bibinfo {author} {\bibfnamefont {G.}~\bibnamefont {Bortel}},
  \bibinfo {author} {\bibfnamefont {G.}~\bibnamefont {Faigel}}, \bibinfo
  {author} {\bibfnamefont {M.}~\bibnamefont {Tegze}}, \bibinfo {author}
  {\bibfnamefont {A.}~\bibnamefont {J\'anossy}}, \bibinfo {author}
  {\bibfnamefont {S.}~\bibnamefont {Pekker}}, \bibinfo {author} {\bibfnamefont
  {G.}~\bibnamefont {Oszlanyi}}, \ and\ \bibinfo {author} {\bibfnamefont
  {L.}~\bibnamefont {Forr\'o}},\ }\href {https://doi.org/10.1038/370636a0}
  {\bibfield  {journal} {\bibinfo  {journal} {Nature}\ }\textbf {\bibinfo
  {volume} {370}},\ \bibinfo {pages} {636} (\bibinfo {year}
  {1994})}\BibitemShut {NoStop}%
\bibitem [{\citenamefont {Gunnarsson}(1997)}]{Gunnarsson1997}%
  \BibitemOpen
  \bibfield  {author} {\bibinfo {author} {\bibfnamefont {O.}~\bibnamefont
  {Gunnarsson}},\ }\href {\doibase 10.1103/RevModPhys.69.575} {\bibfield
  {journal} {\bibinfo  {journal} {Rev. Mod. Phys.}\ }\textbf {\bibinfo {volume}
  {69}},\ \bibinfo {pages} {575} (\bibinfo {year} {1997})}\BibitemShut
  {NoStop}%
\bibitem [{\citenamefont {Huq}\ \emph {et~al.}(2001)\citenamefont {Huq},
  \citenamefont {Stephens}, \citenamefont {Bendele},\ and\ \citenamefont
  {Ibberson}}]{Huq2001}%
  \BibitemOpen
  \bibfield  {author} {\bibinfo {author} {\bibfnamefont {A.}~\bibnamefont
  {Huq}}, \bibinfo {author} {\bibfnamefont {P.}~\bibnamefont {Stephens}},
  \bibinfo {author} {\bibfnamefont {G.~M.}\ \bibnamefont {Bendele}}, \ and\
  \bibinfo {author} {\bibfnamefont {R.}~\bibnamefont {Ibberson}},\ }\href
  {\doibase 10.1016/S0009-2614(01)01019-3} {\bibfield  {journal} {\bibinfo
  {journal} {Chemical Physics Letters}\ }\textbf {\bibinfo {volume} {347}},\
  \bibinfo {pages} {13} (\bibinfo {year} {2001})}\BibitemShut {NoStop}%
\bibitem [{\citenamefont {Hou}\ \emph {et~al.}(2022)\citenamefont {Hou},
  \citenamefont {Cui}, \citenamefont {Guan}, \citenamefont {Wang},
  \citenamefont {Li}, \citenamefont {Liu}, \citenamefont {Zhu},\ and\
  \citenamefont {Zheng}}]{Hou2022}%
  \BibitemOpen
  \bibfield  {author} {\bibinfo {author} {\bibfnamefont {L.}~\bibnamefont
  {Hou}}, \bibinfo {author} {\bibfnamefont {X.}~\bibnamefont {Cui}}, \bibinfo
  {author} {\bibfnamefont {B.}~\bibnamefont {Guan}}, \bibinfo {author}
  {\bibfnamefont {S.}~\bibnamefont {Wang}}, \bibinfo {author} {\bibfnamefont
  {R.}~\bibnamefont {Li}}, \bibinfo {author} {\bibfnamefont {Y.}~\bibnamefont
  {Liu}}, \bibinfo {author} {\bibfnamefont {D.}~\bibnamefont {Zhu}}, \ and\
  \bibinfo {author} {\bibfnamefont {J.}~\bibnamefont {Zheng}},\ }\href
  {\doibase 10.1038/s41586-022-04771-5} {\bibfield  {journal} {\bibinfo
  {journal} {Nature}\ }\textbf {\bibinfo {volume} {606}},\ \bibinfo {pages}
  {507} (\bibinfo {year} {2022})}\BibitemShut {NoStop}%
\bibitem [{\citenamefont {Peng}(2022)}]{Peng2022c}%
  \BibitemOpen
  \bibfield  {author} {\bibinfo {author} {\bibfnamefont {B.}~\bibnamefont
  {Peng}},\ }\href {\doibase 10.1021/jacs.2c08054} {\bibfield  {journal}
  {\bibinfo  {journal} {J. Am. Chem. Soc.}\ }\textbf {\bibinfo {volume}
  {144}},\ \bibinfo {pages} {19921} (\bibinfo {year} {2022})}\BibitemShut
  {NoStop}%
\bibitem [{\citenamefont {Jones}\ and\ \citenamefont {Peng}(2023)}]{Jones2023}%
  \BibitemOpen
  \bibfield  {author} {\bibinfo {author} {\bibfnamefont {C.}~\bibnamefont
  {Jones}}\ and\ \bibinfo {author} {\bibfnamefont {B.}~\bibnamefont {Peng}},\
  }\href {\doibase 10.1021/acs.jpclett.3c02578} {\bibfield  {journal} {\bibinfo
   {journal} {J. Phys. Chem. Lett.}\ }\textbf {\bibinfo {volume} {14}},\
  \bibinfo {pages} {11768} (\bibinfo {year} {2023})}\BibitemShut {NoStop}%
\bibitem [{\citenamefont {Wu}\ and\ \citenamefont {Peng}(2025)}]{Wu2025}%
  \BibitemOpen
  \bibfield  {author} {\bibinfo {author} {\bibfnamefont {J.}~\bibnamefont
  {Wu}}\ and\ \bibinfo {author} {\bibfnamefont {B.}~\bibnamefont {Peng}},\
  }\href {\doibase 10.1021/jacs.4c13167} {\bibfield  {journal} {\bibinfo
  {journal} {J. Am. Chem. Soc.}\ }\textbf {\bibinfo {volume} {147}},\ \bibinfo
  {pages} {1749} (\bibinfo {year} {2025})}\BibitemShut {NoStop}%
\bibitem [{\citenamefont {Shearsby}\ \emph {et~al.}(2025)\citenamefont
  {Shearsby}, \citenamefont {Wu}, \citenamefont {Yang},\ and\ \citenamefont
  {Peng}}]{Shearsby2025}%
  \BibitemOpen
  \bibfield  {author} {\bibinfo {author} {\bibfnamefont {D.}~\bibnamefont
  {Shearsby}}, \bibinfo {author} {\bibfnamefont {J.}~\bibnamefont {Wu}},
  \bibinfo {author} {\bibfnamefont {D.}~\bibnamefont {Yang}}, \ and\ \bibinfo
  {author} {\bibfnamefont {B.}~\bibnamefont {Peng}},\ }\href {\doibase
  10.1039/D4NR04540H} {\bibfield  {journal} {\bibinfo  {journal} {Nanoscale}\
  }\textbf {\bibinfo {volume} {17}},\ \bibinfo {pages} {2616} (\bibinfo {year}
  {2025})}\BibitemShut {NoStop}%
\bibitem [{\citenamefont {Meirzadeh}\ \emph {et~al.}(2023)\citenamefont
  {Meirzadeh}, \citenamefont {Evans}, \citenamefont {Rezaee}, \citenamefont
  {Milich}, \citenamefont {Dionne}, \citenamefont {Darlington}, \citenamefont
  {Bao}, \citenamefont {Bartholomew}, \citenamefont {Handa}, \citenamefont
  {Rizzo}, \citenamefont {Wiscons}, \citenamefont {Reza}, \citenamefont
  {Zangiabadi}, \citenamefont {Fardian-Melamed}, \citenamefont {Crowther},
  \citenamefont {Schuck}, \citenamefont {Basov}, \citenamefont {Zhu},
  \citenamefont {Giri}, \citenamefont {Hopkins}, \citenamefont {Kim},
  \citenamefont {Steigerwald}, \citenamefont {Yang}, \citenamefont {Nuckolls},\
  and\ \citenamefont {Roy}}]{Meirzadeh2023}%
  \BibitemOpen
  \bibfield  {author} {\bibinfo {author} {\bibfnamefont {E.}~\bibnamefont
  {Meirzadeh}}, \bibinfo {author} {\bibfnamefont {A.~M.}\ \bibnamefont
  {Evans}}, \bibinfo {author} {\bibfnamefont {M.}~\bibnamefont {Rezaee}},
  \bibinfo {author} {\bibfnamefont {M.}~\bibnamefont {Milich}}, \bibinfo
  {author} {\bibfnamefont {C.~J.}\ \bibnamefont {Dionne}}, \bibinfo {author}
  {\bibfnamefont {T.~P.}\ \bibnamefont {Darlington}}, \bibinfo {author}
  {\bibfnamefont {S.~T.}\ \bibnamefont {Bao}}, \bibinfo {author} {\bibfnamefont
  {A.~K.}\ \bibnamefont {Bartholomew}}, \bibinfo {author} {\bibfnamefont
  {T.}~\bibnamefont {Handa}}, \bibinfo {author} {\bibfnamefont {D.~J.}\
  \bibnamefont {Rizzo}}, \bibinfo {author} {\bibfnamefont {R.~A.}\ \bibnamefont
  {Wiscons}}, \bibinfo {author} {\bibfnamefont {M.}~\bibnamefont {Reza}},
  \bibinfo {author} {\bibfnamefont {A.}~\bibnamefont {Zangiabadi}}, \bibinfo
  {author} {\bibfnamefont {N.}~\bibnamefont {Fardian-Melamed}}, \bibinfo
  {author} {\bibfnamefont {A.~C.}\ \bibnamefont {Crowther}}, \bibinfo {author}
  {\bibfnamefont {P.~J.}\ \bibnamefont {Schuck}}, \bibinfo {author}
  {\bibfnamefont {D.~N.}\ \bibnamefont {Basov}}, \bibinfo {author}
  {\bibfnamefont {X.}~\bibnamefont {Zhu}}, \bibinfo {author} {\bibfnamefont
  {A.}~\bibnamefont {Giri}}, \bibinfo {author} {\bibfnamefont {P.~E.}\
  \bibnamefont {Hopkins}}, \bibinfo {author} {\bibfnamefont {P.}~\bibnamefont
  {Kim}}, \bibinfo {author} {\bibfnamefont {M.~L.}\ \bibnamefont
  {Steigerwald}}, \bibinfo {author} {\bibfnamefont {J.}~\bibnamefont {Yang}},
  \bibinfo {author} {\bibfnamefont {C.}~\bibnamefont {Nuckolls}}, \ and\
  \bibinfo {author} {\bibfnamefont {X.}~\bibnamefont {Roy}},\ }\href {\doibase
  10.1038/s41586-022-05401-w} {\bibfield  {journal} {\bibinfo  {journal}
  {Nature}\ }\textbf {\bibinfo {volume} {613}},\ \bibinfo {pages} {71}
  (\bibinfo {year} {2023})}\BibitemShut {NoStop}%
\bibitem [{\citenamefont {Hohenberg}\ and\ \citenamefont
  {Kohn}(1964)}]{Hohenberg1964}%
  \BibitemOpen
  \bibfield  {author} {\bibinfo {author} {\bibfnamefont {P.}~\bibnamefont
  {Hohenberg}}\ and\ \bibinfo {author} {\bibfnamefont {W.}~\bibnamefont
  {Kohn}},\ }\href {\doibase 10.1103/PhysRev.136.B864} {\bibfield  {journal}
  {\bibinfo  {journal} {Phys. Rev.}\ }\textbf {\bibinfo {volume} {136}},\
  \bibinfo {pages} {B864} (\bibinfo {year} {1964})}\BibitemShut {NoStop}%
\bibitem [{\citenamefont {Kohn}\ and\ \citenamefont {Sham}(1965)}]{Kohn1965}%
  \BibitemOpen
  \bibfield  {author} {\bibinfo {author} {\bibfnamefont {W.}~\bibnamefont
  {Kohn}}\ and\ \bibinfo {author} {\bibfnamefont {L.~J.}\ \bibnamefont
  {Sham}},\ }\href {\doibase 10.1103/PhysRev.140.A1133} {\bibfield  {journal}
  {\bibinfo  {journal} {Phys. Rev.}\ }\textbf {\bibinfo {volume} {140}},\
  \bibinfo {pages} {A1133} (\bibinfo {year} {1965})}\BibitemShut {NoStop}%
\bibitem [{\citenamefont {Kroto}(1987)}]{Kroto1987}%
  \BibitemOpen
  \bibfield  {author} {\bibinfo {author} {\bibfnamefont {H.~W.}\ \bibnamefont
  {Kroto}},\ }\href {\doibase 10.1038/329529a0} {\bibfield  {journal} {\bibinfo
   {journal} {Nature}\ }\textbf {\bibinfo {volume} {329}},\ \bibinfo {pages}
  {529} (\bibinfo {year} {1987})}\BibitemShut {NoStop}%
\bibitem [{\citenamefont {Goroff}(1996)}]{Goroff1996}%
  \BibitemOpen
  \bibfield  {author} {\bibinfo {author} {\bibfnamefont {N.~S.}\ \bibnamefont
  {Goroff}},\ }\href {\doibase 10.1021/ar950162d} {\bibfield  {journal}
  {\bibinfo  {journal} {Acc. Chem. Res.}\ }\textbf {\bibinfo {volume} {29}},\
  \bibinfo {pages} {77} (\bibinfo {year} {1996})}\BibitemShut {NoStop}%
\bibitem [{\citenamefont {Bernal}\ \emph {et~al.}(2019)\citenamefont {Bernal},
  \citenamefont {Haenecour}, \citenamefont {Howe}, \citenamefont {Zega},
  \citenamefont {Amari},\ and\ \citenamefont {Ziurys}}]{Bernal2019}%
  \BibitemOpen
  \bibfield  {author} {\bibinfo {author} {\bibfnamefont {J.~J.}\ \bibnamefont
  {Bernal}}, \bibinfo {author} {\bibfnamefont {P.}~\bibnamefont {Haenecour}},
  \bibinfo {author} {\bibfnamefont {J.}~\bibnamefont {Howe}}, \bibinfo {author}
  {\bibfnamefont {T.~J.}\ \bibnamefont {Zega}}, \bibinfo {author}
  {\bibfnamefont {S.}~\bibnamefont {Amari}}, \ and\ \bibinfo {author}
  {\bibfnamefont {L.~M.}\ \bibnamefont {Ziurys}},\ }\href {\doibase
  10.3847/2041-8213/ab4206} {\bibfield  {journal} {\bibinfo  {journal} {The
  Astrophysical Journal Letters}\ }\textbf {\bibinfo {volume} {883}},\ \bibinfo
  {pages} {L43} (\bibinfo {year} {2019})}\BibitemShut {NoStop}%
\bibitem [{\citenamefont {Wang}\ \emph {et~al.}(2023)\citenamefont {Wang},
  \citenamefont {Zhang}, \citenamefont {Wu}, \citenamefont {Chen},
  \citenamefont {Yang}, \citenamefont {Lu},\ and\ \citenamefont
  {Du}}]{Wang2023}%
  \BibitemOpen
  \bibfield  {author} {\bibinfo {author} {\bibfnamefont {T.}~\bibnamefont
  {Wang}}, \bibinfo {author} {\bibfnamefont {L.}~\bibnamefont {Zhang}},
  \bibinfo {author} {\bibfnamefont {J.}~\bibnamefont {Wu}}, \bibinfo {author}
  {\bibfnamefont {M.}~\bibnamefont {Chen}}, \bibinfo {author} {\bibfnamefont
  {S.}~\bibnamefont {Yang}}, \bibinfo {author} {\bibfnamefont {Y.}~\bibnamefont
  {Lu}}, \ and\ \bibinfo {author} {\bibfnamefont {P.}~\bibnamefont {Du}},\
  }\href {\doibase 10.1002/anie.202311352} {\bibfield  {journal} {\bibinfo
  {journal} {Angew. Chem. Int. Ed.}\ }\textbf {\bibinfo {volume} {62}},\
  \bibinfo {pages} {e202311352} (\bibinfo {year} {2023})}\BibitemShut {NoStop}%
\bibitem [{\citenamefont {Chen}\ \emph {et~al.}(2024)\citenamefont {Chen},
  \citenamefont {Mu}, \citenamefont {Jin}, \citenamefont {Wei}, \citenamefont
  {Hao}, \citenamefont {Wang}, \citenamefont {Caro},\ and\ \citenamefont
  {Huang}}]{Chen2024}%
  \BibitemOpen
  \bibfield  {author} {\bibinfo {author} {\bibfnamefont {X.}~\bibnamefont
  {Chen}}, \bibinfo {author} {\bibfnamefont {Y.}~\bibnamefont {Mu}}, \bibinfo
  {author} {\bibfnamefont {C.}~\bibnamefont {Jin}}, \bibinfo {author}
  {\bibfnamefont {Y.}~\bibnamefont {Wei}}, \bibinfo {author} {\bibfnamefont
  {J.}~\bibnamefont {Hao}}, \bibinfo {author} {\bibfnamefont {H.}~\bibnamefont
  {Wang}}, \bibinfo {author} {\bibfnamefont {J.}~\bibnamefont {Caro}}, \ and\
  \bibinfo {author} {\bibfnamefont {A.}~\bibnamefont {Huang}},\ }\href
  {\doibase 10.1002/anie.202401747} {\bibfield  {journal} {\bibinfo  {journal}
  {Angew. Chem. Int. Ed.}\ }\textbf {\bibinfo {volume} {63}},\ \bibinfo {pages}
  {e202401747} (\bibinfo {year} {2024})}\BibitemShut {NoStop}%
\bibitem [{\citenamefont {Zhang}\ \emph {et~al.}(2025)\citenamefont {Zhang},
  \citenamefont {Xie}, \citenamefont {Mei}, \citenamefont {Yu}, \citenamefont
  {Li}, \citenamefont {He}, \citenamefont {Fan}, \citenamefont {Zhang},
  \citenamefont {Ricciardulli}, \citenamefont {Samor{\`i}}, \citenamefont
  {Li},\ and\ \citenamefont {Yang}}]{Zhang2025}%
  \BibitemOpen
  \bibfield  {author} {\bibinfo {author} {\bibfnamefont {Y.}~\bibnamefont
  {Zhang}}, \bibinfo {author} {\bibfnamefont {Y.}~\bibnamefont {Xie}}, \bibinfo
  {author} {\bibfnamefont {H.}~\bibnamefont {Mei}}, \bibinfo {author}
  {\bibfnamefont {H.}~\bibnamefont {Yu}}, \bibinfo {author} {\bibfnamefont
  {M.}~\bibnamefont {Li}}, \bibinfo {author} {\bibfnamefont {Z.}~\bibnamefont
  {He}}, \bibinfo {author} {\bibfnamefont {W.}~\bibnamefont {Fan}}, \bibinfo
  {author} {\bibfnamefont {P.}~\bibnamefont {Zhang}}, \bibinfo {author}
  {\bibfnamefont {A.~G.}\ \bibnamefont {Ricciardulli}}, \bibinfo {author}
  {\bibfnamefont {P.}~\bibnamefont {Samor{\`i}}}, \bibinfo {author}
  {\bibfnamefont {M.}~\bibnamefont {Li}}, \ and\ \bibinfo {author}
  {\bibfnamefont {S.}~\bibnamefont {Yang}},\ }\href {\doibase
  10.1002/adma.202416741} {\bibfield  {journal} {\bibinfo  {journal} {Adv.
  Mater.}\ }\textbf {\bibinfo {volume} {37}},\ \bibinfo {pages} {2416741}
  (\bibinfo {year} {2025})}\BibitemShut {NoStop}%
\bibitem [{\citenamefont {Yu}\ \emph {et~al.}(2022)\citenamefont {Yu},
  \citenamefont {Xu}, \citenamefont {Peng}, \citenamefont {Qin},\ and\
  \citenamefont {Su}}]{Yu2022}%
  \BibitemOpen
  \bibfield  {author} {\bibinfo {author} {\bibfnamefont {L.}~\bibnamefont
  {Yu}}, \bibinfo {author} {\bibfnamefont {J.}~\bibnamefont {Xu}}, \bibinfo
  {author} {\bibfnamefont {B.}~\bibnamefont {Peng}}, \bibinfo {author}
  {\bibfnamefont {G.}~\bibnamefont {Qin}}, \ and\ \bibinfo {author}
  {\bibfnamefont {G.}~\bibnamefont {Su}},\ }\href {\doibase
  10.1021/acs.jpclett.2c02702} {\bibfield  {journal} {\bibinfo  {journal} {J.
  Phys. Chem. Lett.}\ }\textbf {\bibinfo {volume} {13}},\ \bibinfo {pages}
  {11622} (\bibinfo {year} {2022})}\BibitemShut {NoStop}%
\bibitem [{\citenamefont {Tromer}\ \emph {et~al.}(2022)\citenamefont {Tromer},
  \citenamefont {Ribeiro},\ and\ \citenamefont {Galv{\~a}o}}]{Tromer2022}%
  \BibitemOpen
  \bibfield  {author} {\bibinfo {author} {\bibfnamefont {R.~M.}\ \bibnamefont
  {Tromer}}, \bibinfo {author} {\bibfnamefont {L.~A.}\ \bibnamefont {Ribeiro}},
  \ and\ \bibinfo {author} {\bibfnamefont {D.~S.}\ \bibnamefont {Galv{\~a}o}},\
  }\href {\doibase 10.1016/j.cplett.2022.139925} {\bibfield  {journal}
  {\bibinfo  {journal} {Chemical Physics Letters}\ }\textbf {\bibinfo {volume}
  {804}},\ \bibinfo {pages} {139925} (\bibinfo {year} {2022})}\BibitemShut
  {NoStop}%
\bibitem [{\citenamefont {Ying}\ \emph {et~al.}(2023)\citenamefont {Ying},
  \citenamefont {Dong}, \citenamefont {Liang}, \citenamefont {Fan},
  \citenamefont {Zhong},\ and\ \citenamefont {Zhang}}]{Ying2023}%
  \BibitemOpen
  \bibfield  {author} {\bibinfo {author} {\bibfnamefont {P.}~\bibnamefont
  {Ying}}, \bibinfo {author} {\bibfnamefont {H.}~\bibnamefont {Dong}}, \bibinfo
  {author} {\bibfnamefont {T.}~\bibnamefont {Liang}}, \bibinfo {author}
  {\bibfnamefont {Z.}~\bibnamefont {Fan}}, \bibinfo {author} {\bibfnamefont
  {Z.}~\bibnamefont {Zhong}}, \ and\ \bibinfo {author} {\bibfnamefont
  {J.}~\bibnamefont {Zhang}},\ }\href {\doibase 10.1016/j.eml.2022.101929}
  {\bibfield  {journal} {\bibinfo  {journal} {Extreme Mechanics Letters}\
  }\textbf {\bibinfo {volume} {58}},\ \bibinfo {pages} {101929} (\bibinfo
  {year} {2023})}\BibitemShut {NoStop}%
\bibitem [{\citenamefont {Ribeiro}\ \emph {et~al.}(2022)\citenamefont
  {Ribeiro}, \citenamefont {Pereira}, \citenamefont {Giozza}, \citenamefont
  {Tromer},\ and\ \citenamefont {Galv{\~a}o}}]{Ribeiro2022}%
  \BibitemOpen
  \bibfield  {author} {\bibinfo {author} {\bibfnamefont {L.}~\bibnamefont
  {Ribeiro}}, \bibinfo {author} {\bibfnamefont {M.}~\bibnamefont {Pereira}},
  \bibinfo {author} {\bibfnamefont {W.}~\bibnamefont {Giozza}}, \bibinfo
  {author} {\bibfnamefont {R.}~\bibnamefont {Tromer}}, \ and\ \bibinfo {author}
  {\bibfnamefont {D.~S.}\ \bibnamefont {Galv{\~a}o}},\ }\href {\doibase
  10.1016/j.cplett.2022.140075} {\bibfield  {journal} {\bibinfo  {journal}
  {Chemical Physics Letters}\ }\textbf {\bibinfo {volume} {807}},\ \bibinfo
  {pages} {140075} (\bibinfo {year} {2022})}\BibitemShut {NoStop}%
\bibitem [{\citenamefont {Zhang}\ \emph {et~al.}(2012)\citenamefont {Zhang},
  \citenamefont {Legut}, \citenamefont {Lin}, \citenamefont {Zhao},
  \citenamefont {Mao},\ and\ \citenamefont {Veprek}}]{Zhang2012}%
  \BibitemOpen
  \bibfield  {author} {\bibinfo {author} {\bibfnamefont {R.~F.}\ \bibnamefont
  {Zhang}}, \bibinfo {author} {\bibfnamefont {D.}~\bibnamefont {Legut}},
  \bibinfo {author} {\bibfnamefont {Z.~J.}\ \bibnamefont {Lin}}, \bibinfo
  {author} {\bibfnamefont {Y.~S.}\ \bibnamefont {Zhao}}, \bibinfo {author}
  {\bibfnamefont {H.~K.}\ \bibnamefont {Mao}}, \ and\ \bibinfo {author}
  {\bibfnamefont {S.}~\bibnamefont {Veprek}},\ }\href {\doibase
  10.1103/PhysRevLett.108.255502} {\bibfield  {journal} {\bibinfo  {journal}
  {Phys. Rev. Lett.}\ }\textbf {\bibinfo {volume} {108}},\ \bibinfo {pages}
  {255502} (\bibinfo {year} {2012})}\BibitemShut {NoStop}%
\bibitem [{\citenamefont {Peng}(2023)}]{Peng2023}%
  \BibitemOpen
  \bibfield  {author} {\bibinfo {author} {\bibfnamefont {B.}~\bibnamefont
  {Peng}},\ }\href {\doibase 10.1021/acs.nanolett.2c04497} {\bibfield
  {journal} {\bibinfo  {journal} {Nano Lett.}\ }\textbf {\bibinfo {volume}
  {23}},\ \bibinfo {pages} {652} (\bibinfo {year} {2023})}\BibitemShut
  {NoStop}%
\bibitem [{\citenamefont {Malyi}\ \emph {et~al.}(2019)\citenamefont {Malyi},
  \citenamefont {Sopiha},\ and\ \citenamefont {Persson}}]{Malyi2019}%
  \BibitemOpen
  \bibfield  {author} {\bibinfo {author} {\bibfnamefont {O.~I.}\ \bibnamefont
  {Malyi}}, \bibinfo {author} {\bibfnamefont {K.~V.}\ \bibnamefont {Sopiha}}, \
  and\ \bibinfo {author} {\bibfnamefont {C.}~\bibnamefont {Persson}},\ }\href
  {\doibase 10.1021/acsami.9b01261} {\bibfield  {journal} {\bibinfo  {journal}
  {ACS Appl. Mater. Interfaces}\ }\textbf {\bibinfo {volume} {11}},\ \bibinfo
  {pages} {24876} (\bibinfo {year} {2019})}\BibitemShut {NoStop}%
\bibitem [{\citenamefont {Luo}\ \emph {et~al.}(2022)\citenamefont {Luo},
  \citenamefont {Yin},\ and\ \citenamefont {Dronskowski}}]{Luo2022}%
  \BibitemOpen
  \bibfield  {author} {\bibinfo {author} {\bibfnamefont {D.}~\bibnamefont
  {Luo}}, \bibinfo {author} {\bibfnamefont {K.}~\bibnamefont {Yin}}, \ and\
  \bibinfo {author} {\bibfnamefont {R.}~\bibnamefont {Dronskowski}},\ }\href
  {\doibase 10.1021/jacs.2c00592} {\bibfield  {journal} {\bibinfo  {journal}
  {J. Am. Chem. Soc.}\ }\textbf {\bibinfo {volume} {144}},\ \bibinfo {pages}
  {5155} (\bibinfo {year} {2022})}\BibitemShut {NoStop}%
\bibitem [{\citenamefont {Pallikara}\ \emph {et~al.}(2022)\citenamefont
  {Pallikara}, \citenamefont {Kayastha}, \citenamefont {Skelton},\ and\
  \citenamefont {Whalley}}]{Pallikara2022}%
  \BibitemOpen
  \bibfield  {author} {\bibinfo {author} {\bibfnamefont {I.}~\bibnamefont
  {Pallikara}}, \bibinfo {author} {\bibfnamefont {P.}~\bibnamefont {Kayastha}},
  \bibinfo {author} {\bibfnamefont {J.~M.}\ \bibnamefont {Skelton}}, \ and\
  \bibinfo {author} {\bibfnamefont {L.~D.}\ \bibnamefont {Whalley}},\ }\href
  {\doibase 10.1088/2516-1075/ac78b3} {\bibfield  {journal} {\bibinfo
  {journal} {Electron. Struct.}\ }\textbf {\bibinfo {volume} {4}},\ \bibinfo
  {pages} {033002} (\bibinfo {year} {2022})}\BibitemShut {NoStop}%
\bibitem [{\citenamefont {Baroni}\ \emph {et~al.}(2001)\citenamefont {Baroni},
  \citenamefont {de~Gironcoli}, \citenamefont {Dal~Corso},\ and\ \citenamefont
  {Giannozzi}}]{DFPT}%
  \BibitemOpen
  \bibfield  {author} {\bibinfo {author} {\bibfnamefont {S.}~\bibnamefont
  {Baroni}}, \bibinfo {author} {\bibfnamefont {S.}~\bibnamefont
  {de~Gironcoli}}, \bibinfo {author} {\bibfnamefont {A.}~\bibnamefont
  {Dal~Corso}}, \ and\ \bibinfo {author} {\bibfnamefont {P.}~\bibnamefont
  {Giannozzi}},\ }\href {\doibase 10.1103/RevModPhys.73.515} {\bibfield
  {journal} {\bibinfo  {journal} {Rev. Mod. Phys.}\ }\textbf {\bibinfo {volume}
  {73}},\ \bibinfo {pages} {515} (\bibinfo {year} {2001})}\BibitemShut
  {NoStop}%
\bibitem [{\citenamefont {Gonze}(1995{\natexlab{a}})}]{Gonze1995}%
  \BibitemOpen
  \bibfield  {author} {\bibinfo {author} {\bibfnamefont {X.}~\bibnamefont
  {Gonze}},\ }\href {\doibase 10.1103/PhysRevA.52.1086} {\bibfield  {journal}
  {\bibinfo  {journal} {Phys. Rev. A}\ }\textbf {\bibinfo {volume} {52}},\
  \bibinfo {pages} {1086} (\bibinfo {year} {1995}{\natexlab{a}})}\BibitemShut
  {NoStop}%
\bibitem [{\citenamefont {Gonze}(1995{\natexlab{b}})}]{Gonze1995a}%
  \BibitemOpen
  \bibfield  {author} {\bibinfo {author} {\bibfnamefont {X.}~\bibnamefont
  {Gonze}},\ }\href {\doibase 10.1103/PhysRevA.52.1096} {\bibfield  {journal}
  {\bibinfo  {journal} {Phys. Rev. A}\ }\textbf {\bibinfo {volume} {52}},\
  \bibinfo {pages} {1096} (\bibinfo {year} {1995}{\natexlab{b}})}\BibitemShut
  {NoStop}%
\bibitem [{\citenamefont {Dove}(1993)}]{Dove1993}%
  \BibitemOpen
  \bibfield  {author} {\bibinfo {author} {\bibfnamefont {M.~T.}\ \bibnamefont
  {Dove}},\ }\href@noop {} {\emph {\bibinfo {title} {Introduction to Lattice
  Dynamics}}}\ (\bibinfo  {publisher} {Cambridge University Press},\ \bibinfo
  {year} {1993})\BibitemShut {NoStop}%
\bibitem [{\citenamefont {Togo}\ \emph {et~al.}(2008)\citenamefont {Togo},
  \citenamefont {Oba},\ and\ \citenamefont {Tanaka}}]{Togo2008}%
  \BibitemOpen
  \bibfield  {author} {\bibinfo {author} {\bibfnamefont {A.}~\bibnamefont
  {Togo}}, \bibinfo {author} {\bibfnamefont {F.}~\bibnamefont {Oba}}, \ and\
  \bibinfo {author} {\bibfnamefont {I.}~\bibnamefont {Tanaka}},\ }\href
  {\doibase 10.1103/PhysRevB.78.134106} {\bibfield  {journal} {\bibinfo
  {journal} {Phys. Rev. B}\ }\textbf {\bibinfo {volume} {78}},\ \bibinfo
  {pages} {134106} (\bibinfo {year} {2008})}\BibitemShut {NoStop}%
\bibitem [{\citenamefont {Togo}\ and\ \citenamefont {Tanaka}(2015)}]{Togo2015}%
  \BibitemOpen
  \bibfield  {author} {\bibinfo {author} {\bibfnamefont {A.}~\bibnamefont
  {Togo}}\ and\ \bibinfo {author} {\bibfnamefont {I.}~\bibnamefont {Tanaka}},\
  }\href {\doibase http://dx.doi.org/10.1016/j.scriptamat.2015.07.021}
  {\bibfield  {journal} {\bibinfo  {journal} {Scripta Materialia}\ }\textbf
  {\bibinfo {volume} {108}},\ \bibinfo {pages} {1} (\bibinfo {year}
  {2015})}\BibitemShut {NoStop}%
\bibitem [{\citenamefont {Huang}\ \emph {et~al.}(2015)\citenamefont {Huang},
  \citenamefont {Gong},\ and\ \citenamefont {Zeng}}]{Huang2015a}%
  \BibitemOpen
  \bibfield  {author} {\bibinfo {author} {\bibfnamefont {L.-F.}\ \bibnamefont
  {Huang}}, \bibinfo {author} {\bibfnamefont {P.-L.}\ \bibnamefont {Gong}}, \
  and\ \bibinfo {author} {\bibfnamefont {Z.}~\bibnamefont {Zeng}},\ }\href
  {\doibase 10.1103/PhysRevB.91.205433} {\bibfield  {journal} {\bibinfo
  {journal} {Phys. Rev. B}\ }\textbf {\bibinfo {volume} {91}},\ \bibinfo
  {pages} {205433} (\bibinfo {year} {2015})}\BibitemShut {NoStop}%
\bibitem [{\citenamefont {Huang}\ \emph
  {et~al.}(2016{\natexlab{a}})\citenamefont {Huang}, \citenamefont {Lu},
  \citenamefont {Tennessen},\ and\ \citenamefont {M.Rondinelli}}]{Huang2016}%
  \BibitemOpen
  \bibfield  {author} {\bibinfo {author} {\bibfnamefont {L.-F.}\ \bibnamefont
  {Huang}}, \bibinfo {author} {\bibfnamefont {X.-Z.}\ \bibnamefont {Lu}},
  \bibinfo {author} {\bibfnamefont {E.}~\bibnamefont {Tennessen}}, \ and\
  \bibinfo {author} {\bibfnamefont {J.}~\bibnamefont {M.Rondinelli}},\ }\href
  {\doibase 10.1016/j.commatsci.2016.04.012} {\bibfield  {journal} {\bibinfo
  {journal} {Computational Materials Science}\ }\textbf {\bibinfo {volume}
  {120}},\ \bibinfo {pages} {84} (\bibinfo {year}
  {2016}{\natexlab{a}})}\BibitemShut {NoStop}%
\bibitem [{\citenamefont {Peng}\ \emph
  {et~al.}(2019{\natexlab{b}})\citenamefont {Peng}, \citenamefont
  {Bravi\ifmmode~\acute{c}\else \'{c}\fi{}}, \citenamefont
  {MacManus-Driscoll},\ and\ \citenamefont {Monserrat}}]{Peng2019}%
  \BibitemOpen
  \bibfield  {author} {\bibinfo {author} {\bibfnamefont {B.}~\bibnamefont
  {Peng}}, \bibinfo {author} {\bibfnamefont {I.}~\bibnamefont
  {Bravi\ifmmode~\acute{c}\else \'{c}\fi{}}}, \bibinfo {author} {\bibfnamefont
  {J.~L.}\ \bibnamefont {MacManus-Driscoll}}, \ and\ \bibinfo {author}
  {\bibfnamefont {B.}~\bibnamefont {Monserrat}},\ }\href {\doibase
  10.1103/PhysRevB.100.161101} {\bibfield  {journal} {\bibinfo  {journal}
  {Phys. Rev. B}\ }\textbf {\bibinfo {volume} {100}},\ \bibinfo {pages}
  {161101} (\bibinfo {year} {2019}{\natexlab{b}})}\BibitemShut {NoStop}%
\bibitem [{\citenamefont {Shaikh}\ and\ \citenamefont
  {Peng}(2025)}]{Shaikh2025}%
  \BibitemOpen
  \bibfield  {author} {\bibinfo {author} {\bibfnamefont {A.}~\bibnamefont
  {Shaikh}}\ and\ \bibinfo {author} {\bibfnamefont {B.}~\bibnamefont {Peng}},\
  }\href@noop {} {\bibfield  {journal} {\bibinfo  {journal} {arXiv:}\ ,\
  \bibinfo {pages} {2504.02037}} (\bibinfo {year} {2025})}\BibitemShut
  {NoStop}%
\bibitem [{\citenamefont {Peng}\ \emph
  {et~al.}(2018{\natexlab{d}})\citenamefont {Peng}, \citenamefont {Xu},
  \citenamefont {Zhang}, \citenamefont {Ning}, \citenamefont {Shao},
  \citenamefont {Ni}, \citenamefont {Li}, \citenamefont {Zhu}, \citenamefont
  {Zhu},\ and\ \citenamefont {Soukoulis}}]{Peng2018}%
  \BibitemOpen
  \bibfield  {author} {\bibinfo {author} {\bibfnamefont {B.}~\bibnamefont
  {Peng}}, \bibinfo {author} {\bibfnamefont {K.}~\bibnamefont {Xu}}, \bibinfo
  {author} {\bibfnamefont {H.}~\bibnamefont {Zhang}}, \bibinfo {author}
  {\bibfnamefont {Z.}~\bibnamefont {Ning}}, \bibinfo {author} {\bibfnamefont
  {H.}~\bibnamefont {Shao}}, \bibinfo {author} {\bibfnamefont {G.}~\bibnamefont
  {Ni}}, \bibinfo {author} {\bibfnamefont {J.}~\bibnamefont {Li}}, \bibinfo
  {author} {\bibfnamefont {Y.}~\bibnamefont {Zhu}}, \bibinfo {author}
  {\bibfnamefont {H.}~\bibnamefont {Zhu}}, \ and\ \bibinfo {author}
  {\bibfnamefont {C.~M.}\ \bibnamefont {Soukoulis}},\ }\href {\doibase
  10.1002/adts.201700005} {\bibfield  {journal} {\bibinfo  {journal} {Adv.
  Theory Simul.}\ }\textbf {\bibinfo {volume} {1}},\ \bibinfo {pages} {1700005}
  (\bibinfo {year} {2018}{\natexlab{d}})}\BibitemShut {NoStop}%
\bibitem [{\citenamefont {Lange}\ \emph {et~al.}(2022)\citenamefont {Lange},
  \citenamefont {Bouhon}, \citenamefont {Monserrat},\ and\ \citenamefont
  {Slager}}]{Lange2022}%
  \BibitemOpen
  \bibfield  {author} {\bibinfo {author} {\bibfnamefont {G.~F.}\ \bibnamefont
  {Lange}}, \bibinfo {author} {\bibfnamefont {A.}~\bibnamefont {Bouhon}},
  \bibinfo {author} {\bibfnamefont {B.}~\bibnamefont {Monserrat}}, \ and\
  \bibinfo {author} {\bibfnamefont {R.-J.}\ \bibnamefont {Slager}},\ }\href
  {\doibase 10.1103/PhysRevB.105.064301} {\bibfield  {journal} {\bibinfo
  {journal} {Phys. Rev. B}\ }\textbf {\bibinfo {volume} {105}},\ \bibinfo
  {pages} {064301} (\bibinfo {year} {2022})}\BibitemShut {NoStop}%
\bibitem [{\citenamefont {Pavone}\ \emph {et~al.}(1998)\citenamefont {Pavone},
  \citenamefont {Baroni},\ and\ \citenamefont {de~Gironcoli}}]{Pavone1998}%
  \BibitemOpen
  \bibfield  {author} {\bibinfo {author} {\bibfnamefont {P.}~\bibnamefont
  {Pavone}}, \bibinfo {author} {\bibfnamefont {S.}~\bibnamefont {Baroni}}, \
  and\ \bibinfo {author} {\bibfnamefont {S.}~\bibnamefont {de~Gironcoli}},\
  }\href {\doibase 10.1103/PhysRevB.57.10421} {\bibfield  {journal} {\bibinfo
  {journal} {Phys. Rev. B}\ }\textbf {\bibinfo {volume} {57}},\ \bibinfo
  {pages} {10421} (\bibinfo {year} {1998})}\BibitemShut {NoStop}%
\bibitem [{\citenamefont {Pavone}(2001)}]{Pavone2001}%
  \BibitemOpen
  \bibfield  {author} {\bibinfo {author} {\bibfnamefont {P.}~\bibnamefont
  {Pavone}},\ }\href {\doibase 10.1088/0953-8984/13/34/308} {\bibfield
  {journal} {\bibinfo  {journal} {Journal of Physics: Condensed Matter}\
  }\textbf {\bibinfo {volume} {13}},\ \bibinfo {pages} {7593} (\bibinfo {year}
  {2001})}\BibitemShut {NoStop}%
\bibitem [{\citenamefont {Masago}\ \emph {et~al.}(2006)\citenamefont {Masago},
  \citenamefont {Shirai},\ and\ \citenamefont {Katayama-Yoshida}}]{Masago2006}%
  \BibitemOpen
  \bibfield  {author} {\bibinfo {author} {\bibfnamefont {A.}~\bibnamefont
  {Masago}}, \bibinfo {author} {\bibfnamefont {K.}~\bibnamefont {Shirai}}, \
  and\ \bibinfo {author} {\bibfnamefont {H.}~\bibnamefont {Katayama-Yoshida}},\
  }\href {\doibase 10.1103/PhysRevB.73.104102} {\bibfield  {journal} {\bibinfo
  {journal} {Phys. Rev. B}\ }\textbf {\bibinfo {volume} {73}},\ \bibinfo
  {pages} {104102} (\bibinfo {year} {2006})}\BibitemShut {NoStop}%
\bibitem [{\citenamefont {van Setten}\ \emph {et~al.}(2007)\citenamefont {van
  Setten}, \citenamefont {Uijttewaal}, \citenamefont {de~Wijs},\ and\
  \citenamefont {de~Groot}}]{Setten2007}%
  \BibitemOpen
  \bibfield  {author} {\bibinfo {author} {\bibfnamefont {M.~J.}\ \bibnamefont
  {van Setten}}, \bibinfo {author} {\bibfnamefont {M.~A.}\ \bibnamefont
  {Uijttewaal}}, \bibinfo {author} {\bibfnamefont {G.~A.}\ \bibnamefont
  {de~Wijs}}, \ and\ \bibinfo {author} {\bibfnamefont {R.~A.}\ \bibnamefont
  {de~Groot}},\ }\href {\doibase 10.1021/ja0631246} {\bibfield  {journal}
  {\bibinfo  {journal} {J. Am. Chem. Soc.}\ }\textbf {\bibinfo {volume}
  {129}},\ \bibinfo {pages} {2458} (\bibinfo {year} {2007})}\BibitemShut
  {NoStop}%
\bibitem [{\citenamefont {Stoffel}\ \emph {et~al.}(2010)\citenamefont
  {Stoffel}, \citenamefont {Wessel}, \citenamefont {Lumey},\ and\ \citenamefont
  {Dronskowski}}]{Stoffel2010}%
  \BibitemOpen
  \bibfield  {author} {\bibinfo {author} {\bibfnamefont {R.~P.}\ \bibnamefont
  {Stoffel}}, \bibinfo {author} {\bibfnamefont {C.}~\bibnamefont {Wessel}},
  \bibinfo {author} {\bibfnamefont {M.-W.}\ \bibnamefont {Lumey}}, \ and\
  \bibinfo {author} {\bibfnamefont {R.}~\bibnamefont {Dronskowski}},\ }\href
  {\doibase 10.1002/anie.200906780} {\bibfield  {journal} {\bibinfo  {journal}
  {Angew. Chem. Int. Ed.}\ }\textbf {\bibinfo {volume} {49}},\ \bibinfo {pages}
  {5242} (\bibinfo {year} {2010})}\BibitemShut {NoStop}%
\bibitem [{\citenamefont {Zhang}\ \emph
  {et~al.}(2011{\natexlab{a}})\citenamefont {Zhang}, \citenamefont {Lin},
  \citenamefont {Mao},\ and\ \citenamefont {Zhao}}]{Zhang2011a}%
  \BibitemOpen
  \bibfield  {author} {\bibinfo {author} {\bibfnamefont {R.~F.}\ \bibnamefont
  {Zhang}}, \bibinfo {author} {\bibfnamefont {Z.~J.}\ \bibnamefont {Lin}},
  \bibinfo {author} {\bibfnamefont {H.-K.}\ \bibnamefont {Mao}}, \ and\
  \bibinfo {author} {\bibfnamefont {Y.}~\bibnamefont {Zhao}},\ }\href {\doibase
  10.1103/PhysRevB.83.060101} {\bibfield  {journal} {\bibinfo  {journal} {Phys.
  Rev. B}\ }\textbf {\bibinfo {volume} {83}},\ \bibinfo {pages} {060101}
  (\bibinfo {year} {2011}{\natexlab{a}})}\BibitemShut {NoStop}%
\bibitem [{\citenamefont {Deringer}\ \emph
  {et~al.}(2014{\natexlab{a}})\citenamefont {Deringer}, \citenamefont
  {Stoffel},\ and\ \citenamefont {Dronskowski}}]{Deringer2014}%
  \BibitemOpen
  \bibfield  {author} {\bibinfo {author} {\bibfnamefont {V.~L.}\ \bibnamefont
  {Deringer}}, \bibinfo {author} {\bibfnamefont {R.~P.}\ \bibnamefont
  {Stoffel}}, \ and\ \bibinfo {author} {\bibfnamefont {R.}~\bibnamefont
  {Dronskowski}},\ }\href {\doibase 10.1021/cg401822g} {\bibfield  {journal}
  {\bibinfo  {journal} {Crystal Growth \& Design}\ }\textbf {\bibinfo {volume}
  {14}},\ \bibinfo {pages} {871} (\bibinfo {year}
  {2014}{\natexlab{a}})}\BibitemShut {NoStop}%
\bibitem [{\citenamefont {Deringer}\ \emph
  {et~al.}(2014{\natexlab{b}})\citenamefont {Deringer}, \citenamefont
  {Stoffel},\ and\ \citenamefont {Dronskowski}}]{Deringer2014a}%
  \BibitemOpen
  \bibfield  {author} {\bibinfo {author} {\bibfnamefont {V.~L.}\ \bibnamefont
  {Deringer}}, \bibinfo {author} {\bibfnamefont {R.~P.}\ \bibnamefont
  {Stoffel}}, \ and\ \bibinfo {author} {\bibfnamefont {R.}~\bibnamefont
  {Dronskowski}},\ }\href {\doibase 10.1103/PhysRevB.89.094303} {\bibfield
  {journal} {\bibinfo  {journal} {Phys. Rev. B}\ }\textbf {\bibinfo {volume}
  {89}},\ \bibinfo {pages} {094303} (\bibinfo {year}
  {2014}{\natexlab{b}})}\BibitemShut {NoStop}%
\bibitem [{\citenamefont {White}\ \emph {et~al.}(2015)\citenamefont {White},
  \citenamefont {Cerqueira}, \citenamefont {Whitman}, \citenamefont {Johnson},\
  and\ \citenamefont {Ogitsu}}]{White2015}%
  \BibitemOpen
  \bibfield  {author} {\bibinfo {author} {\bibfnamefont {M.~A.}\ \bibnamefont
  {White}}, \bibinfo {author} {\bibfnamefont {A.~B.}\ \bibnamefont
  {Cerqueira}}, \bibinfo {author} {\bibfnamefont {C.~A.}\ \bibnamefont
  {Whitman}}, \bibinfo {author} {\bibfnamefont {M.~B.}\ \bibnamefont
  {Johnson}}, \ and\ \bibinfo {author} {\bibfnamefont {T.}~\bibnamefont
  {Ogitsu}},\ }\href {\doibase 10.1002/anie.201409169} {\bibfield  {journal}
  {\bibinfo  {journal} {Angew. Chem. Int. Ed.}\ }\textbf {\bibinfo {volume}
  {54}},\ \bibinfo {pages} {3626} (\bibinfo {year} {2015})}\BibitemShut
  {NoStop}%
\bibitem [{\citenamefont {Nyman}\ \emph {et~al.}(2016)\citenamefont {Nyman},
  \citenamefont {Pundyke},\ and\ \citenamefont {Day}}]{Nyman2016}%
  \BibitemOpen
  \bibfield  {author} {\bibinfo {author} {\bibfnamefont {J.}~\bibnamefont
  {Nyman}}, \bibinfo {author} {\bibfnamefont {O.~S.}\ \bibnamefont {Pundyke}},
  \ and\ \bibinfo {author} {\bibfnamefont {G.~M.}\ \bibnamefont {Day}},\ }\href
  {\doibase 10.1039/C6CP02261H} {\bibfield  {journal} {\bibinfo  {journal}
  {Phys. Chem. Chem. Phys.}\ }\textbf {\bibinfo {volume} {18}},\ \bibinfo
  {pages} {15828} (\bibinfo {year} {2016})}\BibitemShut {NoStop}%
\bibitem [{\citenamefont {Skelton}\ \emph {et~al.}(2017)\citenamefont
  {Skelton}, \citenamefont {Burton}, \citenamefont {Oba},\ and\ \citenamefont
  {Walsh}}]{Skelton2017}%
  \BibitemOpen
  \bibfield  {author} {\bibinfo {author} {\bibfnamefont {J.~M.}\ \bibnamefont
  {Skelton}}, \bibinfo {author} {\bibfnamefont {L.~A.}\ \bibnamefont {Burton}},
  \bibinfo {author} {\bibfnamefont {F.}~\bibnamefont {Oba}}, \ and\ \bibinfo
  {author} {\bibfnamefont {A.}~\bibnamefont {Walsh}},\ }\href {\doibase
  10.1021/acs.jpcc.6b12581} {\bibfield  {journal} {\bibinfo  {journal} {J.
  Phys. Chem. C}\ }\textbf {\bibinfo {volume} {121}},\ \bibinfo {pages} {6446}
  (\bibinfo {year} {2017})}\BibitemShut {NoStop}%
\bibitem [{\citenamefont {Pallikara}\ and\ \citenamefont
  {Skelton}(2021)}]{Pallikara2021}%
  \BibitemOpen
  \bibfield  {author} {\bibinfo {author} {\bibfnamefont {I.}~\bibnamefont
  {Pallikara}}\ and\ \bibinfo {author} {\bibfnamefont {J.~M.}\ \bibnamefont
  {Skelton}},\ }\href {\doibase 10.1039/D1CP02597J} {\bibfield  {journal}
  {\bibinfo  {journal} {Phys. Chem. Chem. Phys.}\ }\textbf {\bibinfo {volume}
  {23}},\ \bibinfo {pages} {19219} (\bibinfo {year} {2021})}\BibitemShut
  {NoStop}%
\bibitem [{\citenamefont {Bartel}(2022)}]{Bartel2022}%
  \BibitemOpen
  \bibfield  {author} {\bibinfo {author} {\bibfnamefont {C.~J.}\ \bibnamefont
  {Bartel}},\ }\href {\doibase 10.1007/s10853-022-06915-4} {\bibfield
  {journal} {\bibinfo  {journal} {Journal of Materials Science}\ }\textbf
  {\bibinfo {volume} {57}},\ \bibinfo {pages} {10475} (\bibinfo {year}
  {2022})}\BibitemShut {NoStop}%
\bibitem [{\citenamefont {Le~Page}\ and\ \citenamefont
  {Saxe}(2002)}]{LePage2002}%
  \BibitemOpen
  \bibfield  {author} {\bibinfo {author} {\bibfnamefont {Y.}~\bibnamefont
  {Le~Page}}\ and\ \bibinfo {author} {\bibfnamefont {P.}~\bibnamefont {Saxe}},\
  }\href {\doibase 10.1103/PhysRevB.65.104104} {\bibfield  {journal} {\bibinfo
  {journal} {Phys. Rev. B}\ }\textbf {\bibinfo {volume} {65}},\ \bibinfo
  {pages} {104104} (\bibinfo {year} {2002})}\BibitemShut {NoStop}%
\bibitem [{\citenamefont {Wu}\ \emph {et~al.}(2005)\citenamefont {Wu},
  \citenamefont {Vanderbilt},\ and\ \citenamefont {Hamann}}]{Wu2005}%
  \BibitemOpen
  \bibfield  {author} {\bibinfo {author} {\bibfnamefont {X.}~\bibnamefont
  {Wu}}, \bibinfo {author} {\bibfnamefont {D.}~\bibnamefont {Vanderbilt}}, \
  and\ \bibinfo {author} {\bibfnamefont {D.~R.}\ \bibnamefont {Hamann}},\
  }\href {\doibase 10.1103/PhysRevB.72.035105} {\bibfield  {journal} {\bibinfo
  {journal} {Phys. Rev. B}\ }\textbf {\bibinfo {volume} {72}},\ \bibinfo
  {pages} {035105} (\bibinfo {year} {2005})}\BibitemShut {NoStop}%
\bibitem [{\citenamefont {Born}\ and\ \citenamefont {Huang}(1954)}]{Born1954}%
  \BibitemOpen
  \bibfield  {author} {\bibinfo {author} {\bibfnamefont {M.}~\bibnamefont
  {Born}}\ and\ \bibinfo {author} {\bibfnamefont {K.}~\bibnamefont {Huang}},\
  }\href@noop {} {\emph {\bibinfo {title} {Dynamical theory of crystal
  lattices}}}\ (\bibinfo  {publisher} {Clarendon Press},\ \bibinfo {address}
  {Oxford},\ \bibinfo {year} {1954})\BibitemShut {NoStop}%
\bibitem [{\citenamefont {Wu}\ \emph {et~al.}(2007)\citenamefont {Wu},
  \citenamefont {Zhao}, \citenamefont {Xiang}, \citenamefont {Hao},
  \citenamefont {Liu},\ and\ \citenamefont {Meng}}]{Wu2007}%
  \BibitemOpen
  \bibfield  {author} {\bibinfo {author} {\bibfnamefont {Z.-j.}\ \bibnamefont
  {Wu}}, \bibinfo {author} {\bibfnamefont {E.-j.}\ \bibnamefont {Zhao}},
  \bibinfo {author} {\bibfnamefont {H.-p.}\ \bibnamefont {Xiang}}, \bibinfo
  {author} {\bibfnamefont {X.-f.}\ \bibnamefont {Hao}}, \bibinfo {author}
  {\bibfnamefont {X.-j.}\ \bibnamefont {Liu}}, \ and\ \bibinfo {author}
  {\bibfnamefont {J.}~\bibnamefont {Meng}},\ }\href {\doibase
  10.1103/PhysRevB.76.054115} {\bibfield  {journal} {\bibinfo  {journal} {Phys.
  Rev. B}\ }\textbf {\bibinfo {volume} {76}},\ \bibinfo {pages} {054115}
  (\bibinfo {year} {2007})}\BibitemShut {NoStop}%
\bibitem [{\citenamefont {Goodwin}\ \emph {et~al.}(2008)\citenamefont
  {Goodwin}, \citenamefont {Calleja}, \citenamefont {Conterio}, \citenamefont
  {Dove}, \citenamefont {Evans}, \citenamefont {Keen}, \citenamefont {Peters},\
  and\ \citenamefont {Tucker}}]{Goodwin2008}%
  \BibitemOpen
  \bibfield  {author} {\bibinfo {author} {\bibfnamefont {A.~L.}\ \bibnamefont
  {Goodwin}}, \bibinfo {author} {\bibfnamefont {M.}~\bibnamefont {Calleja}},
  \bibinfo {author} {\bibfnamefont {M.~J.}\ \bibnamefont {Conterio}}, \bibinfo
  {author} {\bibfnamefont {M.~T.}\ \bibnamefont {Dove}}, \bibinfo {author}
  {\bibfnamefont {J.~S.~O.}\ \bibnamefont {Evans}}, \bibinfo {author}
  {\bibfnamefont {D.~A.}\ \bibnamefont {Keen}}, \bibinfo {author}
  {\bibfnamefont {L.}~\bibnamefont {Peters}}, \ and\ \bibinfo {author}
  {\bibfnamefont {M.~G.}\ \bibnamefont {Tucker}},\ }\href {\doibase
  10.1126/science.1151442} {\bibfield  {journal} {\bibinfo  {journal}
  {Science}\ }\textbf {\bibinfo {volume} {319}},\ \bibinfo {pages} {794}
  (\bibinfo {year} {2008})}\BibitemShut {NoStop}%
\bibitem [{\citenamefont {Dove}\ and\ \citenamefont {Fang}(2016)}]{Dove2016}%
  \BibitemOpen
  \bibfield  {author} {\bibinfo {author} {\bibfnamefont {M.~T.}\ \bibnamefont
  {Dove}}\ and\ \bibinfo {author} {\bibfnamefont {H.}~\bibnamefont {Fang}},\
  }\href {\doibase 10.1088/0034-4885/79/6/066503} {\bibfield  {journal}
  {\bibinfo  {journal} {Reports on Progress in Physics}\ }\textbf {\bibinfo
  {volume} {79}},\ \bibinfo {pages} {066503} (\bibinfo {year}
  {2016})}\BibitemShut {NoStop}%
\bibitem [{\citenamefont {Pryde}\ \emph {et~al.}(1996)\citenamefont {Pryde},
  \citenamefont {Hammonds}, \citenamefont {Dove}, \citenamefont {Heine},
  \citenamefont {Gale},\ and\ \citenamefont {Warren}}]{Pryde1996}%
  \BibitemOpen
  \bibfield  {author} {\bibinfo {author} {\bibfnamefont {A.~K.~A.}\
  \bibnamefont {Pryde}}, \bibinfo {author} {\bibfnamefont {K.~D.}\ \bibnamefont
  {Hammonds}}, \bibinfo {author} {\bibfnamefont {M.~T.}\ \bibnamefont {Dove}},
  \bibinfo {author} {\bibfnamefont {V.}~\bibnamefont {Heine}}, \bibinfo
  {author} {\bibfnamefont {J.~D.}\ \bibnamefont {Gale}}, \ and\ \bibinfo
  {author} {\bibfnamefont {M.~C.}\ \bibnamefont {Warren}},\ }\href {\doibase
  10.1088/0953-8984/8/50/023} {\bibfield  {journal} {\bibinfo  {journal}
  {Journal of Physics: Condensed Matter}\ }\textbf {\bibinfo {volume} {8}},\
  \bibinfo {pages} {10973} (\bibinfo {year} {1996})}\BibitemShut {NoStop}%
\bibitem [{\citenamefont {Sleight}(1998)}]{Sleight1998}%
  \BibitemOpen
  \bibfield  {author} {\bibinfo {author} {\bibfnamefont {A.~W.}\ \bibnamefont
  {Sleight}},\ }\href {\doibase 10.1021/ic980253h} {\bibfield  {journal}
  {\bibinfo  {journal} {Inorg. Chem.}\ }\textbf {\bibinfo {volume} {37}},\
  \bibinfo {pages} {2854} (\bibinfo {year} {1998})}\BibitemShut {NoStop}%
\bibitem [{\citenamefont {Tucker}\ \emph {et~al.}(2005)\citenamefont {Tucker},
  \citenamefont {Goodwin}, \citenamefont {Dove}, \citenamefont {Keen},
  \citenamefont {Wells},\ and\ \citenamefont {Evans}}]{Tucker2005}%
  \BibitemOpen
  \bibfield  {author} {\bibinfo {author} {\bibfnamefont {M.~G.}\ \bibnamefont
  {Tucker}}, \bibinfo {author} {\bibfnamefont {A.~L.}\ \bibnamefont {Goodwin}},
  \bibinfo {author} {\bibfnamefont {M.~T.}\ \bibnamefont {Dove}}, \bibinfo
  {author} {\bibfnamefont {D.~A.}\ \bibnamefont {Keen}}, \bibinfo {author}
  {\bibfnamefont {S.~A.}\ \bibnamefont {Wells}}, \ and\ \bibinfo {author}
  {\bibfnamefont {J.~S.~O.}\ \bibnamefont {Evans}},\ }\href {\doibase
  10.1103/PhysRevLett.95.255501} {\bibfield  {journal} {\bibinfo  {journal}
  {Phys. Rev. Lett.}\ }\textbf {\bibinfo {volume} {95}},\ \bibinfo {pages}
  {255501} (\bibinfo {year} {2005})}\BibitemShut {NoStop}%
\bibitem [{\citenamefont {Tan}\ \emph {et~al.}(2024)\citenamefont {Tan},
  \citenamefont {Heine}, \citenamefont {Li},\ and\ \citenamefont
  {Dove}}]{Tan2024}%
  \BibitemOpen
  \bibfield  {author} {\bibinfo {author} {\bibfnamefont {L.}~\bibnamefont
  {Tan}}, \bibinfo {author} {\bibfnamefont {V.}~\bibnamefont {Heine}}, \bibinfo
  {author} {\bibfnamefont {G.}~\bibnamefont {Li}}, \ and\ \bibinfo {author}
  {\bibfnamefont {M.~T.}\ \bibnamefont {Dove}},\ }\href {\doibase
  10.1088/1361-6633/acc7b7} {\bibfield  {journal} {\bibinfo  {journal} {Reports
  on Progress in Physics}\ }\textbf {\bibinfo {volume} {87}},\ \bibinfo {pages}
  {126501} (\bibinfo {year} {2024})}\BibitemShut {NoStop}%
\bibitem [{\citenamefont {Hancock}\ \emph {et~al.}(2004)\citenamefont
  {Hancock}, \citenamefont {Turpen}, \citenamefont {Schlesinger}, \citenamefont
  {Kowach},\ and\ \citenamefont {Ramirez}}]{Hancock2004}%
  \BibitemOpen
  \bibfield  {author} {\bibinfo {author} {\bibfnamefont {J.~N.}\ \bibnamefont
  {Hancock}}, \bibinfo {author} {\bibfnamefont {C.}~\bibnamefont {Turpen}},
  \bibinfo {author} {\bibfnamefont {Z.}~\bibnamefont {Schlesinger}}, \bibinfo
  {author} {\bibfnamefont {G.~R.}\ \bibnamefont {Kowach}}, \ and\ \bibinfo
  {author} {\bibfnamefont {A.~P.}\ \bibnamefont {Ramirez}},\ }\href {\doibase
  10.1103/PhysRevLett.93.225501} {\bibfield  {journal} {\bibinfo  {journal}
  {Phys. Rev. Lett.}\ }\textbf {\bibinfo {volume} {93}},\ \bibinfo {pages}
  {225501} (\bibinfo {year} {2004})}\BibitemShut {NoStop}%
\bibitem [{\citenamefont {Goodwin}\ and\ \citenamefont
  {Kepert}(2005)}]{Goodwin2005}%
  \BibitemOpen
  \bibfield  {author} {\bibinfo {author} {\bibfnamefont {A.~L.}\ \bibnamefont
  {Goodwin}}\ and\ \bibinfo {author} {\bibfnamefont {C.~J.}\ \bibnamefont
  {Kepert}},\ }\href {\doibase 10.1103/PhysRevB.71.140301} {\bibfield
  {journal} {\bibinfo  {journal} {Phys. Rev. B}\ }\textbf {\bibinfo {volume}
  {71}},\ \bibinfo {pages} {140301} (\bibinfo {year} {2005})}\BibitemShut
  {NoStop}%
\bibitem [{\citenamefont {Huang}\ \emph
  {et~al.}(2016{\natexlab{b}})\citenamefont {Huang}, \citenamefont {Lu},\ and\
  \citenamefont {Rondinelli}}]{Huang2016c}%
  \BibitemOpen
  \bibfield  {author} {\bibinfo {author} {\bibfnamefont {L.-F.}\ \bibnamefont
  {Huang}}, \bibinfo {author} {\bibfnamefont {X.-Z.}\ \bibnamefont {Lu}}, \
  and\ \bibinfo {author} {\bibfnamefont {J.~M.}\ \bibnamefont {Rondinelli}},\
  }\href {\doibase 10.1103/PhysRevLett.117.115901} {\bibfield  {journal}
  {\bibinfo  {journal} {Phys. Rev. Lett.}\ }\textbf {\bibinfo {volume} {117}},\
  \bibinfo {pages} {115901} (\bibinfo {year} {2016}{\natexlab{b}})}\BibitemShut
  {NoStop}%
\bibitem [{\citenamefont {Koocher}\ \emph {et~al.}(2021)\citenamefont
  {Koocher}, \citenamefont {Huang},\ and\ \citenamefont
  {Rondinelli}}]{Koocher2021}%
  \BibitemOpen
  \bibfield  {author} {\bibinfo {author} {\bibfnamefont {N.~Z.}\ \bibnamefont
  {Koocher}}, \bibinfo {author} {\bibfnamefont {L.-F.}\ \bibnamefont {Huang}},
  \ and\ \bibinfo {author} {\bibfnamefont {J.~M.}\ \bibnamefont {Rondinelli}},\
  }\href {\doibase 10.1103/PhysRevMaterials.5.053601} {\bibfield  {journal}
  {\bibinfo  {journal} {Phys. Rev. Mater.}\ }\textbf {\bibinfo {volume} {5}},\
  \bibinfo {pages} {053601} (\bibinfo {year} {2021})}\BibitemShut {NoStop}%
\bibitem [{\citenamefont {Fujishima}\ and\ \citenamefont
  {Honda}(1972)}]{Fujishima1972}%
  \BibitemOpen
  \bibfield  {author} {\bibinfo {author} {\bibfnamefont {A.}~\bibnamefont
  {Fujishima}}\ and\ \bibinfo {author} {\bibfnamefont {K.}~\bibnamefont
  {Honda}},\ }\href {\doibase 10.1038/238037a0} {\bibfield  {journal} {\bibinfo
   {journal} {Nature}\ }\textbf {\bibinfo {volume} {238}},\ \bibinfo {pages}
  {37} (\bibinfo {year} {1972})}\BibitemShut {NoStop}%
\bibitem [{\citenamefont {N{\o}rskov}\ \emph {et~al.}(2004)\citenamefont
  {N{\o}rskov}, \citenamefont {Rossmeisl}, \citenamefont {Logadottir},
  \citenamefont {Lindqvist}, \citenamefont {Kitchin}, \citenamefont
  {Bligaard},\ and\ \citenamefont {J{\'o}nsson}}]{Norskov2004}%
  \BibitemOpen
  \bibfield  {author} {\bibinfo {author} {\bibfnamefont {J.~K.}\ \bibnamefont
  {N{\o}rskov}}, \bibinfo {author} {\bibfnamefont {J.}~\bibnamefont
  {Rossmeisl}}, \bibinfo {author} {\bibfnamefont {A.}~\bibnamefont
  {Logadottir}}, \bibinfo {author} {\bibfnamefont {L.}~\bibnamefont
  {Lindqvist}}, \bibinfo {author} {\bibfnamefont {J.~R.}\ \bibnamefont
  {Kitchin}}, \bibinfo {author} {\bibfnamefont {T.}~\bibnamefont {Bligaard}}, \
  and\ \bibinfo {author} {\bibfnamefont {H.}~\bibnamefont {J{\'o}nsson}},\
  }\href {\doibase 10.1021/jp047349j} {\bibfield  {journal} {\bibinfo
  {journal} {J. Phys. Chem. B}\ }\textbf {\bibinfo {volume} {108}},\ \bibinfo
  {pages} {17886} (\bibinfo {year} {2004})}\BibitemShut {NoStop}%
\bibitem [{\citenamefont {Rossmeisl}\ \emph {et~al.}(2007)\citenamefont
  {Rossmeisl}, \citenamefont {Qu}, \citenamefont {Zhu}, \citenamefont {Kroes},\
  and\ \citenamefont {N{\o}rskov}}]{Rossmeisl2007}%
  \BibitemOpen
  \bibfield  {author} {\bibinfo {author} {\bibfnamefont {J.}~\bibnamefont
  {Rossmeisl}}, \bibinfo {author} {\bibfnamefont {Z.-W.}\ \bibnamefont {Qu}},
  \bibinfo {author} {\bibfnamefont {H.}~\bibnamefont {Zhu}}, \bibinfo {author}
  {\bibfnamefont {G.-J.}\ \bibnamefont {Kroes}}, \ and\ \bibinfo {author}
  {\bibfnamefont {J.}~\bibnamefont {N{\o}rskov}},\ }\href {\doibase
  10.1016/j.jelechem.2006.11.008} {\bibfield  {journal} {\bibinfo  {journal}
  {Journal of Electroanalytical Chemistry}\ }\textbf {\bibinfo {volume}
  {607}},\ \bibinfo {pages} {83} (\bibinfo {year} {2007})}\BibitemShut
  {NoStop}%
\bibitem [{\citenamefont {Zhang}\ \emph {et~al.}(2007)\citenamefont {Zhang},
  \citenamefont {Wang}, \citenamefont {Zhou},\ and\ \citenamefont
  {Xu}}]{Zhang2007}%
  \BibitemOpen
  \bibfield  {author} {\bibinfo {author} {\bibfnamefont {L.}~\bibnamefont
  {Zhang}}, \bibinfo {author} {\bibfnamefont {W.}~\bibnamefont {Wang}},
  \bibinfo {author} {\bibfnamefont {L.}~\bibnamefont {Zhou}}, \ and\ \bibinfo
  {author} {\bibfnamefont {H.}~\bibnamefont {Xu}},\ }\href {\doibase
  10.1002/smll.200700043} {\bibfield  {journal} {\bibinfo  {journal} {Small}\
  }\textbf {\bibinfo {volume} {3}},\ \bibinfo {pages} {1618} (\bibinfo {year}
  {2007})}\BibitemShut {NoStop}%
\bibitem [{\citenamefont {Zhang}\ \emph
  {et~al.}(2011{\natexlab{b}})\citenamefont {Zhang}, \citenamefont {Yu},
  \citenamefont {Yu}, \citenamefont {Zhou},\ and\ \citenamefont
  {Lu}}]{Zhang2011}%
  \BibitemOpen
  \bibfield  {author} {\bibinfo {author} {\bibfnamefont {Y.}~\bibnamefont
  {Zhang}}, \bibinfo {author} {\bibfnamefont {J.}~\bibnamefont {Yu}}, \bibinfo
  {author} {\bibfnamefont {D.}~\bibnamefont {Yu}}, \bibinfo {author}
  {\bibfnamefont {X.}~\bibnamefont {Zhou}}, \ and\ \bibinfo {author}
  {\bibfnamefont {W.}~\bibnamefont {Lu}},\ }\href {\doibase
  10.1007/s12598-011-0267-8} {\bibfield  {journal} {\bibinfo  {journal} {Rare
  Metals}\ }\textbf {\bibinfo {volume} {30}},\ \bibinfo {pages} {192} (\bibinfo
  {year} {2011}{\natexlab{b}})}\BibitemShut {NoStop}%
\bibitem [{\citenamefont {Suzuki}\ \emph {et~al.}(2012)\citenamefont {Suzuki},
  \citenamefont {Hisatomi}, \citenamefont {Teramura}, \citenamefont
  {Shimodaira}, \citenamefont {Kobayashi},\ and\ \citenamefont
  {Domen}}]{Suzuki2012}%
  \BibitemOpen
  \bibfield  {author} {\bibinfo {author} {\bibfnamefont {T.}~\bibnamefont
  {Suzuki}}, \bibinfo {author} {\bibfnamefont {T.}~\bibnamefont {Hisatomi}},
  \bibinfo {author} {\bibfnamefont {K.}~\bibnamefont {Teramura}}, \bibinfo
  {author} {\bibfnamefont {Y.}~\bibnamefont {Shimodaira}}, \bibinfo {author}
  {\bibfnamefont {H.}~\bibnamefont {Kobayashi}}, \ and\ \bibinfo {author}
  {\bibfnamefont {K.}~\bibnamefont {Domen}},\ }\href {\doibase
  10.1039/C2CP43132G} {\bibfield  {journal} {\bibinfo  {journal} {Phys. Chem.
  Chem. Phys.}\ }\textbf {\bibinfo {volume} {14}},\ \bibinfo {pages} {15475}
  (\bibinfo {year} {2012})}\BibitemShut {NoStop}%
\bibitem [{\citenamefont {Jiang}\ \emph {et~al.}(2013)\citenamefont {Jiang},
  \citenamefont {Liu}, \citenamefont {Cheng}, \citenamefont {Sun},\ and\
  \citenamefont {Lin}}]{Jiang2013}%
  \BibitemOpen
  \bibfield  {author} {\bibinfo {author} {\bibfnamefont {H.-Y.}\ \bibnamefont
  {Jiang}}, \bibinfo {author} {\bibfnamefont {J.}~\bibnamefont {Liu}}, \bibinfo
  {author} {\bibfnamefont {K.}~\bibnamefont {Cheng}}, \bibinfo {author}
  {\bibfnamefont {W.}~\bibnamefont {Sun}}, \ and\ \bibinfo {author}
  {\bibfnamefont {J.}~\bibnamefont {Lin}},\ }\href {\doibase 10.1021/jp406834d}
  {\bibfield  {journal} {\bibinfo  {journal} {J. Phys. Chem. C}\ }\textbf
  {\bibinfo {volume} {117}},\ \bibinfo {pages} {20029} (\bibinfo {year}
  {2013})}\BibitemShut {NoStop}%
\bibitem [{\citenamefont {Xu}\ \emph {et~al.}(2013)\citenamefont {Xu},
  \citenamefont {Li}, \citenamefont {Guo}, \citenamefont {Zhang},\ and\
  \citenamefont {Meng}}]{Xu2013b}%
  \BibitemOpen
  \bibfield  {author} {\bibinfo {author} {\bibfnamefont {J.}~\bibnamefont
  {Xu}}, \bibinfo {author} {\bibfnamefont {L.}~\bibnamefont {Li}}, \bibinfo
  {author} {\bibfnamefont {C.}~\bibnamefont {Guo}}, \bibinfo {author}
  {\bibfnamefont {Y.}~\bibnamefont {Zhang}}, \ and\ \bibinfo {author}
  {\bibfnamefont {W.}~\bibnamefont {Meng}},\ }\href {\doibase
  10.1016/j.apcatb.2012.11.013} {\bibfield  {journal} {\bibinfo  {journal}
  {Applied Catalysis B: Environmental}\ }\textbf {\bibinfo {volume}
  {130-131}},\ \bibinfo {pages} {285} (\bibinfo {year} {2013})}\BibitemShut
  {NoStop}%
\bibitem [{\citenamefont {Zhu}\ \emph {et~al.}(2013)\citenamefont {Zhu},
  \citenamefont {Yin}, \citenamefont {Yang}, \citenamefont {Sun}, \citenamefont
  {Yu}, \citenamefont {Hoster}, \citenamefont {Hng}, \citenamefont {Zhang},\
  and\ \citenamefont {Yan}}]{Zhu2013}%
  \BibitemOpen
  \bibfield  {author} {\bibinfo {author} {\bibfnamefont {J.}~\bibnamefont
  {Zhu}}, \bibinfo {author} {\bibfnamefont {Z.}~\bibnamefont {Yin}}, \bibinfo
  {author} {\bibfnamefont {D.}~\bibnamefont {Yang}}, \bibinfo {author}
  {\bibfnamefont {T.}~\bibnamefont {Sun}}, \bibinfo {author} {\bibfnamefont
  {H.}~\bibnamefont {Yu}}, \bibinfo {author} {\bibfnamefont {H.~E.}\
  \bibnamefont {Hoster}}, \bibinfo {author} {\bibfnamefont {H.~H.}\
  \bibnamefont {Hng}}, \bibinfo {author} {\bibfnamefont {H.}~\bibnamefont
  {Zhang}}, \ and\ \bibinfo {author} {\bibfnamefont {Q.}~\bibnamefont {Yan}},\
  }\href {\doibase 10.1039/C2EE24148J} {\bibfield  {journal} {\bibinfo
  {journal} {Energy Environ. Sci.}\ }\textbf {\bibinfo {volume} {6}},\ \bibinfo
  {pages} {987} (\bibinfo {year} {2013})}\BibitemShut {NoStop}%
\bibitem [{\citenamefont {Le~Bahers}\ \emph {et~al.}(2014)\citenamefont
  {Le~Bahers}, \citenamefont {R\'{e}rat},\ and\ \citenamefont
  {Sautet}}]{LeBahers2014}%
  \BibitemOpen
  \bibfield  {author} {\bibinfo {author} {\bibfnamefont {T.}~\bibnamefont
  {Le~Bahers}}, \bibinfo {author} {\bibfnamefont {M.}~\bibnamefont
  {R\'{e}rat}}, \ and\ \bibinfo {author} {\bibfnamefont {P.}~\bibnamefont
  {Sautet}},\ }\href {\doibase 10.1021/jp409724c} {\bibfield  {journal}
  {\bibinfo  {journal} {J. Phys. Chem. C}\ }\textbf {\bibinfo {volume} {118}},\
  \bibinfo {pages} {5997} (\bibinfo {year} {2014})}\BibitemShut {NoStop}%
\bibitem [{\citenamefont {Zheng}\ \emph {et~al.}(2015)\citenamefont {Zheng},
  \citenamefont {Jiao}, \citenamefont {Jaroniec},\ and\ \citenamefont
  {Qiao}}]{Zheng2015}%
  \BibitemOpen
  \bibfield  {author} {\bibinfo {author} {\bibfnamefont {Y.}~\bibnamefont
  {Zheng}}, \bibinfo {author} {\bibfnamefont {Y.}~\bibnamefont {Jiao}},
  \bibinfo {author} {\bibfnamefont {M.}~\bibnamefont {Jaroniec}}, \ and\
  \bibinfo {author} {\bibfnamefont {S.~Z.}\ \bibnamefont {Qiao}},\ }\href
  {\doibase 10.1002/anie.201407031} {\bibfield  {journal} {\bibinfo  {journal}
  {Angew. Chem. Int. Ed.}\ }\textbf {\bibinfo {volume} {54}},\ \bibinfo {pages}
  {52} (\bibinfo {year} {2015})}\BibitemShut {NoStop}%
\bibitem [{\citenamefont {Qiao}\ \emph {et~al.}(2018)\citenamefont {Qiao},
  \citenamefont {Liu}, \citenamefont {Wang}, \citenamefont {Li},\ and\
  \citenamefont {Chen}}]{Qiao2018}%
  \BibitemOpen
  \bibfield  {author} {\bibinfo {author} {\bibfnamefont {M.}~\bibnamefont
  {Qiao}}, \bibinfo {author} {\bibfnamefont {J.}~\bibnamefont {Liu}}, \bibinfo
  {author} {\bibfnamefont {Y.}~\bibnamefont {Wang}}, \bibinfo {author}
  {\bibfnamefont {Y.}~\bibnamefont {Li}}, \ and\ \bibinfo {author}
  {\bibfnamefont {Z.}~\bibnamefont {Chen}},\ }\href {\doibase
  10.1021/jacs.8b07855} {\bibfield  {journal} {\bibinfo  {journal} {J. Am.
  Chem. Soc.}\ }\textbf {\bibinfo {volume} {140}},\ \bibinfo {pages} {12256}
  (\bibinfo {year} {2018})}\BibitemShut {NoStop}%
\bibitem [{\citenamefont {Yang}\ \emph {et~al.}(2019)\citenamefont {Yang},
  \citenamefont {Ma}, \citenamefont {Zhang}, \citenamefont {Jin}, \citenamefont
  {Huang},\ and\ \citenamefont {Dai}}]{Yang2019}%
  \BibitemOpen
  \bibfield  {author} {\bibinfo {author} {\bibfnamefont {H.}~\bibnamefont
  {Yang}}, \bibinfo {author} {\bibfnamefont {Y.}~\bibnamefont {Ma}}, \bibinfo
  {author} {\bibfnamefont {S.}~\bibnamefont {Zhang}}, \bibinfo {author}
  {\bibfnamefont {H.}~\bibnamefont {Jin}}, \bibinfo {author} {\bibfnamefont
  {B.}~\bibnamefont {Huang}}, \ and\ \bibinfo {author} {\bibfnamefont
  {Y.}~\bibnamefont {Dai}},\ }\href {\doibase 10.1039/C9TA02716E} {\bibfield
  {journal} {\bibinfo  {journal} {J. Mater. Chem. A}\ }\textbf {\bibinfo
  {volume} {7}},\ \bibinfo {pages} {12060} (\bibinfo {year}
  {2019})}\BibitemShut {NoStop}%
\bibitem [{\citenamefont {Ju}\ \emph {et~al.}(2020)\citenamefont {Ju},
  \citenamefont {Shang}, \citenamefont {Tang},\ and\ \citenamefont
  {Kou}}]{Ju2020}%
  \BibitemOpen
  \bibfield  {author} {\bibinfo {author} {\bibfnamefont {L.}~\bibnamefont
  {Ju}}, \bibinfo {author} {\bibfnamefont {J.}~\bibnamefont {Shang}}, \bibinfo
  {author} {\bibfnamefont {X.}~\bibnamefont {Tang}}, \ and\ \bibinfo {author}
  {\bibfnamefont {L.}~\bibnamefont {Kou}},\ }\href {\doibase
  10.1021/jacs.9b11614} {\bibfield  {journal} {\bibinfo  {journal} {J. Am.
  Chem. Soc.}\ }\textbf {\bibinfo {volume} {142}},\ \bibinfo {pages} {1492}
  (\bibinfo {year} {2020})}\BibitemShut {NoStop}%
\bibitem [{\citenamefont {Nakada}\ \emph {et~al.}(2021)\citenamefont {Nakada},
  \citenamefont {Kato}, \citenamefont {Nelson}, \citenamefont {Takahira},
  \citenamefont {Yabuuchi}, \citenamefont {Higashi}, \citenamefont {Suzuki},
  \citenamefont {Kirsanova}, \citenamefont {Kakudou}, \citenamefont {Tassel},
  \citenamefont {Yamamoto}, \citenamefont {Brown}, \citenamefont {Dronskowski},
  \citenamefont {Saeki}, \citenamefont {Abakumov}, \citenamefont {Kageyama},\
  and\ \citenamefont {Abe}}]{Nakada2021}%
  \BibitemOpen
  \bibfield  {author} {\bibinfo {author} {\bibfnamefont {A.}~\bibnamefont
  {Nakada}}, \bibinfo {author} {\bibfnamefont {D.}~\bibnamefont {Kato}},
  \bibinfo {author} {\bibfnamefont {R.}~\bibnamefont {Nelson}}, \bibinfo
  {author} {\bibfnamefont {H.}~\bibnamefont {Takahira}}, \bibinfo {author}
  {\bibfnamefont {M.}~\bibnamefont {Yabuuchi}}, \bibinfo {author}
  {\bibfnamefont {M.}~\bibnamefont {Higashi}}, \bibinfo {author} {\bibfnamefont
  {H.}~\bibnamefont {Suzuki}}, \bibinfo {author} {\bibfnamefont
  {M.}~\bibnamefont {Kirsanova}}, \bibinfo {author} {\bibfnamefont
  {N.}~\bibnamefont {Kakudou}}, \bibinfo {author} {\bibfnamefont
  {C.}~\bibnamefont {Tassel}}, \bibinfo {author} {\bibfnamefont
  {T.}~\bibnamefont {Yamamoto}}, \bibinfo {author} {\bibfnamefont {C.~M.}\
  \bibnamefont {Brown}}, \bibinfo {author} {\bibfnamefont {R.}~\bibnamefont
  {Dronskowski}}, \bibinfo {author} {\bibfnamefont {A.}~\bibnamefont {Saeki}},
  \bibinfo {author} {\bibfnamefont {A.}~\bibnamefont {Abakumov}}, \bibinfo
  {author} {\bibfnamefont {H.}~\bibnamefont {Kageyama}}, \ and\ \bibinfo
  {author} {\bibfnamefont {R.}~\bibnamefont {Abe}},\ }\href {\doibase
  10.1021/jacs.0c10288} {\bibfield  {journal} {\bibinfo  {journal} {J. Am.
  Chem. Soc.}\ }\textbf {\bibinfo {volume} {143}},\ \bibinfo {pages} {2491}
  (\bibinfo {year} {2021})}\BibitemShut {NoStop}%
\bibitem [{\citenamefont {Wang}\ \emph {et~al.}(2022)\citenamefont {Wang},
  \citenamefont {Chang}, \citenamefont {Tang}, \citenamefont {Xie},\ and\
  \citenamefont {Ang}}]{Wang2022}%
  \BibitemOpen
  \bibfield  {author} {\bibinfo {author} {\bibfnamefont {G.}~\bibnamefont
  {Wang}}, \bibinfo {author} {\bibfnamefont {J.}~\bibnamefont {Chang}},
  \bibinfo {author} {\bibfnamefont {W.}~\bibnamefont {Tang}}, \bibinfo {author}
  {\bibfnamefont {W.}~\bibnamefont {Xie}}, \ and\ \bibinfo {author}
  {\bibfnamefont {Y.~S.}\ \bibnamefont {Ang}},\ }\href {\doibase
  10.1088/1361-6463/ac5771} {\bibfield  {journal} {\bibinfo  {journal} {Journal
  of Physics D: Applied Physics}\ }\textbf {\bibinfo {volume} {55}},\ \bibinfo
  {pages} {293002} (\bibinfo {year} {2022})}\BibitemShut {NoStop}%
\bibitem [{\citenamefont {Fu}\ \emph {et~al.}(2022)\citenamefont {Fu},
  \citenamefont {Wang}, \citenamefont {Huang}, \citenamefont {Chen},
  \citenamefont {Yuan}, \citenamefont {Ang},\ and\ \citenamefont
  {Chen}}]{Fu2022}%
  \BibitemOpen
  \bibfield  {author} {\bibinfo {author} {\bibfnamefont {C.}~\bibnamefont
  {Fu}}, \bibinfo {author} {\bibfnamefont {G.}~\bibnamefont {Wang}}, \bibinfo
  {author} {\bibfnamefont {Y.}~\bibnamefont {Huang}}, \bibinfo {author}
  {\bibfnamefont {Y.}~\bibnamefont {Chen}}, \bibinfo {author} {\bibfnamefont
  {H.}~\bibnamefont {Yuan}}, \bibinfo {author} {\bibfnamefont {Y.~S.}\
  \bibnamefont {Ang}}, \ and\ \bibinfo {author} {\bibfnamefont
  {H.}~\bibnamefont {Chen}},\ }\href {\doibase 10.1039/D1CP04679A} {\bibfield
  {journal} {\bibinfo  {journal} {Phys. Chem. Chem. Phys.}\ }\textbf {\bibinfo
  {volume} {24}},\ \bibinfo {pages} {3826} (\bibinfo {year}
  {2022})}\BibitemShut {NoStop}%
\bibitem [{\citenamefont {Perdew}(1985)}]{Perdew1985}%
  \BibitemOpen
  \bibfield  {author} {\bibinfo {author} {\bibfnamefont {J.~P.}\ \bibnamefont
  {Perdew}},\ }\href {\doibase 10.1002/qua.560280846} {\bibfield  {journal}
  {\bibinfo  {journal} {Int. J. Quantum Chem.}\ }\textbf {\bibinfo {volume}
  {28}},\ \bibinfo {pages} {497} (\bibinfo {year} {1985})}\BibitemShut
  {NoStop}%
\bibitem [{\citenamefont {Perdew}\ \emph
  {et~al.}(1996{\natexlab{a}})\citenamefont {Perdew}, \citenamefont {Burke},\
  and\ \citenamefont {Ernzerhof}}]{Perdew1996}%
  \BibitemOpen
  \bibfield  {author} {\bibinfo {author} {\bibfnamefont {J.~P.}\ \bibnamefont
  {Perdew}}, \bibinfo {author} {\bibfnamefont {K.}~\bibnamefont {Burke}}, \
  and\ \bibinfo {author} {\bibfnamefont {M.}~\bibnamefont {Ernzerhof}},\ }\href
  {\doibase 10.1103/PhysRevLett.77.3865} {\bibfield  {journal} {\bibinfo
  {journal} {Phys. Rev. Lett.}\ }\textbf {\bibinfo {volume} {77}},\ \bibinfo
  {pages} {3865} (\bibinfo {year} {1996}{\natexlab{a}})}\BibitemShut {NoStop}%
\bibitem [{\citenamefont {Perdew}\ \emph {et~al.}(2008)\citenamefont {Perdew},
  \citenamefont {Ruzsinszky}, \citenamefont {Csonka}, \citenamefont {Vydrov},
  \citenamefont {Scuseria}, \citenamefont {Constantin}, \citenamefont {Zhou},\
  and\ \citenamefont {Burke}}]{Perdew2008}%
  \BibitemOpen
  \bibfield  {author} {\bibinfo {author} {\bibfnamefont {J.~P.}\ \bibnamefont
  {Perdew}}, \bibinfo {author} {\bibfnamefont {A.}~\bibnamefont {Ruzsinszky}},
  \bibinfo {author} {\bibfnamefont {G.~I.}\ \bibnamefont {Csonka}}, \bibinfo
  {author} {\bibfnamefont {O.~A.}\ \bibnamefont {Vydrov}}, \bibinfo {author}
  {\bibfnamefont {G.~E.}\ \bibnamefont {Scuseria}}, \bibinfo {author}
  {\bibfnamefont {L.~A.}\ \bibnamefont {Constantin}}, \bibinfo {author}
  {\bibfnamefont {X.}~\bibnamefont {Zhou}}, \ and\ \bibinfo {author}
  {\bibfnamefont {K.}~\bibnamefont {Burke}},\ }\href {\doibase
  10.1103/PhysRevLett.100.136406} {\bibfield  {journal} {\bibinfo  {journal}
  {Phys. Rev. Lett.}\ }\textbf {\bibinfo {volume} {100}},\ \bibinfo {pages}
  {136406} (\bibinfo {year} {2008})}\BibitemShut {NoStop}%
\bibitem [{\citenamefont {Heyd}\ \emph {et~al.}(2003)\citenamefont {Heyd},
  \citenamefont {Scuseria},\ and\ \citenamefont {Ernzerhof}}]{HSE1}%
  \BibitemOpen
  \bibfield  {author} {\bibinfo {author} {\bibfnamefont {J.}~\bibnamefont
  {Heyd}}, \bibinfo {author} {\bibfnamefont {G.~E.}\ \bibnamefont {Scuseria}},
  \ and\ \bibinfo {author} {\bibfnamefont {M.}~\bibnamefont {Ernzerhof}},\
  }\href {\doibase 10.1063/1.1564060} {\bibfield  {journal} {\bibinfo
  {journal} {J. Chem. Phys.}\ }\textbf {\bibinfo {volume} {118}},\ \bibinfo
  {pages} {8207} (\bibinfo {year} {2003})}\BibitemShut {NoStop}%
\bibitem [{\citenamefont {Heyd}\ \emph {et~al.}(2006)\citenamefont {Heyd},
  \citenamefont {Scuseria},\ and\ \citenamefont {Ernzerhof}}]{HSE2}%
  \BibitemOpen
  \bibfield  {author} {\bibinfo {author} {\bibfnamefont {J.}~\bibnamefont
  {Heyd}}, \bibinfo {author} {\bibfnamefont {G.~E.}\ \bibnamefont {Scuseria}},
  \ and\ \bibinfo {author} {\bibfnamefont {M.}~\bibnamefont {Ernzerhof}},\
  }\href {\doibase 10.1063/1.2204597} {\bibfield  {journal} {\bibinfo
  {journal} {J. Chem. Phys.}\ }\textbf {\bibinfo {volume} {124}},\ \bibinfo
  {pages} {219906} (\bibinfo {year} {2006})}\BibitemShut {NoStop}%
\bibitem [{\citenamefont {Peralta}\ \emph {et~al.}(2006)\citenamefont
  {Peralta}, \citenamefont {Heyd}, \citenamefont {Scuseria},\ and\
  \citenamefont {Martin}}]{HSE3}%
  \BibitemOpen
  \bibfield  {author} {\bibinfo {author} {\bibfnamefont {J.~E.}\ \bibnamefont
  {Peralta}}, \bibinfo {author} {\bibfnamefont {J.}~\bibnamefont {Heyd}},
  \bibinfo {author} {\bibfnamefont {G.~E.}\ \bibnamefont {Scuseria}}, \ and\
  \bibinfo {author} {\bibfnamefont {R.~L.}\ \bibnamefont {Martin}},\ }\href
  {\doibase 10.1103/PhysRevB.74.073101} {\bibfield  {journal} {\bibinfo
  {journal} {Phys. Rev. B}\ }\textbf {\bibinfo {volume} {74}},\ \bibinfo
  {pages} {073101} (\bibinfo {year} {2006})}\BibitemShut {NoStop}%
\bibitem [{\citenamefont {Schimka}\ \emph {et~al.}(2011)\citenamefont
  {Schimka}, \citenamefont {Harl},\ and\ \citenamefont {Kresse}}]{Schimka2011}%
  \BibitemOpen
  \bibfield  {author} {\bibinfo {author} {\bibfnamefont {L.}~\bibnamefont
  {Schimka}}, \bibinfo {author} {\bibfnamefont {J.}~\bibnamefont {Harl}}, \
  and\ \bibinfo {author} {\bibfnamefont {G.}~\bibnamefont {Kresse}},\ }\href
  {\doibase 10.1063/1.3524336} {\bibfield  {journal} {\bibinfo  {journal} {J.
  Chem. Phys.}\ }\textbf {\bibinfo {volume} {134}},\ \bibinfo {pages} {024116}
  (\bibinfo {year} {2011})}\BibitemShut {NoStop}%
\bibitem [{\citenamefont {Perdew}\ \emph
  {et~al.}(1996{\natexlab{b}})\citenamefont {Perdew}, \citenamefont
  {Ernzerhof},\ and\ \citenamefont {Burke}}]{Perdew1996a}%
  \BibitemOpen
  \bibfield  {author} {\bibinfo {author} {\bibfnamefont {J.~P.}\ \bibnamefont
  {Perdew}}, \bibinfo {author} {\bibfnamefont {M.}~\bibnamefont {Ernzerhof}}, \
  and\ \bibinfo {author} {\bibfnamefont {K.}~\bibnamefont {Burke}},\ }\href
  {\doibase 10.1063/1.472933} {\bibfield  {journal} {\bibinfo  {journal} {J.
  Chem. Phys.}\ }\textbf {\bibinfo {volume} {105}},\ \bibinfo {pages} {9982}
  (\bibinfo {year} {1996}{\natexlab{b}})}\BibitemShut {NoStop}%
\bibitem [{\citenamefont {Becke}(1996)}]{Becke1996}%
  \BibitemOpen
  \bibfield  {author} {\bibinfo {author} {\bibfnamefont {A.~D.}\ \bibnamefont
  {Becke}},\ }\href {\doibase 10.1063/1.470829} {\bibfield  {journal} {\bibinfo
   {journal} {J. Chem. Phys.}\ }\textbf {\bibinfo {volume} {104}},\ \bibinfo
  {pages} {1040} (\bibinfo {year} {1996})}\BibitemShut {NoStop}%
\bibitem [{\citenamefont {Adamo}\ and\ \citenamefont
  {Barone}(1999)}]{Adamo1999}%
  \BibitemOpen
  \bibfield  {author} {\bibinfo {author} {\bibfnamefont {C.}~\bibnamefont
  {Adamo}}\ and\ \bibinfo {author} {\bibfnamefont {V.}~\bibnamefont {Barone}},\
  }\href {\doibase 10.1063/1.478522} {\bibfield  {journal} {\bibinfo  {journal}
  {J. Chem. Phys.}\ }\textbf {\bibinfo {volume} {110}},\ \bibinfo {pages}
  {6158} (\bibinfo {year} {1999})}\BibitemShut {NoStop}%
\bibitem [{\citenamefont {Ernzerhof}\ and\ \citenamefont
  {Scuseria}(1999)}]{Ernzerhof1999}%
  \BibitemOpen
  \bibfield  {author} {\bibinfo {author} {\bibfnamefont {M.}~\bibnamefont
  {Ernzerhof}}\ and\ \bibinfo {author} {\bibfnamefont {G.~E.}\ \bibnamefont
  {Scuseria}},\ }\href {\doibase 10.1063/1.478401} {\bibfield  {journal}
  {\bibinfo  {journal} {J. Chem. Phys.}\ }\textbf {\bibinfo {volume} {110}},\
  \bibinfo {pages} {5029} (\bibinfo {year} {1999})}\BibitemShut {NoStop}%
\bibitem [{\citenamefont {Bernardi}(2020)}]{Bernardi2020}%
  \BibitemOpen
  \bibfield  {author} {\bibinfo {author} {\bibfnamefont {M.}~\bibnamefont
  {Bernardi}},\ }\href {\doibase 10.1088/1361-648X/ab9409} {\bibfield
  {journal} {\bibinfo  {journal} {J. Phys.: Condens. Matter}\ }\textbf
  {\bibinfo {volume} {32}},\ \bibinfo {pages} {385501} (\bibinfo {year}
  {2020})}\BibitemShut {NoStop}%
\bibitem [{\citenamefont {Champagne}\ \emph {et~al.}(2024)\citenamefont
  {Champagne}, \citenamefont {Camarasa-G{\'o}mez}, \citenamefont {Ricci},
  \citenamefont {Kronik},\ and\ \citenamefont {Neaton}}]{Champagne2024}%
  \BibitemOpen
  \bibfield  {author} {\bibinfo {author} {\bibfnamefont {A.}~\bibnamefont
  {Champagne}}, \bibinfo {author} {\bibfnamefont {M.}~\bibnamefont
  {Camarasa-G{\'o}mez}}, \bibinfo {author} {\bibfnamefont {F.}~\bibnamefont
  {Ricci}}, \bibinfo {author} {\bibfnamefont {L.}~\bibnamefont {Kronik}}, \
  and\ \bibinfo {author} {\bibfnamefont {J.~B.}\ \bibnamefont {Neaton}},\
  }\href {\doibase 10.1021/acs.nanolett.4c01497} {\bibfield  {journal}
  {\bibinfo  {journal} {Nano Lett.}\ }\textbf {\bibinfo {volume} {24}},\
  \bibinfo {pages} {7033} (\bibinfo {year} {2024})}\BibitemShut {NoStop}%
\bibitem [{\citenamefont {Chakrapani}\ \emph {et~al.}(2007)\citenamefont
  {Chakrapani}, \citenamefont {Angus}, \citenamefont {Anderson}, \citenamefont
  {Wolter}, \citenamefont {Stoner},\ and\ \citenamefont
  {Sumanasekera}}]{Chakrapani2007}%
  \BibitemOpen
  \bibfield  {author} {\bibinfo {author} {\bibfnamefont {V.}~\bibnamefont
  {Chakrapani}}, \bibinfo {author} {\bibfnamefont {J.~C.}\ \bibnamefont
  {Angus}}, \bibinfo {author} {\bibfnamefont {A.~B.}\ \bibnamefont {Anderson}},
  \bibinfo {author} {\bibfnamefont {S.~D.}\ \bibnamefont {Wolter}}, \bibinfo
  {author} {\bibfnamefont {B.~R.}\ \bibnamefont {Stoner}}, \ and\ \bibinfo
  {author} {\bibfnamefont {G.~U.}\ \bibnamefont {Sumanasekera}},\ }\href
  {\doibase 10.1126/science.1148841} {\bibfield  {journal} {\bibinfo  {journal}
  {Science}\ }\textbf {\bibinfo {volume} {318}},\ \bibinfo {pages} {1424}
  (\bibinfo {year} {2007})}\BibitemShut {NoStop}%
\bibitem [{\citenamefont {Zhang}\ \emph {et~al.}(2019)\citenamefont {Zhang},
  \citenamefont {Chen}, \citenamefont {Zhang}, \citenamefont {Jiao},\ and\
  \citenamefont {Zhou}}]{Zhang2019d}%
  \BibitemOpen
  \bibfield  {author} {\bibinfo {author} {\bibfnamefont {X.}~\bibnamefont
  {Zhang}}, \bibinfo {author} {\bibfnamefont {A.}~\bibnamefont {Chen}},
  \bibinfo {author} {\bibfnamefont {Z.}~\bibnamefont {Zhang}}, \bibinfo
  {author} {\bibfnamefont {M.}~\bibnamefont {Jiao}}, \ and\ \bibinfo {author}
  {\bibfnamefont {Z.}~\bibnamefont {Zhou}},\ }\href {\doibase
  10.1039/C8NA00084K} {\bibfield  {journal} {\bibinfo  {journal} {Nanoscale
  Adv.}\ }\textbf {\bibinfo {volume} {1}},\ \bibinfo {pages} {154} (\bibinfo
  {year} {2019})}\BibitemShut {NoStop}%
\bibitem [{\citenamefont {Chen}\ \emph {et~al.}(2022)\citenamefont {Chen},
  \citenamefont {Zhao}, \citenamefont {Wang}, \citenamefont {Chen},
  \citenamefont {Zhang},\ and\ \citenamefont {Hua}}]{Chen2022}%
  \BibitemOpen
  \bibfield  {author} {\bibinfo {author} {\bibfnamefont {H.}~\bibnamefont
  {Chen}}, \bibinfo {author} {\bibfnamefont {J.}~\bibnamefont {Zhao}}, \bibinfo
  {author} {\bibfnamefont {X.}~\bibnamefont {Wang}}, \bibinfo {author}
  {\bibfnamefont {X.}~\bibnamefont {Chen}}, \bibinfo {author} {\bibfnamefont
  {Z.}~\bibnamefont {Zhang}}, \ and\ \bibinfo {author} {\bibfnamefont
  {M.}~\bibnamefont {Hua}},\ }\href {\doibase 10.1039/D2NR00466F} {\bibfield
  {journal} {\bibinfo  {journal} {Nanoscale}\ }\textbf {\bibinfo {volume}
  {14}},\ \bibinfo {pages} {5551} (\bibinfo {year} {2022})}\BibitemShut
  {NoStop}%
\bibitem [{\citenamefont {Scanlon}\ \emph {et~al.}(2013)\citenamefont
  {Scanlon}, \citenamefont {Dunnill}, \citenamefont {Buckeridge}, \citenamefont
  {Shevlin}, \citenamefont {Logsdail}, \citenamefont {Woodley}, \citenamefont
  {Catlow}, \citenamefont {Powell}, \citenamefont {Palgrave}, \citenamefont
  {Parkin}, \citenamefont {Watson}, \citenamefont {Keal}, \citenamefont
  {Sherwood}, \citenamefont {Walsh},\ and\ \citenamefont
  {Sokol}}]{Scanlon2013}%
  \BibitemOpen
  \bibfield  {author} {\bibinfo {author} {\bibfnamefont {D.~O.}\ \bibnamefont
  {Scanlon}}, \bibinfo {author} {\bibfnamefont {C.~W.}\ \bibnamefont
  {Dunnill}}, \bibinfo {author} {\bibfnamefont {J.}~\bibnamefont {Buckeridge}},
  \bibinfo {author} {\bibfnamefont {S.~A.}\ \bibnamefont {Shevlin}}, \bibinfo
  {author} {\bibfnamefont {A.~J.}\ \bibnamefont {Logsdail}}, \bibinfo {author}
  {\bibfnamefont {S.~M.}\ \bibnamefont {Woodley}}, \bibinfo {author}
  {\bibfnamefont {C.~R.~A.}\ \bibnamefont {Catlow}}, \bibinfo {author}
  {\bibfnamefont {M.~J.}\ \bibnamefont {Powell}}, \bibinfo {author}
  {\bibfnamefont {R.~G.}\ \bibnamefont {Palgrave}}, \bibinfo {author}
  {\bibfnamefont {I.~P.}\ \bibnamefont {Parkin}}, \bibinfo {author}
  {\bibfnamefont {G.~W.}\ \bibnamefont {Watson}}, \bibinfo {author}
  {\bibfnamefont {T.~W.}\ \bibnamefont {Keal}}, \bibinfo {author}
  {\bibfnamefont {P.}~\bibnamefont {Sherwood}}, \bibinfo {author}
  {\bibfnamefont {A.}~\bibnamefont {Walsh}}, \ and\ \bibinfo {author}
  {\bibfnamefont {A.~A.}\ \bibnamefont {Sokol}},\ }\href {\doibase
  10.1038/nmat3697} {\bibfield  {journal} {\bibinfo  {journal} {Nature
  Materials}\ }\textbf {\bibinfo {volume} {12}},\ \bibinfo {pages} {798}
  (\bibinfo {year} {2013})}\BibitemShut {NoStop}%
\bibitem [{\citenamefont {\"Oz\c{c}elik}\ \emph {et~al.}(2016)\citenamefont
  {\"Oz\c{c}elik}, \citenamefont {Azadani}, \citenamefont {Yang}, \citenamefont
  {Koester},\ and\ \citenamefont {Low}}]{Ozcelik2016}%
  \BibitemOpen
  \bibfield  {author} {\bibinfo {author} {\bibfnamefont {V.~O.}\ \bibnamefont
  {\"Oz\c{c}elik}}, \bibinfo {author} {\bibfnamefont {J.~G.}\ \bibnamefont
  {Azadani}}, \bibinfo {author} {\bibfnamefont {C.}~\bibnamefont {Yang}},
  \bibinfo {author} {\bibfnamefont {S.~J.}\ \bibnamefont {Koester}}, \ and\
  \bibinfo {author} {\bibfnamefont {T.}~\bibnamefont {Low}},\ }\href {\doibase
  10.1103/PhysRevB.94.035125} {\bibfield  {journal} {\bibinfo  {journal} {Phys.
  Rev. B}\ }\textbf {\bibinfo {volume} {94}},\ \bibinfo {pages} {035125}
  (\bibinfo {year} {2016})}\BibitemShut {NoStop}%
\bibitem [{\citenamefont {Hu}\ and\ \citenamefont {Yang}(2017)}]{Hu2017}%
  \BibitemOpen
  \bibfield  {author} {\bibinfo {author} {\bibfnamefont {W.}~\bibnamefont
  {Hu}}\ and\ \bibinfo {author} {\bibfnamefont {J.}~\bibnamefont {Yang}},\
  }\href {\doibase 10.1039/C7TC04697A} {\bibfield  {journal} {\bibinfo
  {journal} {J. Mater. Chem. C}\ }\textbf {\bibinfo {volume} {5}},\ \bibinfo
  {pages} {12289} (\bibinfo {year} {2017})}\BibitemShut {NoStop}%
\bibitem [{\citenamefont {Yang}\ \emph {et~al.}(2020)\citenamefont {Yang},
  \citenamefont {Singh},\ and\ \citenamefont {Ahuja}}]{Yang2020a}%
  \BibitemOpen
  \bibfield  {author} {\bibinfo {author} {\bibfnamefont {X.}~\bibnamefont
  {Yang}}, \bibinfo {author} {\bibfnamefont {D.}~\bibnamefont {Singh}}, \ and\
  \bibinfo {author} {\bibfnamefont {R.}~\bibnamefont {Ahuja}},\ }\href
  {\doibase 10.3390/catal10101111} {\bibfield  {journal} {\bibinfo  {journal}
  {Catalysts}\ }\textbf {\bibinfo {volume} {10}},\ \bibinfo {pages} {1111}
  (\bibinfo {year} {2020})}\BibitemShut {NoStop}%
\bibitem [{\citenamefont {Rode}(1975)}]{Rode1975}%
  \BibitemOpen
  \bibfield  {author} {\bibinfo {author} {\bibfnamefont {D.}~\bibnamefont
  {Rode}},\ }in\ \href {\doibase 10.1016/S0080-8784(08)60331-2} {\emph
  {\bibinfo {booktitle} {Semiconductors and Semimetals}}},\ Vol.~\bibinfo
  {volume} {10},\ \bibinfo {editor} {edited by\ \bibinfo {editor}
  {\bibfnamefont {R.}~\bibnamefont {Willardson}}\ and\ \bibinfo {editor}
  {\bibfnamefont {A.~C.}\ \bibnamefont {Beer}}}\ (\bibinfo  {publisher}
  {Elsevier},\ \bibinfo {year} {1975})\ pp.\ \bibinfo {pages}
  {1--89}\BibitemShut {NoStop}%
\bibitem [{\citenamefont {Faghaninia}\ \emph {et~al.}(2015)\citenamefont
  {Faghaninia}, \citenamefont {Ager},\ and\ \citenamefont
  {Lo}}]{Faghaninia2015}%
  \BibitemOpen
  \bibfield  {author} {\bibinfo {author} {\bibfnamefont {A.}~\bibnamefont
  {Faghaninia}}, \bibinfo {author} {\bibfnamefont {J.~W.}\ \bibnamefont
  {Ager}}, \ and\ \bibinfo {author} {\bibfnamefont {C.~S.}\ \bibnamefont
  {Lo}},\ }\href {\doibase 10.1103/PhysRevB.91.235123} {\bibfield  {journal}
  {\bibinfo  {journal} {Phys. Rev. B}\ }\textbf {\bibinfo {volume} {91}},\
  \bibinfo {pages} {235123} (\bibinfo {year} {2015})}\BibitemShut {NoStop}%
\bibitem [{\citenamefont {Ganose}\ \emph {et~al.}(2021)\citenamefont {Ganose},
  \citenamefont {Park}, \citenamefont {Faghaninia}, \citenamefont
  {Woods-Robinson}, \citenamefont {Persson},\ and\ \citenamefont
  {Jain}}]{Ganose2021}%
  \BibitemOpen
  \bibfield  {author} {\bibinfo {author} {\bibfnamefont {A.~M.}\ \bibnamefont
  {Ganose}}, \bibinfo {author} {\bibfnamefont {J.}~\bibnamefont {Park}},
  \bibinfo {author} {\bibfnamefont {A.}~\bibnamefont {Faghaninia}}, \bibinfo
  {author} {\bibfnamefont {R.}~\bibnamefont {Woods-Robinson}}, \bibinfo
  {author} {\bibfnamefont {K.~A.}\ \bibnamefont {Persson}}, \ and\ \bibinfo
  {author} {\bibfnamefont {A.}~\bibnamefont {Jain}},\ }\href {\doibase
  10.1038/s41467-021-22440-5} {\bibfield  {journal} {\bibinfo  {journal}
  {Nature Communications}\ }\textbf {\bibinfo {volume} {12}},\ \bibinfo {pages}
  {2222} (\bibinfo {year} {2021})}\BibitemShut {NoStop}%
\bibitem [{\citenamefont {Zhao}\ \emph {et~al.}(2005)\citenamefont {Zhao},
  \citenamefont {Kim}, \citenamefont {Dillon}, \citenamefont {Heben},\ and\
  \citenamefont {Zhang}}]{Zhao2005}%
  \BibitemOpen
  \bibfield  {author} {\bibinfo {author} {\bibfnamefont {Y.}~\bibnamefont
  {Zhao}}, \bibinfo {author} {\bibfnamefont {Y.-H.}\ \bibnamefont {Kim}},
  \bibinfo {author} {\bibfnamefont {A.~C.}\ \bibnamefont {Dillon}}, \bibinfo
  {author} {\bibfnamefont {M.~J.}\ \bibnamefont {Heben}}, \ and\ \bibinfo
  {author} {\bibfnamefont {S.~B.}\ \bibnamefont {Zhang}},\ }\href {\doibase
  10.1103/PhysRevLett.94.155504} {\bibfield  {journal} {\bibinfo  {journal}
  {Phys. Rev. Lett.}\ }\textbf {\bibinfo {volume} {94}},\ \bibinfo {pages}
  {155504} (\bibinfo {year} {2005})}\BibitemShut {NoStop}%
\bibitem [{\citenamefont {Kim}\ \emph {et~al.}(2006)\citenamefont {Kim},
  \citenamefont {Zhao}, \citenamefont {Williamson}, \citenamefont {Heben},\
  and\ \citenamefont {Zhang}}]{Kim2006}%
  \BibitemOpen
  \bibfield  {author} {\bibinfo {author} {\bibfnamefont {Y.-H.}\ \bibnamefont
  {Kim}}, \bibinfo {author} {\bibfnamefont {Y.}~\bibnamefont {Zhao}}, \bibinfo
  {author} {\bibfnamefont {A.}~\bibnamefont {Williamson}}, \bibinfo {author}
  {\bibfnamefont {M.~J.}\ \bibnamefont {Heben}}, \ and\ \bibinfo {author}
  {\bibfnamefont {S.~B.}\ \bibnamefont {Zhang}},\ }\href {\doibase
  10.1103/PhysRevLett.96.016102} {\bibfield  {journal} {\bibinfo  {journal}
  {Phys. Rev. Lett.}\ }\textbf {\bibinfo {volume} {96}},\ \bibinfo {pages}
  {016102} (\bibinfo {year} {2006})}\BibitemShut {NoStop}%
\bibitem [{\citenamefont {Yoon}\ \emph {et~al.}(2008)\citenamefont {Yoon},
  \citenamefont {Yang}, \citenamefont {Hicke}, \citenamefont {Wang},
  \citenamefont {Geohegan},\ and\ \citenamefont {Zhang}}]{Yoon2008}%
  \BibitemOpen
  \bibfield  {author} {\bibinfo {author} {\bibfnamefont {M.}~\bibnamefont
  {Yoon}}, \bibinfo {author} {\bibfnamefont {S.}~\bibnamefont {Yang}}, \bibinfo
  {author} {\bibfnamefont {C.}~\bibnamefont {Hicke}}, \bibinfo {author}
  {\bibfnamefont {E.}~\bibnamefont {Wang}}, \bibinfo {author} {\bibfnamefont
  {D.}~\bibnamefont {Geohegan}}, \ and\ \bibinfo {author} {\bibfnamefont
  {Z.}~\bibnamefont {Zhang}},\ }\href {\doibase 10.1103/PhysRevLett.100.206806}
  {\bibfield  {journal} {\bibinfo  {journal} {Phys. Rev. Lett.}\ }\textbf
  {\bibinfo {volume} {100}},\ \bibinfo {pages} {206806} (\bibinfo {year}
  {2008})}\BibitemShut {NoStop}%
\bibitem [{\citenamefont {Pupysheva}\ \emph {et~al.}(2008)\citenamefont
  {Pupysheva}, \citenamefont {Farajian},\ and\ \citenamefont
  {Yakobson}}]{Pupysheva2008}%
  \BibitemOpen
  \bibfield  {author} {\bibinfo {author} {\bibfnamefont {O.~V.}\ \bibnamefont
  {Pupysheva}}, \bibinfo {author} {\bibfnamefont {A.~A.}\ \bibnamefont
  {Farajian}}, \ and\ \bibinfo {author} {\bibfnamefont {B.~I.}\ \bibnamefont
  {Yakobson}},\ }\href {\doibase 10.1021/nl071436g} {\bibfield  {journal}
  {\bibinfo  {journal} {Nano Lett.}\ }\textbf {\bibinfo {volume} {8}},\
  \bibinfo {pages} {767} (\bibinfo {year} {2008})}\BibitemShut {NoStop}%
\bibitem [{\citenamefont {Wang}\ \emph {et~al.}(2009)\citenamefont {Wang},
  \citenamefont {Sun}, \citenamefont {Jena},\ and\ \citenamefont
  {Kawazoe}}]{Wang2009}%
  \BibitemOpen
  \bibfield  {author} {\bibinfo {author} {\bibfnamefont {Q.}~\bibnamefont
  {Wang}}, \bibinfo {author} {\bibfnamefont {Q.}~\bibnamefont {Sun}}, \bibinfo
  {author} {\bibfnamefont {P.}~\bibnamefont {Jena}}, \ and\ \bibinfo {author}
  {\bibfnamefont {Y.}~\bibnamefont {Kawazoe}},\ }\href {\doibase
  10.1021/ct800373g} {\bibfield  {journal} {\bibinfo  {journal} {J. Chem.
  Theory Comput.}\ }\textbf {\bibinfo {volume} {5}},\ \bibinfo {pages} {374}
  (\bibinfo {year} {2009})}\BibitemShut {NoStop}%
\bibitem [{\citenamefont {Sun}\ \emph {et~al.}(2009)\citenamefont {Sun},
  \citenamefont {Wang},\ and\ \citenamefont {Jena}}]{Sun2009}%
  \BibitemOpen
  \bibfield  {author} {\bibinfo {author} {\bibfnamefont {Q.}~\bibnamefont
  {Sun}}, \bibinfo {author} {\bibfnamefont {Q.}~\bibnamefont {Wang}}, \ and\
  \bibinfo {author} {\bibfnamefont {P.}~\bibnamefont {Jena}},\ }\href {\doibase
  10.1063/1.3058678} {\bibfield  {journal} {\bibinfo  {journal} {Appl. Phys.
  Lett.}\ }\textbf {\bibinfo {volume} {94}},\ \bibinfo {pages} {013111}
  (\bibinfo {year} {2009})}\BibitemShut {NoStop}%
\bibitem [{\citenamefont {Durbin}\ \emph {et~al.}(2016)\citenamefont {Durbin},
  \citenamefont {Allan},\ and\ \citenamefont {Malardier-Jugroot}}]{Durbin2016}%
  \BibitemOpen
  \bibfield  {author} {\bibinfo {author} {\bibfnamefont {D.}~\bibnamefont
  {Durbin}}, \bibinfo {author} {\bibfnamefont {N.}~\bibnamefont {Allan}}, \
  and\ \bibinfo {author} {\bibfnamefont {C.}~\bibnamefont
  {Malardier-Jugroot}},\ }\href {\doibase 10.1016/j.ijhydene.2016.05.001}
  {\bibfield  {journal} {\bibinfo  {journal} {International Journal of Hydrogen
  Energy}\ }\textbf {\bibinfo {volume} {41}},\ \bibinfo {pages} {13116}
  (\bibinfo {year} {2016})}\BibitemShut {NoStop}%
\bibitem [{\citenamefont {Sankar~De}\ \emph {et~al.}(2018)\citenamefont
  {Sankar~De}, \citenamefont {Flores-Livas}, \citenamefont {Saha},
  \citenamefont {Genovese},\ and\ \citenamefont {Goedecker}}]{SankarDe2018}%
  \BibitemOpen
  \bibfield  {author} {\bibinfo {author} {\bibfnamefont {D.}~\bibnamefont
  {Sankar~De}}, \bibinfo {author} {\bibfnamefont {J.~A.}\ \bibnamefont
  {Flores-Livas}}, \bibinfo {author} {\bibfnamefont {S.}~\bibnamefont {Saha}},
  \bibinfo {author} {\bibfnamefont {L.}~\bibnamefont {Genovese}}, \ and\
  \bibinfo {author} {\bibfnamefont {S.}~\bibnamefont {Goedecker}},\ }\href
  {\doibase 10.1016/j.carbon.2017.11.086} {\bibfield  {journal} {\bibinfo
  {journal} {Carbon}\ }\textbf {\bibinfo {volume} {129}},\ \bibinfo {pages}
  {847} (\bibinfo {year} {2018})}\BibitemShut {NoStop}%
\bibitem [{\citenamefont {Ren}\ \emph {et~al.}(2023)\citenamefont {Ren},
  \citenamefont {Lu},\ and\ \citenamefont {Zhang}}]{Ren2023}%
  \BibitemOpen
  \bibfield  {author} {\bibinfo {author} {\bibfnamefont {Y.}~\bibnamefont
  {Ren}}, \bibinfo {author} {\bibfnamefont {Y.}~\bibnamefont {Lu}}, \ and\
  \bibinfo {author} {\bibfnamefont {D.}~\bibnamefont {Zhang}},\ }\href
  {\doibase 10.1021/acs.jpclett.3c02488} {\bibfield  {journal} {\bibinfo
  {journal} {J. Phys. Chem. Lett.}\ }\textbf {\bibinfo {volume} {14}},\
  \bibinfo {pages} {11051} (\bibinfo {year} {2023})}\BibitemShut {NoStop}%
\bibitem [{\citenamefont {Guan}\ \emph {et~al.}(2018)\citenamefont {Guan},
  \citenamefont {Wu}, \citenamefont {Jiang}, \citenamefont {Zhu}, \citenamefont
  {Guan}, \citenamefont {Lei}, \citenamefont {Du}, \citenamefont {Zeng},\ and\
  \citenamefont {Yang}}]{Guan2018}%
  \BibitemOpen
  \bibfield  {author} {\bibinfo {author} {\bibfnamefont {J.}~\bibnamefont
  {Guan}}, \bibinfo {author} {\bibfnamefont {J.}~\bibnamefont {Wu}}, \bibinfo
  {author} {\bibfnamefont {D.}~\bibnamefont {Jiang}}, \bibinfo {author}
  {\bibfnamefont {X.}~\bibnamefont {Zhu}}, \bibinfo {author} {\bibfnamefont
  {R.}~\bibnamefont {Guan}}, \bibinfo {author} {\bibfnamefont {X.}~\bibnamefont
  {Lei}}, \bibinfo {author} {\bibfnamefont {P.}~\bibnamefont {Du}}, \bibinfo
  {author} {\bibfnamefont {H.}~\bibnamefont {Zeng}}, \ and\ \bibinfo {author}
  {\bibfnamefont {S.}~\bibnamefont {Yang}},\ }\href {\doibase
  10.1016/j.ijhydene.2018.03.148} {\bibfield  {journal} {\bibinfo  {journal}
  {International Journal of Hydrogen Energy}\ }\textbf {\bibinfo {volume}
  {43}},\ \bibinfo {pages} {8698} (\bibinfo {year} {2018})}\BibitemShut
  {NoStop}%
\bibitem [{\citenamefont {Kayley}\ and\ \citenamefont
  {Peng}(2025)}]{Kayley2025}%
  \BibitemOpen
  \bibfield  {author} {\bibinfo {author} {\bibfnamefont {D.}~\bibnamefont
  {Kayley}}\ and\ \bibinfo {author} {\bibfnamefont {B.}~\bibnamefont {Peng}},\
  }\href {\doibase 10.1016/j.commt.2025.100030} {\bibfield  {journal} {\bibinfo
   {journal} {Computational Materials Today}\ }\textbf {\bibinfo {volume}
  {6}},\ \bibinfo {pages} {100030} (\bibinfo {year} {2025})}\BibitemShut
  {NoStop}%
\bibitem [{\citenamefont {Peng}\ and\ \citenamefont
  {Pizzochero}(2025)}]{Peng2025}%
  \BibitemOpen
  \bibfield  {author} {\bibinfo {author} {\bibfnamefont {B.}~\bibnamefont
  {Peng}}\ and\ \bibinfo {author} {\bibfnamefont {M.}~\bibnamefont
  {Pizzochero}},\ }\href@noop {} {\bibfield  {journal} {\bibinfo  {journal}
  {arXiv:}\ ,\ \bibinfo {pages} {2504.07790}} (\bibinfo {year}
  {2025})}\BibitemShut {NoStop}%
\bibitem [{\citenamefont {Wang}\ \emph {et~al.}(2025)\citenamefont {Wang},
  \citenamefont {Ren}, \citenamefont {Qiu}, \citenamefont {Zhang},
  \citenamefont {Li}, \citenamefont {Gao}, \citenamefont {Gao},\ and\
  \citenamefont {Zhao}}]{Wang2025}%
  \BibitemOpen
  \bibfield  {author} {\bibinfo {author} {\bibfnamefont {X.}~\bibnamefont
  {Wang}}, \bibinfo {author} {\bibfnamefont {Y.}~\bibnamefont {Ren}}, \bibinfo
  {author} {\bibfnamefont {S.}~\bibnamefont {Qiu}}, \bibinfo {author}
  {\bibfnamefont {F.}~\bibnamefont {Zhang}}, \bibinfo {author} {\bibfnamefont
  {X.}~\bibnamefont {Li}}, \bibinfo {author} {\bibfnamefont {J.}~\bibnamefont
  {Gao}}, \bibinfo {author} {\bibfnamefont {W.}~\bibnamefont {Gao}}, \ and\
  \bibinfo {author} {\bibfnamefont {J.}~\bibnamefont {Zhao}},\ }\href {\doibase
  10.1038/s41524-024-01511-3} {\bibfield  {journal} {\bibinfo  {journal} {npj
  Computational Materials}\ }\textbf {\bibinfo {volume} {11}},\ \bibinfo
  {pages} {5} (\bibinfo {year} {2025})}\BibitemShut {NoStop}%
\bibitem [{\citenamefont {De\'{a}k}\ \emph {et~al.}(2011)\citenamefont
  {De\'{a}k}, \citenamefont {Aradi},\ and\ \citenamefont
  {Frauenheim}}]{Deak2011}%
  \BibitemOpen
  \bibfield  {author} {\bibinfo {author} {\bibfnamefont {P.}~\bibnamefont
  {De\'{a}k}}, \bibinfo {author} {\bibfnamefont {B.}~\bibnamefont {Aradi}}, \
  and\ \bibinfo {author} {\bibfnamefont {T.}~\bibnamefont {Frauenheim}},\
  }\href {\doibase 10.1021/jp1115492} {\bibfield  {journal} {\bibinfo
  {journal} {J. Phys. Chem. C}\ }\textbf {\bibinfo {volume} {115}},\ \bibinfo
  {pages} {3443} (\bibinfo {year} {2011})}\BibitemShut {NoStop}%
\bibitem [{\citenamefont {Pfeifer}\ \emph {et~al.}(2013)\citenamefont
  {Pfeifer}, \citenamefont {Erhart}, \citenamefont {Li}, \citenamefont
  {Rachut}, \citenamefont {Morasch}, \citenamefont {Br\"otz}, \citenamefont
  {Reckers}, \citenamefont {Mayer}, \citenamefont {R\"uhle}, \citenamefont
  {Zaban}, \citenamefont {Mora~Ser\'o}, \citenamefont {Bisquert}, \citenamefont
  {Jaegermann},\ and\ \citenamefont {Klein}}]{Pfeifer2013}%
  \BibitemOpen
  \bibfield  {author} {\bibinfo {author} {\bibfnamefont {V.}~\bibnamefont
  {Pfeifer}}, \bibinfo {author} {\bibfnamefont {P.}~\bibnamefont {Erhart}},
  \bibinfo {author} {\bibfnamefont {S.}~\bibnamefont {Li}}, \bibinfo {author}
  {\bibfnamefont {K.}~\bibnamefont {Rachut}}, \bibinfo {author} {\bibfnamefont
  {J.}~\bibnamefont {Morasch}}, \bibinfo {author} {\bibfnamefont
  {J.}~\bibnamefont {Br\"otz}}, \bibinfo {author} {\bibfnamefont
  {P.}~\bibnamefont {Reckers}}, \bibinfo {author} {\bibfnamefont
  {T.}~\bibnamefont {Mayer}}, \bibinfo {author} {\bibfnamefont
  {S.}~\bibnamefont {R\"uhle}}, \bibinfo {author} {\bibfnamefont
  {A.}~\bibnamefont {Zaban}}, \bibinfo {author} {\bibfnamefont
  {I.}~\bibnamefont {Mora~Ser\'o}}, \bibinfo {author} {\bibfnamefont
  {J.}~\bibnamefont {Bisquert}}, \bibinfo {author} {\bibfnamefont
  {W.}~\bibnamefont {Jaegermann}}, \ and\ \bibinfo {author} {\bibfnamefont
  {A.}~\bibnamefont {Klein}},\ }\href {\doibase 10.1021/jz402165b} {\bibfield
  {journal} {\bibinfo  {journal} {J. Phys. Chem. Lett.}\ }\textbf {\bibinfo
  {volume} {4}},\ \bibinfo {pages} {4182} (\bibinfo {year} {2013})}\BibitemShut
  {NoStop}%
\bibitem [{\citenamefont {Ju}\ \emph {et~al.}(2014)\citenamefont {Ju},
  \citenamefont {Sun}, \citenamefont {Wang}, \citenamefont {Meng},\ and\
  \citenamefont {Liang}}]{Ju2014}%
  \BibitemOpen
  \bibfield  {author} {\bibinfo {author} {\bibfnamefont {M.-G.}\ \bibnamefont
  {Ju}}, \bibinfo {author} {\bibfnamefont {G.}~\bibnamefont {Sun}}, \bibinfo
  {author} {\bibfnamefont {J.}~\bibnamefont {Wang}}, \bibinfo {author}
  {\bibfnamefont {Q.}~\bibnamefont {Meng}}, \ and\ \bibinfo {author}
  {\bibfnamefont {W.}~\bibnamefont {Liang}},\ }\href {\doibase
  10.1021/am502830m} {\bibfield  {journal} {\bibinfo  {journal} {ACS Appl.
  Mater. Interfaces}\ }\textbf {\bibinfo {volume} {6}},\ \bibinfo {pages}
  {12885} (\bibinfo {year} {2014})}\BibitemShut {NoStop}%
\bibitem [{\citenamefont {Mi}\ and\ \citenamefont {Weng}(2015)}]{Mi2015}%
  \BibitemOpen
  \bibfield  {author} {\bibinfo {author} {\bibfnamefont {Y.}~\bibnamefont
  {Mi}}\ and\ \bibinfo {author} {\bibfnamefont {Y.}~\bibnamefont {Weng}},\
  }\href {\doibase 10.1038/srep11482} {\bibfield  {journal} {\bibinfo
  {journal} {Sci. Rep.}\ }\textbf {\bibinfo {volume} {5}},\ \bibinfo {pages}
  {11482} (\bibinfo {year} {2015})}\BibitemShut {NoStop}%
\bibitem [{\citenamefont {Zhang}\ \emph
  {et~al.}(2015{\natexlab{b}})\citenamefont {Zhang}, \citenamefont {Yang},\
  and\ \citenamefont {Dong}}]{Zhang2015n}%
  \BibitemOpen
  \bibfield  {author} {\bibinfo {author} {\bibfnamefont {D.}~\bibnamefont
  {Zhang}}, \bibinfo {author} {\bibfnamefont {M.}~\bibnamefont {Yang}}, \ and\
  \bibinfo {author} {\bibfnamefont {S.}~\bibnamefont {Dong}},\ }\href {\doibase
  10.1039/C5CP04495B} {\bibfield  {journal} {\bibinfo  {journal} {Phys. Chem.
  Chem. Phys.}\ }\textbf {\bibinfo {volume} {17}},\ \bibinfo {pages} {29079}
  (\bibinfo {year} {2015}{\natexlab{b}})}\BibitemShut {NoStop}%
\bibitem [{\citenamefont {De\'{a}k}\ \emph {et~al.}(2016)\citenamefont
  {De\'{a}k}, \citenamefont {Kullgren}, \citenamefont {Aradi}, \citenamefont
  {Frauenheim},\ and\ \citenamefont {Kavan}}]{Deak2016}%
  \BibitemOpen
  \bibfield  {author} {\bibinfo {author} {\bibfnamefont {P.}~\bibnamefont
  {De\'{a}k}}, \bibinfo {author} {\bibfnamefont {J.}~\bibnamefont {Kullgren}},
  \bibinfo {author} {\bibfnamefont {B.}~\bibnamefont {Aradi}}, \bibinfo
  {author} {\bibfnamefont {T.}~\bibnamefont {Frauenheim}}, \ and\ \bibinfo
  {author} {\bibfnamefont {L.}~\bibnamefont {Kavan}},\ }\href {\doibase
  10.1016/j.electacta.2016.03.122} {\bibfield  {journal} {\bibinfo  {journal}
  {Electrochimica Acta}\ }\textbf {\bibinfo {volume} {199}},\ \bibinfo {pages}
  {27} (\bibinfo {year} {2016})}\BibitemShut {NoStop}%
\bibitem [{\citenamefont {Chiodo}\ \emph {et~al.}(2010)\citenamefont {Chiodo},
  \citenamefont {Garc\'{\i}a-Lastra}, \citenamefont {Iacomino}, \citenamefont
  {Ossicini}, \citenamefont {Zhao}, \citenamefont {Petek},\ and\ \citenamefont
  {Rubio}}]{Chiodo2010}%
  \BibitemOpen
  \bibfield  {author} {\bibinfo {author} {\bibfnamefont {L.}~\bibnamefont
  {Chiodo}}, \bibinfo {author} {\bibfnamefont {J.~M.}\ \bibnamefont
  {Garc\'{\i}a-Lastra}}, \bibinfo {author} {\bibfnamefont {A.}~\bibnamefont
  {Iacomino}}, \bibinfo {author} {\bibfnamefont {S.}~\bibnamefont {Ossicini}},
  \bibinfo {author} {\bibfnamefont {J.}~\bibnamefont {Zhao}}, \bibinfo {author}
  {\bibfnamefont {H.}~\bibnamefont {Petek}}, \ and\ \bibinfo {author}
  {\bibfnamefont {A.}~\bibnamefont {Rubio}},\ }\href {\doibase
  10.1103/PhysRevB.82.045207} {\bibfield  {journal} {\bibinfo  {journal} {Phys.
  Rev. B}\ }\textbf {\bibinfo {volume} {82}},\ \bibinfo {pages} {045207}
  (\bibinfo {year} {2010})}\BibitemShut {NoStop}%
\bibitem [{\citenamefont {Li}\ \emph {et~al.}(2020)\citenamefont {Li},
  \citenamefont {Wu},\ and\ \citenamefont {Gao}}]{Li2020a}%
  \BibitemOpen
  \bibfield  {author} {\bibinfo {author} {\bibfnamefont {B.}~\bibnamefont
  {Li}}, \bibinfo {author} {\bibfnamefont {S.}~\bibnamefont {Wu}}, \ and\
  \bibinfo {author} {\bibfnamefont {X.}~\bibnamefont {Gao}},\ }\href {\doibase
  10.1515/ntrev-2020-0085} {\bibfield  {journal} {\bibinfo  {journal}
  {Nanotechnology Reviews}\ }\textbf {\bibinfo {volume} {9}},\ \bibinfo {pages}
  {1080} (\bibinfo {year} {2020})}\BibitemShut {NoStop}%
\bibitem [{\citenamefont {Danilovic}\ \emph {et~al.}(2012)\citenamefont
  {Danilovic}, \citenamefont {Subbaraman}, \citenamefont {Strmcnik},
  \citenamefont {Chang}, \citenamefont {Paulikas}, \citenamefont
  {Stamenkovic},\ and\ \citenamefont {Markovic}}]{Danilovic2012}%
  \BibitemOpen
  \bibfield  {author} {\bibinfo {author} {\bibfnamefont {N.}~\bibnamefont
  {Danilovic}}, \bibinfo {author} {\bibfnamefont {R.}~\bibnamefont
  {Subbaraman}}, \bibinfo {author} {\bibfnamefont {D.}~\bibnamefont
  {Strmcnik}}, \bibinfo {author} {\bibfnamefont {K.-C.}\ \bibnamefont {Chang}},
  \bibinfo {author} {\bibfnamefont {A.~P.}\ \bibnamefont {Paulikas}}, \bibinfo
  {author} {\bibfnamefont {V.~R.}\ \bibnamefont {Stamenkovic}}, \ and\ \bibinfo
  {author} {\bibfnamefont {N.~M.}\ \bibnamefont {Markovic}},\ }\href {\doibase
  10.1002/anie.201204842} {\bibfield  {journal} {\bibinfo  {journal} {Angew.
  Chem. Int. Ed.}\ }\textbf {\bibinfo {volume} {51}},\ \bibinfo {pages} {12495}
  (\bibinfo {year} {2012})}\BibitemShut {NoStop}%
\bibitem [{\citenamefont {Shirley}\ and\ \citenamefont
  {Louie}(1993)}]{Shirley1993}%
  \BibitemOpen
  \bibfield  {author} {\bibinfo {author} {\bibfnamefont {E.~L.}\ \bibnamefont
  {Shirley}}\ and\ \bibinfo {author} {\bibfnamefont {S.~G.}\ \bibnamefont
  {Louie}},\ }\href {\doibase 10.1103/PhysRevLett.71.133} {\bibfield  {journal}
  {\bibinfo  {journal} {Phys. Rev. Lett.}\ }\textbf {\bibinfo {volume} {71}},\
  \bibinfo {pages} {133} (\bibinfo {year} {1993})}\BibitemShut {NoStop}%
\bibitem [{\citenamefont {Shirley}\ \emph {et~al.}(1996)\citenamefont
  {Shirley}, \citenamefont {Benedict},\ and\ \citenamefont
  {Louie}}]{Shirley1996}%
  \BibitemOpen
  \bibfield  {author} {\bibinfo {author} {\bibfnamefont {E.~L.}\ \bibnamefont
  {Shirley}}, \bibinfo {author} {\bibfnamefont {L.~X.}\ \bibnamefont
  {Benedict}}, \ and\ \bibinfo {author} {\bibfnamefont {S.~G.}\ \bibnamefont
  {Louie}},\ }\href {\doibase 10.1103/PhysRevB.54.10970} {\bibfield  {journal}
  {\bibinfo  {journal} {Phys. Rev. B}\ }\textbf {\bibinfo {volume} {54}},\
  \bibinfo {pages} {10970} (\bibinfo {year} {1996})}\BibitemShut {NoStop}%
\bibitem [{\citenamefont {Weaver}\ \emph {et~al.}(1991)\citenamefont {Weaver},
  \citenamefont {Martins}, \citenamefont {Komeda}, \citenamefont {Chen},
  \citenamefont {Ohno}, \citenamefont {Kroll}, \citenamefont {Troullier},
  \citenamefont {Haufler},\ and\ \citenamefont {Smalley}}]{Weaver1991}%
  \BibitemOpen
  \bibfield  {author} {\bibinfo {author} {\bibfnamefont {J.~H.}\ \bibnamefont
  {Weaver}}, \bibinfo {author} {\bibfnamefont {J.~L.}\ \bibnamefont {Martins}},
  \bibinfo {author} {\bibfnamefont {T.}~\bibnamefont {Komeda}}, \bibinfo
  {author} {\bibfnamefont {Y.}~\bibnamefont {Chen}}, \bibinfo {author}
  {\bibfnamefont {T.~R.}\ \bibnamefont {Ohno}}, \bibinfo {author}
  {\bibfnamefont {G.~H.}\ \bibnamefont {Kroll}}, \bibinfo {author}
  {\bibfnamefont {N.}~\bibnamefont {Troullier}}, \bibinfo {author}
  {\bibfnamefont {R.~E.}\ \bibnamefont {Haufler}}, \ and\ \bibinfo {author}
  {\bibfnamefont {R.~E.}\ \bibnamefont {Smalley}},\ }\href {\doibase
  10.1103/PhysRevLett.66.1741} {\bibfield  {journal} {\bibinfo  {journal}
  {Phys. Rev. Lett.}\ }\textbf {\bibinfo {volume} {66}},\ \bibinfo {pages}
  {1741} (\bibinfo {year} {1991})}\BibitemShut {NoStop}%
\bibitem [{\citenamefont {Lof}\ \emph {et~al.}(1992)\citenamefont {Lof},
  \citenamefont {van Veenendaal}, \citenamefont {Koopmans}, \citenamefont
  {Jonkman},\ and\ \citenamefont {Sawatzky}}]{Lof1992}%
  \BibitemOpen
  \bibfield  {author} {\bibinfo {author} {\bibfnamefont {R.~W.}\ \bibnamefont
  {Lof}}, \bibinfo {author} {\bibfnamefont {M.~A.}\ \bibnamefont {van
  Veenendaal}}, \bibinfo {author} {\bibfnamefont {B.}~\bibnamefont {Koopmans}},
  \bibinfo {author} {\bibfnamefont {H.~T.}\ \bibnamefont {Jonkman}}, \ and\
  \bibinfo {author} {\bibfnamefont {G.~A.}\ \bibnamefont {Sawatzky}},\ }\href
  {\doibase 10.1103/PhysRevLett.68.3924} {\bibfield  {journal} {\bibinfo
  {journal} {Phys. Rev. Lett.}\ }\textbf {\bibinfo {volume} {68}},\ \bibinfo
  {pages} {3924} (\bibinfo {year} {1992})}\BibitemShut {NoStop}%
\bibitem [{\citenamefont {Lof}\ \emph {et~al.}(1995)\citenamefont {Lof},
  \citenamefont {van Veenendaal}, \citenamefont {Jonkman},\ and\ \citenamefont
  {Sawatzky}}]{Lof1995}%
  \BibitemOpen
  \bibfield  {author} {\bibinfo {author} {\bibfnamefont {R.}~\bibnamefont
  {Lof}}, \bibinfo {author} {\bibfnamefont {M.}~\bibnamefont {van Veenendaal}},
  \bibinfo {author} {\bibfnamefont {H.}~\bibnamefont {Jonkman}}, \ and\
  \bibinfo {author} {\bibfnamefont {G.}~\bibnamefont {Sawatzky}},\ }\href
  {\doibase 10.1016/0368-2048(94)02308-5} {\bibfield  {journal} {\bibinfo
  {journal} {Journal of Electron Spectroscopy and Related Phenomena}\ }\textbf
  {\bibinfo {volume} {72}},\ \bibinfo {pages} {83} (\bibinfo {year}
  {1995})}\BibitemShut {NoStop}%
\bibitem [{\citenamefont {Schwedhelm}\ \emph {et~al.}(1998)\citenamefont
  {Schwedhelm}, \citenamefont {Kipp}, \citenamefont {Dallmeyer},\ and\
  \citenamefont {Skibowski}}]{Schwedhelm1998}%
  \BibitemOpen
  \bibfield  {author} {\bibinfo {author} {\bibfnamefont {R.}~\bibnamefont
  {Schwedhelm}}, \bibinfo {author} {\bibfnamefont {L.}~\bibnamefont {Kipp}},
  \bibinfo {author} {\bibfnamefont {A.}~\bibnamefont {Dallmeyer}}, \ and\
  \bibinfo {author} {\bibfnamefont {M.}~\bibnamefont {Skibowski}},\ }\href
  {\doibase 10.1103/PhysRevB.58.13176} {\bibfield  {journal} {\bibinfo
  {journal} {Phys. Rev. B}\ }\textbf {\bibinfo {volume} {58}},\ \bibinfo
  {pages} {13176} (\bibinfo {year} {1998})}\BibitemShut {NoStop}%
\bibitem [{\citenamefont {Xie}\ \emph {et~al.}(2020)\citenamefont {Xie},
  \citenamefont {Peng}, \citenamefont {Bravi\'c}, \citenamefont {Yu},
  \citenamefont {Dong}, \citenamefont {Liang}, \citenamefont {Ou},
  \citenamefont {Monserrat},\ and\ \citenamefont {Zhang}}]{Xie2020}%
  \BibitemOpen
  \bibfield  {author} {\bibinfo {author} {\bibfnamefont {Y.}~\bibnamefont
  {Xie}}, \bibinfo {author} {\bibfnamefont {B.}~\bibnamefont {Peng}}, \bibinfo
  {author} {\bibfnamefont {I.}~\bibnamefont {Bravi\'c}}, \bibinfo {author}
  {\bibfnamefont {Y.}~\bibnamefont {Yu}}, \bibinfo {author} {\bibfnamefont
  {Y.}~\bibnamefont {Dong}}, \bibinfo {author} {\bibfnamefont {R.}~\bibnamefont
  {Liang}}, \bibinfo {author} {\bibfnamefont {Q.}~\bibnamefont {Ou}}, \bibinfo
  {author} {\bibfnamefont {B.}~\bibnamefont {Monserrat}}, \ and\ \bibinfo
  {author} {\bibfnamefont {S.}~\bibnamefont {Zhang}},\ }\href {\doibase
  10.1002/advs.202001698} {\bibfield  {journal} {\bibinfo  {journal} {Adv.
  Sci.}\ }\textbf {\bibinfo {volume} {7}},\ \bibinfo {pages} {2001698}
  (\bibinfo {year} {2020})}\BibitemShut {NoStop}%
\bibitem [{\citenamefont {Peng}\ \emph
  {et~al.}(2022{\natexlab{c}})\citenamefont {Peng}, \citenamefont {Bennett},
  \citenamefont {Bravi\'{c}},\ and\ \citenamefont {Monserrat}}]{Peng2022d}%
  \BibitemOpen
  \bibfield  {author} {\bibinfo {author} {\bibfnamefont {B.}~\bibnamefont
  {Peng}}, \bibinfo {author} {\bibfnamefont {D.}~\bibnamefont {Bennett}},
  \bibinfo {author} {\bibfnamefont {I.}~\bibnamefont {Bravi\'{c}}}, \ and\
  \bibinfo {author} {\bibfnamefont {B.}~\bibnamefont {Monserrat}},\ }\href
  {\doibase 10.1103/PhysRevMaterials.6.L082401} {\bibfield  {journal} {\bibinfo
   {journal} {Phys. Rev. Mater.}\ }\textbf {\bibinfo {volume} {6}},\ \bibinfo
  {pages} {L082401} (\bibinfo {year} {2022}{\natexlab{c}})}\BibitemShut
  {NoStop}%
\bibitem [{\citenamefont {Ehrenfreund}\ and\ \citenamefont
  {Foing}(1997)}]{Ehrenfreund1997}%
  \BibitemOpen
  \bibfield  {author} {\bibinfo {author} {\bibfnamefont {P.}~\bibnamefont
  {Ehrenfreund}}\ and\ \bibinfo {author} {\bibfnamefont {B.}~\bibnamefont
  {Foing}},\ }\href {\doibase 10.1016/S0273-1177(97)00350-5} {\bibfield
  {journal} {\bibinfo  {journal} {Advances in Space Research}\ }\textbf
  {\bibinfo {volume} {19}},\ \bibinfo {pages} {1033} (\bibinfo {year}
  {1997})}\BibitemShut {NoStop}%
\bibitem [{\citenamefont {Cami}\ \emph {et~al.}(2010)\citenamefont {Cami},
  \citenamefont {Bernard-Salas}, \citenamefont {Peeters},\ and\ \citenamefont
  {Malek}}]{Cami2010}%
  \BibitemOpen
  \bibfield  {author} {\bibinfo {author} {\bibfnamefont {J.}~\bibnamefont
  {Cami}}, \bibinfo {author} {\bibfnamefont {J.}~\bibnamefont {Bernard-Salas}},
  \bibinfo {author} {\bibfnamefont {E.}~\bibnamefont {Peeters}}, \ and\
  \bibinfo {author} {\bibfnamefont {S.~E.}\ \bibnamefont {Malek}},\ }\href
  {\doibase 10.1126/science.1192035} {\bibfield  {journal} {\bibinfo  {journal}
  {Science}\ }\textbf {\bibinfo {volume} {329}},\ \bibinfo {pages} {1180}
  (\bibinfo {year} {2010})}\BibitemShut {NoStop}%
\bibitem [{\citenamefont {Woods}(2020)}]{Woods2020}%
  \BibitemOpen
  \bibfield  {author} {\bibinfo {author} {\bibfnamefont {P.}~\bibnamefont
  {Woods}},\ }\href {\doibase 10.1038/s41550-020-1076-5} {\bibfield  {journal}
  {\bibinfo  {journal} {Nature Astronomy}\ }\textbf {\bibinfo {volume} {4}},\
  \bibinfo {pages} {299} (\bibinfo {year} {2020})}\BibitemShut {NoStop}%
\bibitem [{\citenamefont {Nakada}\ \emph {et~al.}(1996)\citenamefont {Nakada},
  \citenamefont {Fujita}, \citenamefont {Dresselhaus},\ and\ \citenamefont
  {Dresselhaus}}]{Nakada1996}%
  \BibitemOpen
  \bibfield  {author} {\bibinfo {author} {\bibfnamefont {K.}~\bibnamefont
  {Nakada}}, \bibinfo {author} {\bibfnamefont {M.}~\bibnamefont {Fujita}},
  \bibinfo {author} {\bibfnamefont {G.}~\bibnamefont {Dresselhaus}}, \ and\
  \bibinfo {author} {\bibfnamefont {M.~S.}\ \bibnamefont {Dresselhaus}},\
  }\href {\doibase 10.1103/PhysRevB.54.17954} {\bibfield  {journal} {\bibinfo
  {journal} {Phys. Rev. B}\ }\textbf {\bibinfo {volume} {54}},\ \bibinfo
  {pages} {17954} (\bibinfo {year} {1996})}\BibitemShut {NoStop}%
\bibitem [{\citenamefont {Yazyev}(2013)}]{Yazyev2013}%
  \BibitemOpen
  \bibfield  {author} {\bibinfo {author} {\bibfnamefont {O.~V.}\ \bibnamefont
  {Yazyev}},\ }\href {\doibase 10.1021/ar3001487} {\bibfield  {journal}
  {\bibinfo  {journal} {Acc. Chem. Res.}\ }\textbf {\bibinfo {volume} {46}},\
  \bibinfo {pages} {2319} (\bibinfo {year} {2013})}\BibitemShut {NoStop}%
\bibitem [{\citenamefont {Chen}\ \emph {et~al.}(2020)\citenamefont {Chen},
  \citenamefont {Narita},\ and\ \citenamefont {M{\"u}llen}}]{Chen2020b}%
  \BibitemOpen
  \bibfield  {author} {\bibinfo {author} {\bibfnamefont {Z.}~\bibnamefont
  {Chen}}, \bibinfo {author} {\bibfnamefont {A.}~\bibnamefont {Narita}}, \ and\
  \bibinfo {author} {\bibfnamefont {K.}~\bibnamefont {M{\"u}llen}},\ }\href
  {\doibase 10.1002/adma.202001893} {\bibfield  {journal} {\bibinfo  {journal}
  {Adv. Mater.}\ }\textbf {\bibinfo {volume} {32}},\ \bibinfo {pages} {2001893}
  (\bibinfo {year} {2020})}\BibitemShut {NoStop}%
\bibitem [{\citenamefont {Wang}\ \emph {et~al.}(2021)\citenamefont {Wang},
  \citenamefont {Wang}, \citenamefont {Ma}, \citenamefont {Chen}, \citenamefont
  {Jiang}, \citenamefont {Chen}, \citenamefont {Xie}, \citenamefont {Li},\ and\
  \citenamefont {Wang}}]{Wang2021c}%
  \BibitemOpen
  \bibfield  {author} {\bibinfo {author} {\bibfnamefont {H.}~\bibnamefont
  {Wang}}, \bibinfo {author} {\bibfnamefont {H.~S.}\ \bibnamefont {Wang}},
  \bibinfo {author} {\bibfnamefont {C.}~\bibnamefont {Ma}}, \bibinfo {author}
  {\bibfnamefont {L.}~\bibnamefont {Chen}}, \bibinfo {author} {\bibfnamefont
  {C.}~\bibnamefont {Jiang}}, \bibinfo {author} {\bibfnamefont
  {C.}~\bibnamefont {Chen}}, \bibinfo {author} {\bibfnamefont {X.}~\bibnamefont
  {Xie}}, \bibinfo {author} {\bibfnamefont {A.-P.}\ \bibnamefont {Li}}, \ and\
  \bibinfo {author} {\bibfnamefont {X.}~\bibnamefont {Wang}},\ }\href {\doibase
  10.1038/s42254-021-00370-x} {\bibfield  {journal} {\bibinfo  {journal}
  {Nature Reviews Physics}\ }\textbf {\bibinfo {volume} {3}},\ \bibinfo {pages}
  {791} (\bibinfo {year} {2021})}\BibitemShut {NoStop}%
\bibitem [{\citenamefont {Son}\ \emph {et~al.}(2006{\natexlab{a}})\citenamefont
  {Son}, \citenamefont {Cohen},\ and\ \citenamefont {Louie}}]{Son2006}%
  \BibitemOpen
  \bibfield  {author} {\bibinfo {author} {\bibfnamefont {Y.-W.}\ \bibnamefont
  {Son}}, \bibinfo {author} {\bibfnamefont {M.~L.}\ \bibnamefont {Cohen}}, \
  and\ \bibinfo {author} {\bibfnamefont {S.~G.}\ \bibnamefont {Louie}},\ }\href
  {\doibase 10.1103/PhysRevLett.97.216803} {\bibfield  {journal} {\bibinfo
  {journal} {Phys. Rev. Lett.}\ }\textbf {\bibinfo {volume} {97}},\ \bibinfo
  {pages} {216803} (\bibinfo {year} {2006}{\natexlab{a}})}\BibitemShut
  {NoStop}%
\bibitem [{\citenamefont {Chen}\ \emph {et~al.}(2015)\citenamefont {Chen},
  \citenamefont {Cao}, \citenamefont {Chen}, \citenamefont {Pedramrazi},
  \citenamefont {Haberer}, \citenamefont {de~Oteyza}, \citenamefont {Fischer},
  \citenamefont {Louie},\ and\ \citenamefont {Crommie}}]{Chen2015b}%
  \BibitemOpen
  \bibfield  {author} {\bibinfo {author} {\bibfnamefont {Y.-C.}\ \bibnamefont
  {Chen}}, \bibinfo {author} {\bibfnamefont {T.}~\bibnamefont {Cao}}, \bibinfo
  {author} {\bibfnamefont {C.}~\bibnamefont {Chen}}, \bibinfo {author}
  {\bibfnamefont {Z.}~\bibnamefont {Pedramrazi}}, \bibinfo {author}
  {\bibfnamefont {D.}~\bibnamefont {Haberer}}, \bibinfo {author} {\bibfnamefont
  {D.~G.}\ \bibnamefont {de~Oteyza}}, \bibinfo {author} {\bibfnamefont {F.~R.}\
  \bibnamefont {Fischer}}, \bibinfo {author} {\bibfnamefont {S.~G.}\
  \bibnamefont {Louie}}, \ and\ \bibinfo {author} {\bibfnamefont {M.~F.}\
  \bibnamefont {Crommie}},\ }\href {\doibase 10.1038/nnano.2014.307} {\bibfield
   {journal} {\bibinfo  {journal} {Nature Nanotechnology}\ }\textbf {\bibinfo
  {volume} {10}},\ \bibinfo {pages} {156} (\bibinfo {year} {2015})}\BibitemShut
  {NoStop}%
\bibitem [{\citenamefont {Nguyen}\ \emph {et~al.}(2017)\citenamefont {Nguyen},
  \citenamefont {Tsai}, \citenamefont {Omrani}, \citenamefont {Marangoni},
  \citenamefont {Wu}, \citenamefont {Rizzo}, \citenamefont {Rodgers},
  \citenamefont {Cloke}, \citenamefont {Durr}, \citenamefont {Sakai},
  \citenamefont {Liou}, \citenamefont {Aikawa}, \citenamefont {Chelikowsky},
  \citenamefont {Louie}, \citenamefont {Fischer},\ and\ \citenamefont
  {Crommie}}]{Nguyen2017}%
  \BibitemOpen
  \bibfield  {author} {\bibinfo {author} {\bibfnamefont {G.~D.}\ \bibnamefont
  {Nguyen}}, \bibinfo {author} {\bibfnamefont {H.-Z.}\ \bibnamefont {Tsai}},
  \bibinfo {author} {\bibfnamefont {A.~A.}\ \bibnamefont {Omrani}}, \bibinfo
  {author} {\bibfnamefont {T.}~\bibnamefont {Marangoni}}, \bibinfo {author}
  {\bibfnamefont {M.}~\bibnamefont {Wu}}, \bibinfo {author} {\bibfnamefont
  {D.~J.}\ \bibnamefont {Rizzo}}, \bibinfo {author} {\bibfnamefont {G.~F.}\
  \bibnamefont {Rodgers}}, \bibinfo {author} {\bibfnamefont {R.~R.}\
  \bibnamefont {Cloke}}, \bibinfo {author} {\bibfnamefont {R.~A.}\ \bibnamefont
  {Durr}}, \bibinfo {author} {\bibfnamefont {Y.}~\bibnamefont {Sakai}},
  \bibinfo {author} {\bibfnamefont {F.}~\bibnamefont {Liou}}, \bibinfo {author}
  {\bibfnamefont {A.~S.}\ \bibnamefont {Aikawa}}, \bibinfo {author}
  {\bibfnamefont {J.~R.}\ \bibnamefont {Chelikowsky}}, \bibinfo {author}
  {\bibfnamefont {S.~G.}\ \bibnamefont {Louie}}, \bibinfo {author}
  {\bibfnamefont {F.~R.}\ \bibnamefont {Fischer}}, \ and\ \bibinfo {author}
  {\bibfnamefont {M.~F.}\ \bibnamefont {Crommie}},\ }\href {\doibase
  10.1038/nnano.2017.155} {\bibfield  {journal} {\bibinfo  {journal} {Nature
  Nanotechnology}\ }\textbf {\bibinfo {volume} {12}},\ \bibinfo {pages} {1077}
  (\bibinfo {year} {2017})}\BibitemShut {NoStop}%
\bibitem [{\citenamefont {{\v{C}}er{\c{n}}evi{\v{c}}s}\ \emph
  {et~al.}(2020)\citenamefont {{\v{C}}er{\c{n}}evi{\v{c}}s}, \citenamefont
  {Yazyev},\ and\ \citenamefont {Pizzochero}}]{Cernevics2020}%
  \BibitemOpen
  \bibfield  {author} {\bibinfo {author} {\bibfnamefont {K.}~\bibnamefont
  {{\v{C}}er{\c{n}}evi{\v{c}}s}}, \bibinfo {author} {\bibfnamefont {O.~V.}\
  \bibnamefont {Yazyev}}, \ and\ \bibinfo {author} {\bibfnamefont
  {M.}~\bibnamefont {Pizzochero}},\ }\href
  {https://link.aps.org/doi/10.1103/PhysRevB.102.201406} {\bibfield  {journal}
  {\bibinfo  {journal} {Phys. Rev. B}\ }\textbf {\bibinfo {volume} {102}},\
  \bibinfo {pages} {201406} (\bibinfo {year} {2020})}\BibitemShut {NoStop}%
\bibitem [{\citenamefont {Raza}\ and\ \citenamefont {Kan}(2008)}]{Raza2008}%
  \BibitemOpen
  \bibfield  {author} {\bibinfo {author} {\bibfnamefont {H.}~\bibnamefont
  {Raza}}\ and\ \bibinfo {author} {\bibfnamefont {E.~C.}\ \bibnamefont {Kan}},\
  }\href {\doibase 10.1103/PhysRevB.77.245434} {\bibfield  {journal} {\bibinfo
  {journal} {Phys. Rev. B}\ }\textbf {\bibinfo {volume} {77}},\ \bibinfo
  {pages} {245434} (\bibinfo {year} {2008})}\BibitemShut {NoStop}%
\bibitem [{\citenamefont {Pizzochero}\ \emph {et~al.}(2021)\citenamefont
  {Pizzochero}, \citenamefont {Tepliakov}, \citenamefont {Mostofi},\ and\
  \citenamefont {Kaxiras}}]{Pizzochero2021b}%
  \BibitemOpen
  \bibfield  {author} {\bibinfo {author} {\bibfnamefont {M.}~\bibnamefont
  {Pizzochero}}, \bibinfo {author} {\bibfnamefont {N.~V.}\ \bibnamefont
  {Tepliakov}}, \bibinfo {author} {\bibfnamefont {A.~A.}\ \bibnamefont
  {Mostofi}}, \ and\ \bibinfo {author} {\bibfnamefont {E.}~\bibnamefont
  {Kaxiras}},\ }\href {\doibase 10.1021/acs.nanolett.1c03596} {\bibfield
  {journal} {\bibinfo  {journal} {Nano Lett.}\ }\textbf {\bibinfo {volume}
  {21}},\ \bibinfo {pages} {9332} (\bibinfo {year} {2021})}\BibitemShut
  {NoStop}%
\bibitem [{\citenamefont {Son}\ \emph {et~al.}(2006{\natexlab{b}})\citenamefont
  {Son}, \citenamefont {Cohen},\ and\ \citenamefont {Louie}}]{Son2006a}%
  \BibitemOpen
  \bibfield  {author} {\bibinfo {author} {\bibfnamefont {Y.-W.}\ \bibnamefont
  {Son}}, \bibinfo {author} {\bibfnamefont {M.~L.}\ \bibnamefont {Cohen}}, \
  and\ \bibinfo {author} {\bibfnamefont {S.~G.}\ \bibnamefont {Louie}},\ }\href
  {\doibase 10.1038/nature05180} {\bibfield  {journal} {\bibinfo  {journal}
  {Nature}\ }\textbf {\bibinfo {volume} {444}},\ \bibinfo {pages} {347}
  (\bibinfo {year} {2006}{\natexlab{b}})}\BibitemShut {NoStop}%
\bibitem [{\citenamefont {Pizzochero}\ and\ \citenamefont
  {Kaxiras}(2022)}]{Pizzochero2022}%
  \BibitemOpen
  \bibfield  {author} {\bibinfo {author} {\bibfnamefont {M.}~\bibnamefont
  {Pizzochero}}\ and\ \bibinfo {author} {\bibfnamefont {E.}~\bibnamefont
  {Kaxiras}},\ }\href {\doibase 10.1021/acs.nanolett.1c04362} {\bibfield
  {journal} {\bibinfo  {journal} {Nano Lett.}\ }\textbf {\bibinfo {volume}
  {22}},\ \bibinfo {pages} {1922} (\bibinfo {year} {2022})}\BibitemShut
  {NoStop}%
\bibitem [{\citenamefont {Yazyev}(2010)}]{Yazyev2010}%
  \BibitemOpen
  \bibfield  {author} {\bibinfo {author} {\bibfnamefont {O.~V.}\ \bibnamefont
  {Yazyev}},\ }\href {\doibase 10.1088/0034-4885/73/5/056501} {\bibfield
  {journal} {\bibinfo  {journal} {Rep. Prog. Phys.}\ }\textbf {\bibinfo
  {volume} {73}},\ \bibinfo {pages} {056501} (\bibinfo {year}
  {2010})}\BibitemShut {NoStop}%
\bibitem [{\citenamefont {Ma}\ \emph {et~al.}(2025)\citenamefont {Ma},
  \citenamefont {Tepliakov}, \citenamefont {Mostofi},\ and\ \citenamefont
  {Pizzochero}}]{Ma2025}%
  \BibitemOpen
  \bibfield  {author} {\bibinfo {author} {\bibfnamefont {R.}~\bibnamefont
  {Ma}}, \bibinfo {author} {\bibfnamefont {N.~V.}\ \bibnamefont {Tepliakov}},
  \bibinfo {author} {\bibfnamefont {A.~A.}\ \bibnamefont {Mostofi}}, \ and\
  \bibinfo {author} {\bibfnamefont {M.}~\bibnamefont {Pizzochero}},\ }\href
  {\doibase 10.1021/acs.jpclett.5c00121} {\bibfield  {journal} {\bibinfo
  {journal} {J. Phys. Chem. Lett.}\ }\textbf {\bibinfo {volume} {16}},\
  \bibinfo {pages} {1680} (\bibinfo {year} {2025})}\BibitemShut {NoStop}%
\bibitem [{\citenamefont {Gr{\"o}ning}\ \emph {et~al.}(2018)\citenamefont
  {Gr{\"o}ning}, \citenamefont {Wang}, \citenamefont {Yao}, \citenamefont
  {Pignedoli}, \citenamefont {Borin~Barin}, \citenamefont {Daniels},
  \citenamefont {Cupo}, \citenamefont {Meunier}, \citenamefont {Feng},
  \citenamefont {Narita}, \citenamefont {M{\"u}llen}, \citenamefont
  {Ruffieux},\ and\ \citenamefont {Fasel}}]{Groning2018}%
  \BibitemOpen
  \bibfield  {author} {\bibinfo {author} {\bibfnamefont {O.}~\bibnamefont
  {Gr{\"o}ning}}, \bibinfo {author} {\bibfnamefont {S.}~\bibnamefont {Wang}},
  \bibinfo {author} {\bibfnamefont {X.}~\bibnamefont {Yao}}, \bibinfo {author}
  {\bibfnamefont {C.~A.}\ \bibnamefont {Pignedoli}}, \bibinfo {author}
  {\bibfnamefont {G.}~\bibnamefont {Borin~Barin}}, \bibinfo {author}
  {\bibfnamefont {C.}~\bibnamefont {Daniels}}, \bibinfo {author} {\bibfnamefont
  {A.}~\bibnamefont {Cupo}}, \bibinfo {author} {\bibfnamefont {V.}~\bibnamefont
  {Meunier}}, \bibinfo {author} {\bibfnamefont {X.}~\bibnamefont {Feng}},
  \bibinfo {author} {\bibfnamefont {A.}~\bibnamefont {Narita}}, \bibinfo
  {author} {\bibfnamefont {K.}~\bibnamefont {M{\"u}llen}}, \bibinfo {author}
  {\bibfnamefont {P.}~\bibnamefont {Ruffieux}}, \ and\ \bibinfo {author}
  {\bibfnamefont {R.}~\bibnamefont {Fasel}},\ }\href {\doibase
  10.1038/s41586-018-0375-9} {\bibfield  {journal} {\bibinfo  {journal}
  {Nature}\ }\textbf {\bibinfo {volume} {560}},\ \bibinfo {pages} {209}
  (\bibinfo {year} {2018})}\BibitemShut {NoStop}%
\bibitem [{\citenamefont {Rizzo}\ \emph {et~al.}(2018)\citenamefont {Rizzo},
  \citenamefont {Veber}, \citenamefont {Cao}, \citenamefont {Bronner},
  \citenamefont {Chen}, \citenamefont {Zhao}, \citenamefont {Rodriguez},
  \citenamefont {Louie}, \citenamefont {Crommie},\ and\ \citenamefont
  {Fischer}}]{Rizzo2018}%
  \BibitemOpen
  \bibfield  {author} {\bibinfo {author} {\bibfnamefont {D.~J.}\ \bibnamefont
  {Rizzo}}, \bibinfo {author} {\bibfnamefont {G.}~\bibnamefont {Veber}},
  \bibinfo {author} {\bibfnamefont {T.}~\bibnamefont {Cao}}, \bibinfo {author}
  {\bibfnamefont {C.}~\bibnamefont {Bronner}}, \bibinfo {author} {\bibfnamefont
  {T.}~\bibnamefont {Chen}}, \bibinfo {author} {\bibfnamefont {F.}~\bibnamefont
  {Zhao}}, \bibinfo {author} {\bibfnamefont {H.}~\bibnamefont {Rodriguez}},
  \bibinfo {author} {\bibfnamefont {S.~G.}\ \bibnamefont {Louie}}, \bibinfo
  {author} {\bibfnamefont {M.~F.}\ \bibnamefont {Crommie}}, \ and\ \bibinfo
  {author} {\bibfnamefont {F.~R.}\ \bibnamefont {Fischer}},\ }\href {\doibase
  10.1038/s41586-018-0376-8} {\bibfield  {journal} {\bibinfo  {journal}
  {Nature}\ }\textbf {\bibinfo {volume} {560}},\ \bibinfo {pages} {204}
  (\bibinfo {year} {2018})}\BibitemShut {NoStop}%
\bibitem [{\citenamefont {Tepliakov}\ \emph {et~al.}(2023)\citenamefont
  {Tepliakov}, \citenamefont {Lischner}, \citenamefont {Kaxiras}, \citenamefont
  {Mostofi},\ and\ \citenamefont {Pizzochero}}]{Tepliakov2023}%
  \BibitemOpen
  \bibfield  {author} {\bibinfo {author} {\bibfnamefont {N.~V.}\ \bibnamefont
  {Tepliakov}}, \bibinfo {author} {\bibfnamefont {J.}~\bibnamefont {Lischner}},
  \bibinfo {author} {\bibfnamefont {E.}~\bibnamefont {Kaxiras}}, \bibinfo
  {author} {\bibfnamefont {A.~A.}\ \bibnamefont {Mostofi}}, \ and\ \bibinfo
  {author} {\bibfnamefont {M.}~\bibnamefont {Pizzochero}},\ }\href {\doibase
  10.1103/PhysRevLett.130.026401} {\bibfield  {journal} {\bibinfo  {journal}
  {Phys. Rev. Lett.}\ }\textbf {\bibinfo {volume} {130}},\ \bibinfo {pages}
  {026401} (\bibinfo {year} {2023})}\BibitemShut {NoStop}%
\bibitem [{\citenamefont {Tong}\ \emph {et~al.}(2023)\citenamefont {Tong},
  \citenamefont {Liu}, \citenamefont {Dai},\ and\ \citenamefont
  {Jiang}}]{Tong2023}%
  \BibitemOpen
  \bibfield  {author} {\bibinfo {author} {\bibfnamefont {Y.}~\bibnamefont
  {Tong}}, \bibinfo {author} {\bibfnamefont {H.}~\bibnamefont {Liu}}, \bibinfo
  {author} {\bibfnamefont {S.}~\bibnamefont {Dai}}, \ and\ \bibinfo {author}
  {\bibfnamefont {D.-e.}\ \bibnamefont {Jiang}},\ }\href {\doibase
  10.1021/acs.nanolett.3c01946} {\bibfield  {journal} {\bibinfo  {journal}
  {Nano Lett.}\ }\textbf {\bibinfo {volume} {23}},\ \bibinfo {pages} {7470}
  (\bibinfo {year} {2023})}\BibitemShut {NoStop}%
\bibitem [{\citenamefont {Tong}\ \emph {et~al.}(2024)\citenamefont {Tong},
  \citenamefont {Liu}, \citenamefont {Mahurin}, \citenamefont {Dai},\ and\
  \citenamefont {Jiang}}]{Tong2024}%
  \BibitemOpen
  \bibfield  {author} {\bibinfo {author} {\bibfnamefont {Y.}~\bibnamefont
  {Tong}}, \bibinfo {author} {\bibfnamefont {H.}~\bibnamefont {Liu}}, \bibinfo
  {author} {\bibfnamefont {S.~M.}\ \bibnamefont {Mahurin}}, \bibinfo {author}
  {\bibfnamefont {S.}~\bibnamefont {Dai}}, \ and\ \bibinfo {author}
  {\bibfnamefont {D.-e.}\ \bibnamefont {Jiang}},\ }\href {\doibase
  10.1016/j.eml.2022.101929} {\bibfield  {journal} {\bibinfo  {journal}
  {Computational Materials Today}\ }\textbf {\bibinfo {volume} {4}},\ \bibinfo
  {pages} {100013} (\bibinfo {year} {2024})}\BibitemShut {NoStop}%
\bibitem [{\citenamefont {Lin}\ \emph {et~al.}(2021)\citenamefont {Lin},
  \citenamefont {Ong}, \citenamefont {Bange}, \citenamefont {Faria~Junior},
  \citenamefont {Peng}, \citenamefont {Ziegler}, \citenamefont {Zipfel},
  \citenamefont {B{\"a}uml}, \citenamefont {Paradiso}, \citenamefont
  {Watanabe}, \citenamefont {Taniguchi}, \citenamefont {Strunk}, \citenamefont
  {Monserrat}, \citenamefont {Fabian}, \citenamefont {Chernikov}, \citenamefont
  {Qiu}, \citenamefont {Louie},\ and\ \citenamefont {Lupton}}]{Lin2021}%
  \BibitemOpen
  \bibfield  {author} {\bibinfo {author} {\bibfnamefont {K.-Q.}\ \bibnamefont
  {Lin}}, \bibinfo {author} {\bibfnamefont {C.~S.}\ \bibnamefont {Ong}},
  \bibinfo {author} {\bibfnamefont {S.}~\bibnamefont {Bange}}, \bibinfo
  {author} {\bibfnamefont {P.~E.}\ \bibnamefont {Faria~Junior}}, \bibinfo
  {author} {\bibfnamefont {B.}~\bibnamefont {Peng}}, \bibinfo {author}
  {\bibfnamefont {J.~D.}\ \bibnamefont {Ziegler}}, \bibinfo {author}
  {\bibfnamefont {J.}~\bibnamefont {Zipfel}}, \bibinfo {author} {\bibfnamefont
  {C.}~\bibnamefont {B{\"a}uml}}, \bibinfo {author} {\bibfnamefont
  {N.}~\bibnamefont {Paradiso}}, \bibinfo {author} {\bibfnamefont
  {K.}~\bibnamefont {Watanabe}}, \bibinfo {author} {\bibfnamefont
  {T.}~\bibnamefont {Taniguchi}}, \bibinfo {author} {\bibfnamefont
  {C.}~\bibnamefont {Strunk}}, \bibinfo {author} {\bibfnamefont
  {B.}~\bibnamefont {Monserrat}}, \bibinfo {author} {\bibfnamefont
  {J.}~\bibnamefont {Fabian}}, \bibinfo {author} {\bibfnamefont
  {A.}~\bibnamefont {Chernikov}}, \bibinfo {author} {\bibfnamefont {D.~Y.}\
  \bibnamefont {Qiu}}, \bibinfo {author} {\bibfnamefont {S.~G.}\ \bibnamefont
  {Louie}}, \ and\ \bibinfo {author} {\bibfnamefont {J.~M.}\ \bibnamefont
  {Lupton}},\ }\href {\doibase 10.1038/s41467-021-25499-2} {\bibfield
  {journal} {\bibinfo  {journal} {Nature Communications}\ }\textbf {\bibinfo
  {volume} {12}},\ \bibinfo {pages} {5500} (\bibinfo {year}
  {2021})}\BibitemShut {NoStop}%
\end{thebibliography}

%

\end{document}